 \documentclass[proof]{pasj00}
 \SetRunningHead{R. Yamada et al.}{PAH 3.3 $\mu$m emission from galaxies with AKARI}
 \Received{2013 February 28}
 \Accepted{2013 June 10}
 \Published{2013 October 25}         
 \begin{document}
 \title{A relation of the PAH 3.3 $\mu$m feature with star-forming activity for galaxies with a wide range of infrared luminosity} 
\author{Rika \textsc{Yamada}, Shinki \textsc{Oyabu}, Hidehiro \textsc{Kaneda}, Mitsuyoshi \textsc{Yamagishi}, \\Daisuke \textsc{Ishihara} }
\affil{Graduate school of science, Nagoya University, Nagoya, Aichi 464-8602}
\email{ryamada@u.phys.nagoya-u.ac.jp}
\and
\author{Ji H. \textsc{Kim}, Myungshin \textsc{Im}}
\affil{Department of Physics and Astronomy, Seoul National University, Seoul 151-747, Korea}
 \maketitle

\begin{abstract}

For star-forming galaxies, we investigate a global relation between polycyclic aromatic hydrocarbon (PAH) emission luminosity at 3.3 $\mu$m, $L_{\mathrm{PAH3.3}}$, and infrared ($8-1000$ $\mu$m) luminosity, $L_{\mathrm{IR}}$, to understand how the PAH 3.3 $\mu$m feature relates to the star formation activity.
With AKARI, we performed near-infrared (2.5$-$5 $\mu$m) spectroscopy of 184 galaxies which have $L_{\mathrm{IR}} \sim10^8 -10^{13}\:\mathrm{L_{\odot}}$.
We classify the samples into infrared galaxies (IRGs; $L_{\mathrm{IR}}<10^{11}\: \mathrm{L_{\odot}}$), luminous infrared galaxies (LIRGs; $L_{\mathrm{IR}} \sim10^{11} -10^{12}\:\mathrm{L_{\odot}}$) and ultra luminous infrared galaxies (ULIRGs; $L_{\mathrm{IR}}>10^{12}\:\mathrm{L_{\odot}}$).
We exclude sources which are likely contaminated by AGN activity, based on the rest-frame equivalent width of the PAH emission feature ($< 40$ nm) and the power-law index representing the slope of continuum emission (${\bf \it{\Gamma}}> 1$; $ F_{\nu}\propto \lambda^{\it{\Gamma}}$). 
Of these samples, 13 IRGs, 67 LIRGs and 20 ULIRGs show PAH emission feature at  $\lambda_{\mathrm{rest}}=$ 3.3 $\mu$m in their spectra.
We find that the $L_{\mathrm{PAH3.3}}/L_{\mathrm{IR}}$ ratio considerably decreases toward the luminous end.
Utilizing the mass and temperature of dust grains as well as the Br$\alpha$ emission for the galaxies, we discuss the cause of the relative decrease in the PAH emission with $L_{\mathrm{IR}}$.

\end{abstract}

\section{INTRODUCTION}

Revealing the star formation history in the universe is one of the most important topics in astrophysics.
For this purpose, accurate measurements of star-formation activity in galaxies are indispensable. 
One popular method is detecting infrared (IR) emission from large dust grains, because dust grains surround star-forming regions, absorb ultraviolet (UV) photons from young massive stars and re-radiate photons in the IR.
Using the method, the Spitzer, AKARI and Herschel satellites have revealed a great deal of IR emission in the universe, showing strong evolution in the IR luminosity density (e.g., \cite{lf 2005, go 2010, ro 2010, ca 2012}).
However, we need to consider the contribution of active galactic nuclei (AGN) to the IR luminosity \citep{h 2003, sc 2006}. 
It is not easy to distinguish the energy sources only by using photometric data in the IR.

Polycyclic aromatic hydrocarbon (PAH) emission features are proposed to be another excellent indicator of star formation activity (e.g., \cite{p 2004, b 2006}). 
PAHs are found to be present in a wide range of objects and the environments.
They are illuminated by UV photons mostly from early-type stars in star forming regions, while they are destroyed by hard radiation from an AGN central engine \citep{v 1992}.
Recent studies with Spitzer revealed that the inter-band ratios of the PAH features at wavelengths of $6-17$ $\mu$m reflect the physical conditions of PAHs such as
ionization state and size distribution (e.g., \cite{ka 2005}, \yearcite{ka 2008}; \cite{sm 2007, g 2008}). 
These studies were performed in the mid-IR bands, such as the 6.2, 7.7, 8.6, 11.3, 12.7 and 17 $\mu$m emission features, while Spitzer did not observe the 3.3 $\mu$m emission feature from galaxies except for high-redshift ones ($z\geq 0.6$) \citep{si 2009}.
Several earlier studies showed that the PAH 3.3 $\mu$m emission feature is prominent in starburst, IR luminous galaxies and obscured AGNs associated with star formation activities \citep{moo 1986, i 2000, rv 2003}.
The 3.3 $\mu$m emission feature is attributed to very small PAHs which have small heat capacities.
Therefore they are easily excited by UV photons or destroyed.
In other words, the PAH 3.3 $\mu$m luminosity, $L_{\mathrm{PAH3.3}}$, seems to reflect the radiation environment more sensitively than the other PAH emission features at longer wavelengths.

AKARI has a slit-less spectroscopic capability in the wavelength range of 2.5 to 5 $\mu$m with a 1\arcmin$\times$1\arcmin~aperture \citep{mu 2007}. 
The capability is crucial to make accurate measurements of the total fluxes of galaxies. 
Using the capability, systematic studies of the PAH 3.3 $\mu$m feature on ultra-luminous infrared galaxies (ULIRGs) and luminous infrared galaxies (LIRGs) were performed by \authorcite{i 2008} (\yearcite{i 2008}, \yearcite{i 2010}). 
\citet{w 2012} detected the PAH 3.3 $\mu$m emission feature from 7 out of 25 intermediate-luminosity type I AGN at z$\sim$0.4 with AKARI. 
\citet{k 2012} pointed out that $L_{\mathrm{PAH3.3}}$ as a proxy for the IR luminosity is hampered at $L_{\mathrm{PAH3.3}}$ $\gtrsim 10^{42}\ \mathrm{erg}\ \mathrm{s}^{-1}$ comparing AKARI near-IR spectra for 20 subsamples of the 5 mJy unbiased Spitzer extragalactic survey with other samples (\cite{rv 2003}; \cite{i 2008}, \yearcite{i 2010}; \cite{sa 2009, lee 2012}).  
These studies treat only high-luminosity samples with the IR luminosity of $\gtrsim 10^{11}\ \mathrm{L}_{\odot}$. 
We investigate a global relation between $L_{\mathrm{PAH3.3}}$ and the IR (8-1000 $\mu$m) luminosity, $L_{\mathrm{IR}}$, derived from IRAS \citep{s 1996}, using a sample which covers a wider luminosity range of $L_{\mathrm{IR}}=10^8 -10^{13}\:\mathrm{L_{\odot}}$, in order to understand how the PAH 3.3 $\mu$m feature relates to the star formation activity.
Throughout this paper, we assume that the universe is flat with $\Omega_{\mathrm{M}}=0.3$, $\Omega_{\mathrm{\Lambda}}=0.7$, and $H_0=70\ \mathrm{km}\ \mathrm{s}^{-1}\ \mathrm{Mpc}^{-1}$ in calculating a distance from a redshift.

\section{Sample selection}

We selected our targets from the two AKARI mission programs, Mid-infrared Search for Active Galactic Nuclei (MSAGN; \cite{o 2011}) and Evolution of ULIRGs and AGNs (AGNUL).
The MSAGN program performed near-IR spectroscopy of 79 galaxies discovered in the AKARI/Infrared Camera (IRC) all-sky survey point source catalog \citep{is 2010}.
In this search, the targets were selected with $F$(9 $\mu$m)/$F$($K_{\mathrm{S}}$) $>2$ as mid-IR-excess sources at high galactic latitudes ($|b| > 30^{\circ}$), where $F$(9 $\mu$m) and $F$($K_{\mathrm{S}}$) are the flux densities at 9 $\mu$m and $K_{\mathrm{S}}$-band taken with AKARI and 2MASS, respectively. 
In case that $F$(9 $\mu$m) was not available, we used a $F$(18 $\mu$m).
The criterion $F$(9 or 18 $\mu$m)/$F$($K_{\mathrm{S}}$) $>$ 2 excludes normal stars and quiescent galaxies from the sample.

The AGNUL program performed the systematic spectral study of ULIRGs and LIRGs in the local universe.
The spectra of all the sample galaxies have been reported by \authorcite{i 2008} (\yearcite{i 2008}, \yearcite{i 2010}).
From the AGNUL sample, we selected 105 galaxies as mid-IR excess galaxies with the same criterion, $F(9\:\mathrm{or}\:18\:\mu\mathrm{m})/F(K_{\mathrm{S}})>2$.
However, some of them are not listed in the point source catalog, because the current catalog contains only bright sources and they are too faint.
Therefore we checked the AKARI mid-IR diffuse map for those not listed in the catalog and performed their photometry to confirm that the criterion $F(9\: \mathrm{or}\: 18 \mu$m)/$F(K_{\mathrm{S}})>2$ was satisfied. 
Note that we exclude VV 114 E/W, Arp 299 (IC 694 + NGC 3690), NGC 6285/6 and NGC 7592 E/W from the AGNUL sample because they are double-nucleus sources and $L_{\mathrm{IR}}$ for each nucleus is not measured.

Both samples combined together, we obtain the AKARI near-IR spectra of the 184 mid-IR-excess galaxies which are classified into three populations of galaxies according to their $L_{\mathrm{IR}}$.
Eighteen galaxies are defined as IR galaxies (IRGs; $L_{\mathrm{IR}} < 10^{11}\: \mathrm{L_{\odot}}$), 89 LIRGs,  and 55 ULIRGs.
The other 22 galaxies have no information on $L_{\mathrm{IR}}$, because they are not detected with IRAS.

The spectroscopic data used in this study were taken by the IRC (\cite{on 2007}) on board AKARI.
All the observations except five were performed in the warm mission phase (Phase 3), while the five observations were performed in the cold mission phase (Phases 1 and 2).
The near-IR grism mode (NG) was adopted for these observations, providing a spectral coverage from 2.5 to 5 $\mu$m and a spectral resolution of $R$ $\sim$ 120 at 3.6 $\mu$m for a point source \citep{oh 2007}. 
With a 1\arcmin~$\times$ 1\arcmin~aperture used for these observations, IR emission from the entire galaxies in our sample is spatially covered. 
We assigned two or three pointings for each target to ensure data redundancy.

For spectral analyses, we used the data reduction package, ``IRC Spectroscopy Toolkit for Phase 3 data Version 20090211'' for Phase 3 data and ``IRC Spectroscopy Toolkit Version 20090211'' for Phases 1 and 2 data. 
Each frame was dark-subtracted and linearity-corrected.
After performing wavelength and flux calibrations, the spectra of the objects were extracted with the aperture of 7\mbox{$.\!\!^{\prime\prime}$}3 in radius.
Fourteen out of the 184 mid-IR-excess galaxies exhibit significantly extended emission, for which we enlarged an aperture size up to 14\mbox{$.\!\!^{\prime\prime}$}6, depending on the actual signal profile of each source.

\section{RESULTS}

We present the AKARI 2.5-5 $\mu$m spectra of the 79 MSAGN sample galaxies in figures \ref{1} and \ref{2}, which have not been reported in any papers except for two
galaxies\footnote{In \citet{o 2011}, J1442347+660604 and J0127539+284750 are reported as LEDA 84274 and IRAS 01250+2832, respectively.}.
In the MSAGN sample, we significantly detect the PAH 3.3 $\mu$m feature from 47 out of the 79 galaxies.
Figure \ref{1} shows the spectra of the 47 galaxies, for all of which aliphatic hydrocarbon emission features (sub-features) are also detected at $\lambda_{\mathrm{rest}}=3.4-3.5\:\mu$m. 
In addition to them, the Br$\alpha$ emission line at 4.05 $\mu$m and the $\mathrm{H_2 O}$ ice absorption feature around $\lambda_{\mathrm{rest}}=3.05\:\mu$m are detected in the 32 and 30 galaxies, respectively. 
Figure \ref{2} shows the spectra of the 32 galaxies with no detection of the PAH 3.3 $\mu$m emission feature.
They show a red continuum typical of AGNs. 
Tables 1 and 2 summarize the properties of the galaxies with and without detection of the PAH feature, respectively, which include the 2MASS coordinates, flux densities in the 9 $\mu$m and 18 $\mu$m bands and $K_{\mathrm{S}}$ magnitudes, and $L_{\mathrm{IR}}$.
As for the 105 AGNUL galaxies, our data reduction provide basically the same spectra as in \authorcite{i 2008} (\yearcite{i 2008}, \yearcite{i 2010}). 
We significantly detect the PAH 3.3 $\mu$m feature from 87 out of the 105 galaxies.

In order to confirm the detection of the lines and features and to measure their strengths, spectral model fitting is performed for the two spectral regions of each spectrum. 
One spectral region is for the PAH 3.3 $\mu$m emission feature and the other is for the Br$\alpha$ emission line.
The first spectral region is $\lambda_{\mathrm{rest}}=2.60$ to $3.98 \: \mu$m, containing the PAH 3.3 $\mu$m emission feature, sub-features and the $\mathrm{H_2 O}$ ice absorption feature as well as a continuum. 
In fitting a spectrum around $\lambda_{\mathrm{rest}}\sim 3.3$ $\mu$m, we use the Drude profile \citep{l 2001}:
\begin{equation}
    I_{\nu}=\frac{b_r {\it{\gamma}_r}^2}{(\lambda /\lambda_r -\lambda_r /\lambda )^2 +{\it{\gamma}_r}^2},
\end{equation} 
where $\lambda_r$ is the central wavelength of the feature, $\it{\gamma}_r$ is the full width at half maximum divided by $\lambda_r$, and $b_r$ is the flux density at the line center. 
The line flux of the Drude profile is given by
\begin{equation}
I \equiv \int I_{\nu} d\nu=\frac{\pi c}{2}\frac{b_r \gamma_r}{\lambda_r},
\end{equation}
where $c$ is a speed of light.
For the $\mathrm{H_2 O}$ ice absorption feature, we use a Gaussian profile with the 1 $\sigma$ width of 0.168 $\mu$m, which was used for the previous study of the $\mathrm{H_2 O}$ ice absorption feature in NGC 253 \citep{y 2011}.
The continuum slope $\it{\Gamma}$ is defined as $I_{\nu}\propto \lambda^{\it{\Gamma}}$.
The region of the sub-features ($\lambda_{\mathrm{rest}}=3.38 - 3.75 \: \mu$m) is not used for the fitting. 
The fitting parameters are $\lambda_r$, $\it{\gamma}_r$, $b_r$ and $\it{\Gamma}$ as well as the central wavelength and the absorption depth of the $\mathrm{H_2 O}$ ice feature. 
The second spectral region is $3.8$ to $4.8 \: \mu$m, containing the Br$\alpha$ emission line and a continuum.
We fit the Br$\alpha$ emission line and continuum with a Gaussian profile and a power-law function, respectively. 
The fitting parameters for the second region are the wavelength and the flux density at the line center for the Br$\alpha$ emission line, and a power-law slope of a continuum. 
The obtained central wavelengths are consistent with the redshifts in tables 1 and 2 within the error of 0.04 $\mu$m which are caused by a statistical and systematic errors in the source extraction.

The spectral properties derived for the IRGs, LIRGs, ULIRGs, and galaxies with no IRAS data are summarized in tables 3, 4, 5, and 6, respectively, which include $L_{\mathrm{IR}}$, $L_{\mathrm{PAH3.3}}$, the luminosities of the Br$\alpha$ emission lines, $L_{\mathrm{Br\alpha}}$, the optical depth of the $\mathrm{H_2 O}$ ice absorption feature, $\tau_{3.1}$, the rest-frame equivalent width of PAH 3.3 $\mu$m feature, $EW_{\mathrm{PAH3.3}}$, and continuum slopes ($\it{\Gamma}$) measured in the first spectral region. 
Below we do not use the data of the galaxies with no IRAS data.
We find that 132 of the 162 mid-IR-excess galaxies exhibit the PAH 3.3 $\mu$m feature with the signal-to-noise ratio higher than 5, which consist of 17 IRGs, 80 LIRGs, 35 ULIRGs. 
The difference between our fitting method and that of \authorcite{i 2008} (\yearcite{i 2008}, \yearcite{i 2010}) lies in modeling a continuum underlying the PAH feature profile.
In \authorcite{i 2008} (\yearcite{i 2008}, \yearcite{i 2010}), continuum levels are determined by a linear interpolation between the data point at $\lambda_{\mathrm{rest}}<2.75\: \mu$m and $\lambda_{\mathrm{rest}}>3.55\: \mu$m, and therefore the ice absorption feature likely affected their the measurements.
Our fitting method accounts for the H$_2$O ice absorption feature so that more accurate measurements may be performed for the PAH 3.3 $\mu$m feature.
As a result, $L_{\mathrm{PAH3.3}}$ values in the present paper are about 1 to 2.5 times larger than those in \authorcite{i 2008} (\yearcite{i 2008}, \yearcite{i 2010}) for the same galaxies, depending on the strength of the $\mathrm{H_2 O}$ ice absorption feature.

Figure \ref{3} shows a comparison of $L_{\mathrm{PAH3.3}}$ with $L_{\mathrm{IR}}$; $L_{\mathrm{PAH3.3}}$ is expected to correlate with $L_{\mathrm{IR}}$.
As can be seen in the figure, many galaxies follow the relationship $L_{\mathrm{PAH3.3}}/L_{\mathrm{IR}}=10^{-3}$, which is a ratio typical of starburst galaxies \citep{mou 1990, i 2002}. 
However, some of them significantly deviate from the relationship.
In order to study the relation between $L_{\mathrm{PAH3.3}}$ and $L_{\mathrm{IR}}$ for star-forming activity, we exclude sources likely contaminated by AGN activity, based on the equivalent width of the PAH emission feature ($EW_{\mathrm{PAH3.3}}< 40$ nm; \cite{moo 1986, i 2000, i 2008}) and the power-law index representing the slope of the continuum emission ($\it{\Gamma}$ $>1$; \cite{r 2006, i 2010}).
Figure \ref{4} plots $\it{\Gamma}$ and $EW_{\mathrm{PAH3.3}}$ for each galaxy, in which the galaxies in the bottom right region are categorized into galaxies with no AGN signatures.
Out of the 162 galaxies, 100 galaxies do not show the AGN signatures, while the other galaxies are considered to have signatures of AGN activities.
The spectral types of the samples are summarized in table 7.
We below discuss only the galaxies with no AGN signatures, which consist of 13 IRGs, 67 LIRGs and 20 ULIRGs.
They have redshifts $z\simeq0.01-0.3$.

Figure \ref{5} shows the relationship between $L_{\mathrm{PAH3.3}}/L_{\mathrm{IR}}$ and $L_{\mathrm{IR}}$.
At low $L_{\mathrm{IR}}$ ($10^8-10^{11.5}\:\mathrm{L_{\odot}}$), $L_{\mathrm{PAH3.3}}$ correlates with $L_{\mathrm{IR}}$, or $L_{\mathrm{PAH}3.3}/L_{\mathrm{IR}}\sim 10^{-3}$.
At higher $L_{\mathrm{IR}}$ ($\gtrsim 10^{11.5}\: \mathrm{L_{\odot}}$), however, $L_{\mathrm{PAH3.3}}/L_{\mathrm{IR}}$ shows a significant decline with $L_{\mathrm{IR}}$.
Although similar trends were already reported by \citet{k 2012}, our results show the trend more clearly because our sample covers a wider luminosity range of $L_{\mathrm{IR}}= 10^8$ to $10^{13}\: \mathrm{L_{\odot}}$, while their sample only covers $L_{\mathrm{IR}} \gtsim 10^{11} \: \mathrm{L_{\odot}}$. 
Thus we cannot utilize $L_{\mathrm{PAH3.3}}$ as indicators of star-formation rates for local ULIRGs.
Figure \ref{6} displays $L_{\mathrm{PAH3.3}}/L_{\mathrm{IR}}$ against the dust temperature and dust mass, which are obtained by the ratio of the IRAS 60 to 100 $\mu$m flux densities with the emissivity power-law index $\beta=1$, and summarized in the tables 3, 4 and 5 for the IRGs, LIRGs and ULIRGs, respectively. 
Figure \ref{6} shows globally decreasing trends in $L_{\mathrm{PAH3.3}}/L_{\mathrm{IR}}$ for both dust temperature and dust mass. 
Within each galaxy population, however, there is no significant correlation of $L_{\mathrm{PAH3.3}}/L_{\mathrm{IR}}$ with either dust temperature or dust mass, although $L_{\mathrm{PAH3.3}}/L_{\mathrm{IR}}$ systematically changes from population to population.
In particular, the LIRG and ULIRG samples span wide ranges of dust temperature, and yet do not show any systematic changes in $L_{\mathrm{PAH3.3}}/L_{\mathrm{IR}}$  with dust temperature.

 \section{DISCUSSION}

The decline of $L_{\mathrm{PAH3.3}}/L_{\mathrm{IR}}$ with $L_{\mathrm{IR}}$ (figure \ref{5}) can be caused by (1) photo-dissociation of PAHs under intense UV radiation due to star-formation activity, (2) dust extinction of the PAH 3.3 $\mu$m emission, (3) effects of hidden AGNs via hard radiation, (4) contribution of embedded YSOs on $L_{\mathrm{IR}}$, and (5) intrinsically low abundance ratios of PAHs to dust.
No correlation between $L_{\mathrm{PAH3.3}}/L_{\mathrm{IR}}$ and the dust temperature (figure 6) in each galaxy population indicates that the UV radiation is not a crucial parameter to explain the observed decline in $L_{\mathrm{PAH3.3}}/L_{\mathrm{IR}}$, because the dust temperature is determined mostly from the intensity of UV radiation. 
This implies that the photo-dissociation of PAHs due to high star-formation activity is not a major cause of the relatively weak PAH emission in the ULIRGs. $L_{\mathrm{PAH3.3}}/L_{\mathrm{IR}}$ can also decrease, if star-forming regions are heavily obscured by a copious amount of dust. In this case, the Br$\alpha$ emission should also be absorbed as well as the PAH 3.3 $\mu$m emission since they appear at similar wavelengths in the near-IR.
Figure \ref{7} shows $L_{\mathrm{Br\alpha}}/L_{\mathrm{IR}}$ plotted against $L_{\mathrm{IR}}$.
Comparing figures \ref{5} and \ref{7}, we find that $L_{\mathrm{Br\alpha}}/L_{\mathrm{IR}}$ does not significantly decrease with $L_{\mathrm{IR}}$, unlike $L_{\mathrm{PAH3.3}}/L_{\mathrm{IR}}$, which implies that the dust extinction is not a major cause, either.

Although we apply stringent criteria to select galaxies with no AGN signatures, a significant fraction of our sample might still contain hidden AGNs at the luminous end. \citet{k 2012} indicated that the deviation of correlation between $L_{\mathrm{PAH3.3}}$ and $L_{\mathrm{IR}}$ in their ULIRG sample is due to the influence of the hard radiation field from hidden AGNs, via contribution to $L_{\mathrm{IR}}$ and destruction of PAHs. Figure \ref{8} shows the relationship between $L_{\mathrm{PAH3.3}}/L_{\mathrm{IR}}$ and $L_{\mathrm{IR}}$ for the galaxies with and without AGN signatures. In fact, the IRGs and LIRGs with AGN signatures show systematically lower $L_{\mathrm{PAH3.3}}/L_{\mathrm{IR}}$ ratios than those with no AGN signatures. However, for the ULIRGs, there is no systematic difference between galaxies with and without AGN signatures; $L_{\mathrm{PAH3.3}}/L_{\mathrm{IR}}$ in our sample ULIRGs is as small as the ULIRGs with AGN signatures. This suggests that the hidden AGNs, if any, may not appreciably decrease $L_{\mathrm{PAH3.3}}/L_{\mathrm{IR}}$ for our sample ULIRGs. As also pointed out by \citet{k 2012}, soft radiation from heavily-embedded YSOs may contribute to increasing $L_{\mathrm{IR}}$, but not much to increasing $L_{\mathrm{PAH3.3}}$ because it is too soft to excite PAHs. In this case, it is likely that the Br$\alpha$ emission is relatively weak, similarly to the PAH 3.3 $\mu$m emission. Figure \ref{7}, however, does not show such a decline in $L_{\mathrm{Br\alpha}}/L_{\mathrm{IR}}$ toward the luminous end as observed in $L_{\mathrm{PAH3.3}}/L_{\mathrm{IR}}$. Therefore we can also rule out the possibility that the contribution of soft radiation field on $L_{\mathrm{IR}}$ is a major cause of the decline in $L_{\mathrm{PAH3.3}}/L_{\mathrm{IR}}$.

Consequently, we conclude that the intrinsically low abundance ratios of PAHs to dust are likely to cause the small $L_{\mathrm{PAH3.3}}/L_{\mathrm{IR}}$ in the population of the ULIRGs. The abundance ratios are systematically different among the IRGs, LIRGs, and ULIRGs. Figure \ref{9} displays $L_{\mathrm{PAH3.3}}/L_{\mathrm{IR}}$ plotted against $L_{\mathrm{PAH3.3}}$, revealing that each galaxy population has a different relationship between $L_{\mathrm{PAH3.3}}/L_{\mathrm{IR}}$ and $L_{\mathrm{PAH3.3}}$. The IRGs do not have correlation between $L_{\mathrm{PAH3.3}}/L_{\mathrm{IR}}$ and $L_{\mathrm{PAH3.3}}$ with the logarithmic correlation coefficient $r=0.46$ for a sample size of $n=13$, while the LIRGs show correlation ($r=0.56$ for $n=$67). The ULIRGs show a relatively tight correlation ($r=0.69$ for $n=20$). A plausible scenario to interpret such dependence of $L_{\mathrm{PAH3.3}}/L_{\mathrm{IR}}$ on the galaxy population is a merging process of galaxies, which is violent enough to destroy small PAHs. Many morphological observations of local ULIRGs show that they have recently experienced mergers (e.g., \cite{c 1996}). Hence some fraction of PAHs are likely destroyed once by a shock during a merging process, whereas large dust grains survive, which can explain the systematically low abundance ratios of PAHs to dust observed for the ULIRGs. The ratio will vary with the scale of the past merger, i.e., minor or major; a large-scale merging process may destroy more PAHs or difference in the stage of the current merging process.  Since the PAH 3.3 $\mu$m emission traces smallest PAHs, $L_{\mathrm{PAH3.3}}/L_{\mathrm{IR}}$ can sensitively reflect such processing of PAHs through evolution of galaxies. Considering a starburst age typical of local ULIRGs ($\sim 10-100$ Myr; \cite{ge 1998}) as well as a lifetime of intermediate-mass stars ($\sim 100-1000$ Myr) responsible for the production of PAHs at their late stages, it is unlikely that PAHs have already been reproduced and replenished by the stars that were newly born after the merger.

 \section{SUMMARY}

We have investigated a global relation between $L_{\mathrm{PAH3.3}}$ and $L_{\mathrm{IR}}$ using the results of the AKARI IRC $2.5-5$ $\mu$m spectroscopy of the 184 mid-IR-excess galaxies which have $L_{\mathrm{IR}} \sim10^8 -10^{13}\:\mathrm{L_{\odot}}$. We excluded the galaxies possessing AGN signatures with the criteria that $EW_{\mathrm{PAH3.3}}<40$ nm and $\it{\Gamma} \mathrm{> 1}$.
As a result, 13 IRGs, 67 LIRGs and 20 ULIRGs ($z\simeq0.01-0.3$) show the PAH 3.3 $\mu$m emission in their spectra. For such sample galaxies, we find that $L_{\mathrm{PAH3.3}}/L_{\mathrm{IR}}$ is almost constant around $10^{-3}$ in the range of $L_{\mathrm{IR}}\sim 10^8 - 10^{11.5}\: \mathrm{L_{\odot}}$, while $L_{\mathrm{PAH3.3}}/L_{\mathrm{IR}}$ decreases by about one order of magnitude toward the luminous end of $L_{\mathrm{IR}}$ above $L_{\mathrm{IR}}\sim 10^{11.5}\mathrm{L_{\odot}}$. Thus we cannot utilize $L_{\mathrm{PAH3.3}}$ as indicators of star-formation rates for local ULIRGs.

We consider possible causes of this decline as follows: (1) photo-dissociation of PAHs under intense UV radiation due to star-formation activity, (2) dust extinction of the PAH 3.3 $\mu$m emission, (3) effects of hidden AGNs via hard radiation, (4) contribution of embedded YSOs on $L_{\mathrm{IR}}$, and (5) intrinsically low abundance ratios of PAHs to dust. From our discussion based on the dependence of $L_{\mathrm{PAH3.3}}/L_{\mathrm{IR}}$ and $L_{\mathrm{Br\alpha}}/L_{\mathrm{IR}}$ on various parameters, we conclude that local ULIRGs intrinsically possesses smaller amounts of PAHs relative to dust grains, as a result of PAH processing through recent mergers. 
Some fraction of PAHs may have been destroyed once by a shock during a merging process, whereas large dust grains survive, which can explain the systematically low abundance ratios of PAHs to dust observed for the ULIRGs. 
Hence our result is consistent with the observational fact that local ULIRGs are merging galaxies.

 \bigskip
 We thank an anonymous referee for her/his careful reading our manuscript and giving useful comments. 
We would like to thank all the AKARI team members for their continuous efforts.
This work is based on the observations made with AKARI, a JAXA project, with the participation of ESA.
This research also makes use of data products from the Two Micron All Sky Survey, which is a joint project of the University of Massachusetts and the Infrared Processing and Analysis Center/California Institute of Technology, funded by the National Aeronautics and Space Administration and the National Science Foundation.

\begin{figure}
 \begin{center}
     \FigureFile(41mm,26mm){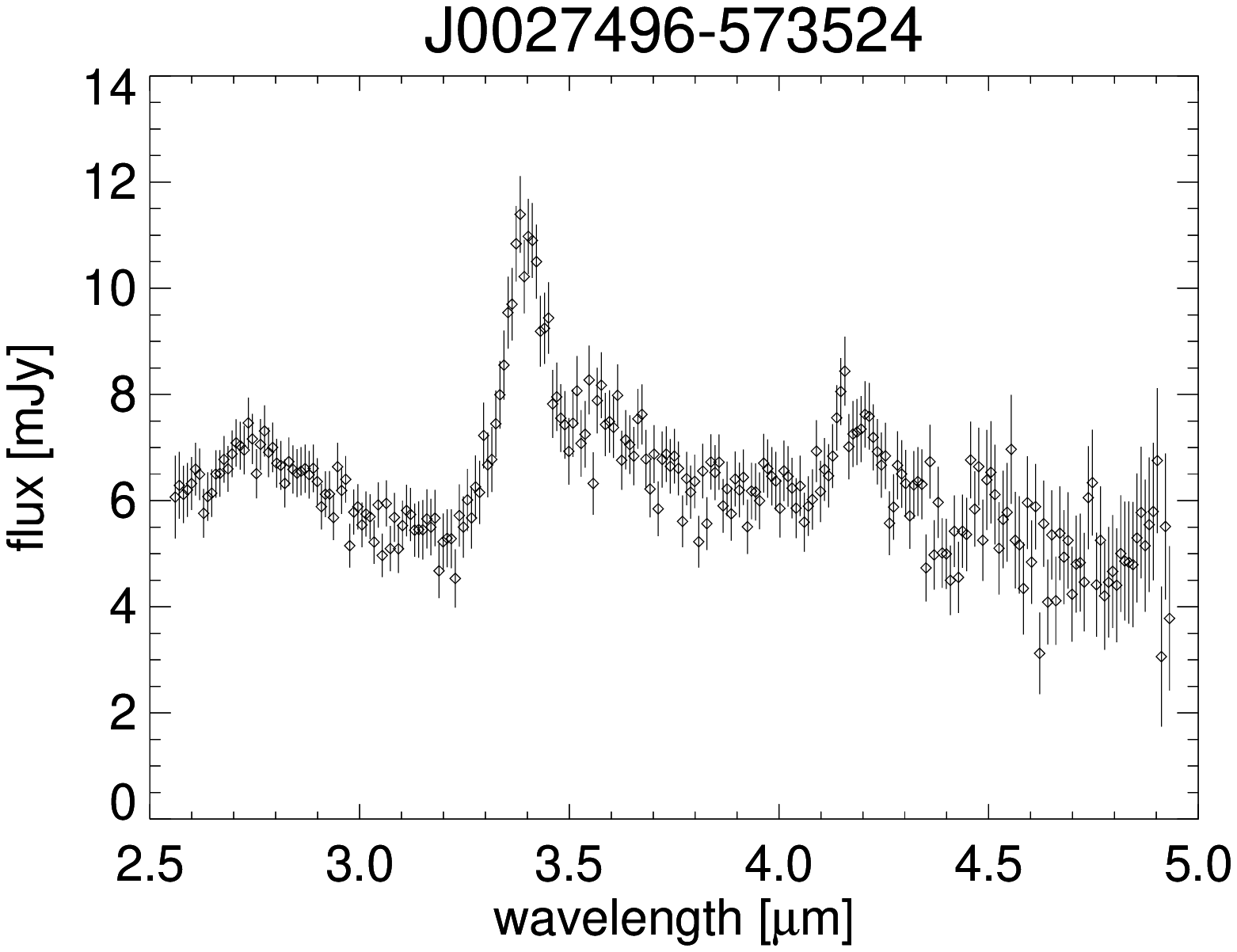}
     \FigureFile(41mm,26mm){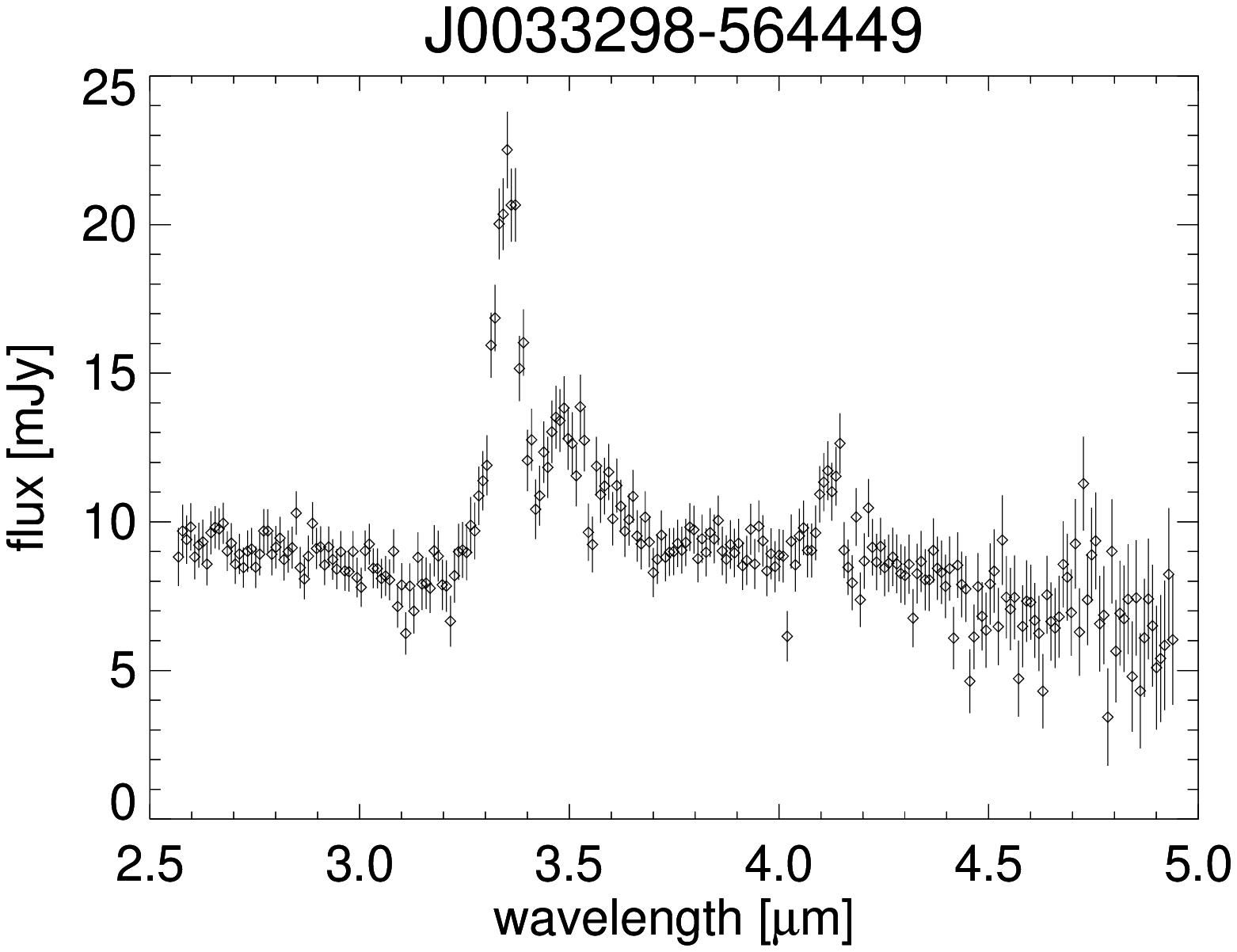}
     \FigureFile(41mm,26mm){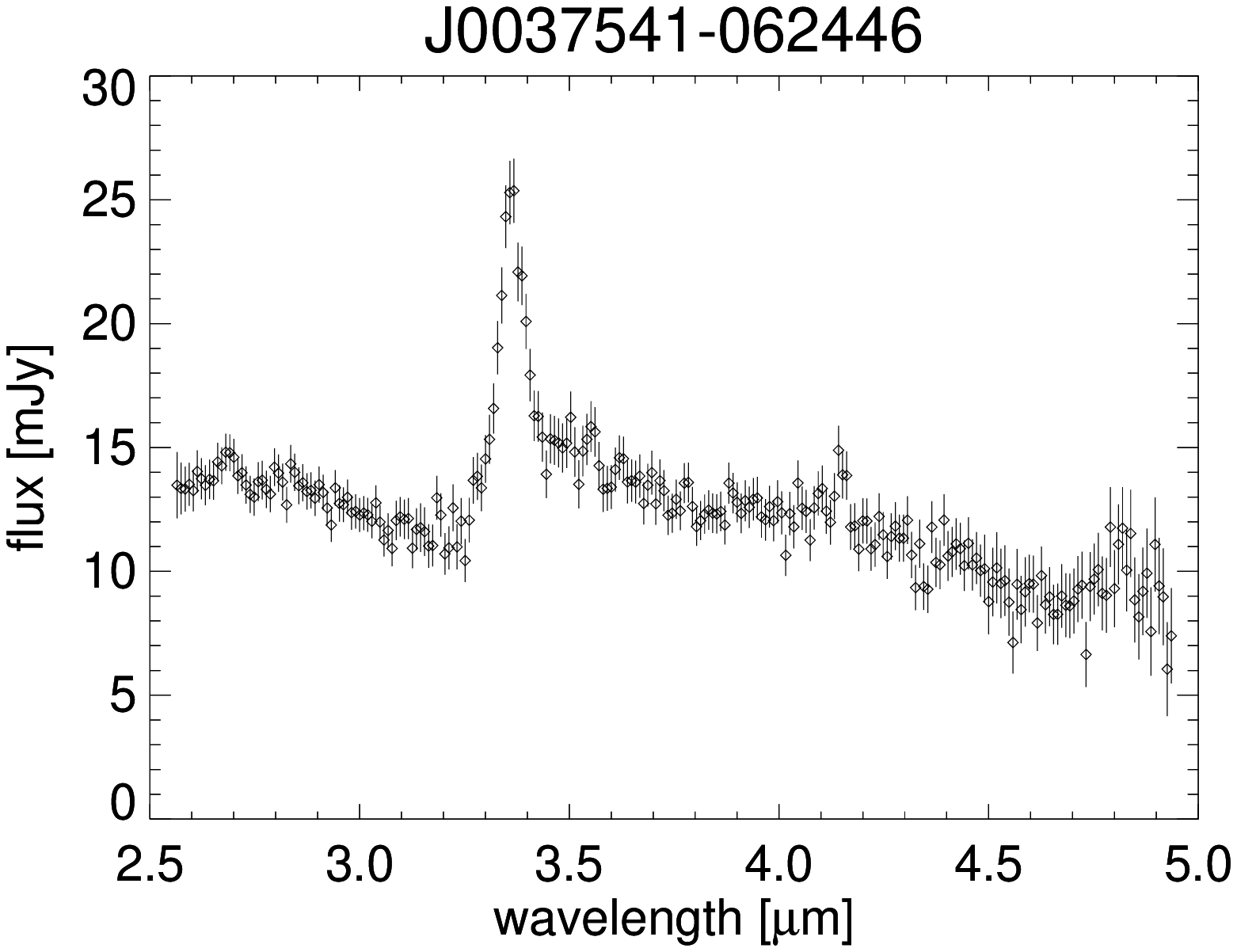}
     \FigureFile(41mm,26mm){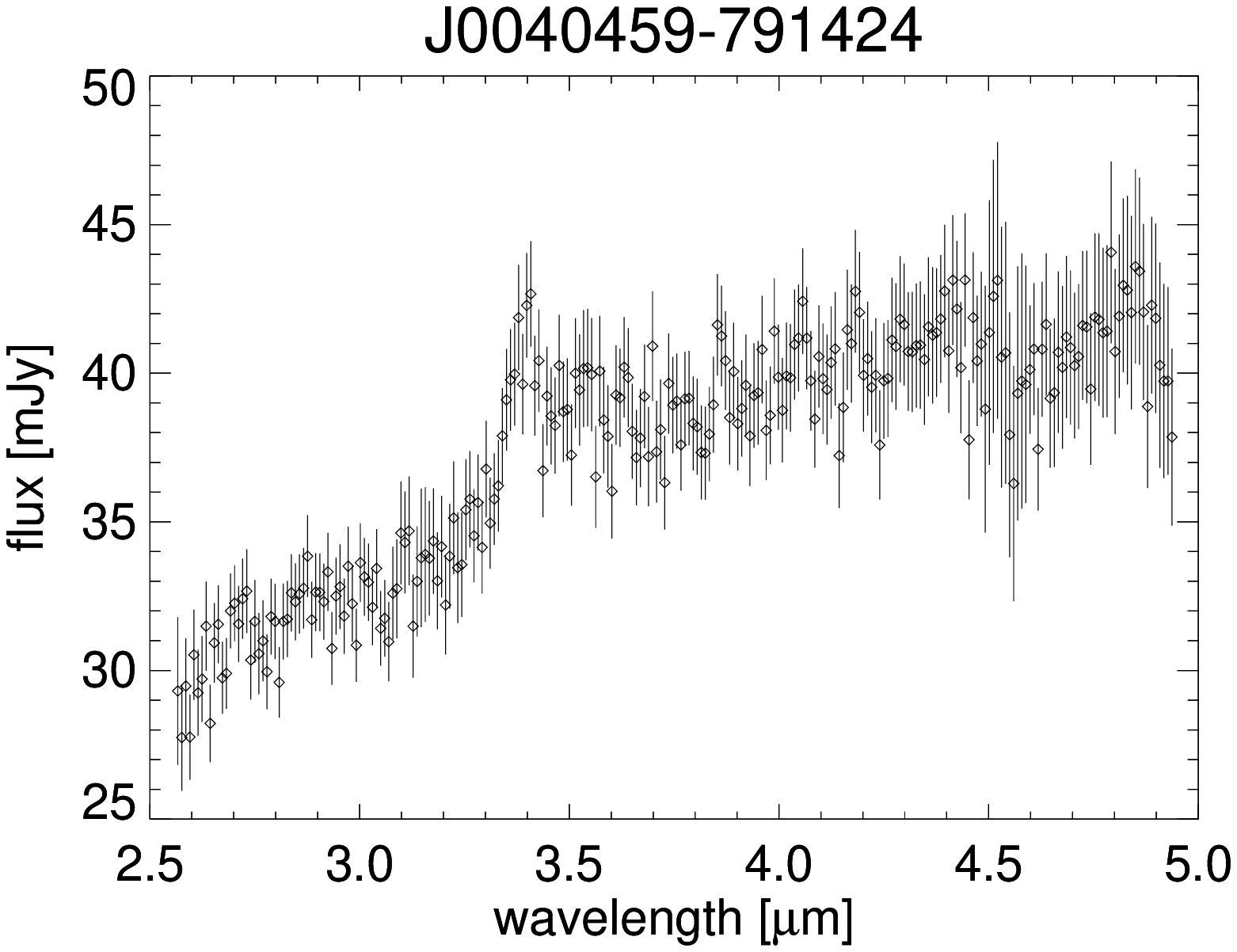}

     \FigureFile(41mm,26mm){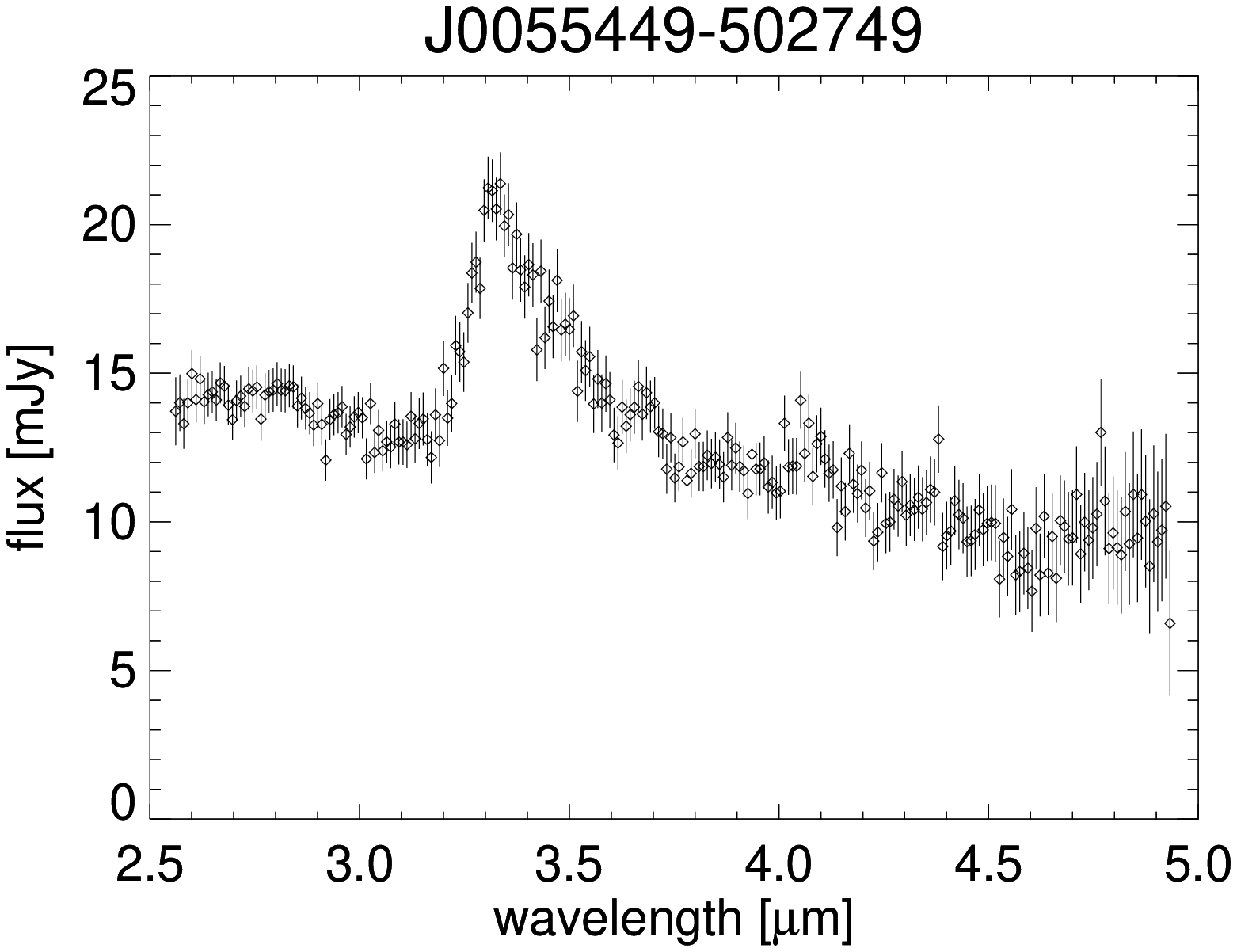}
     \FigureFile(41mm,26mm){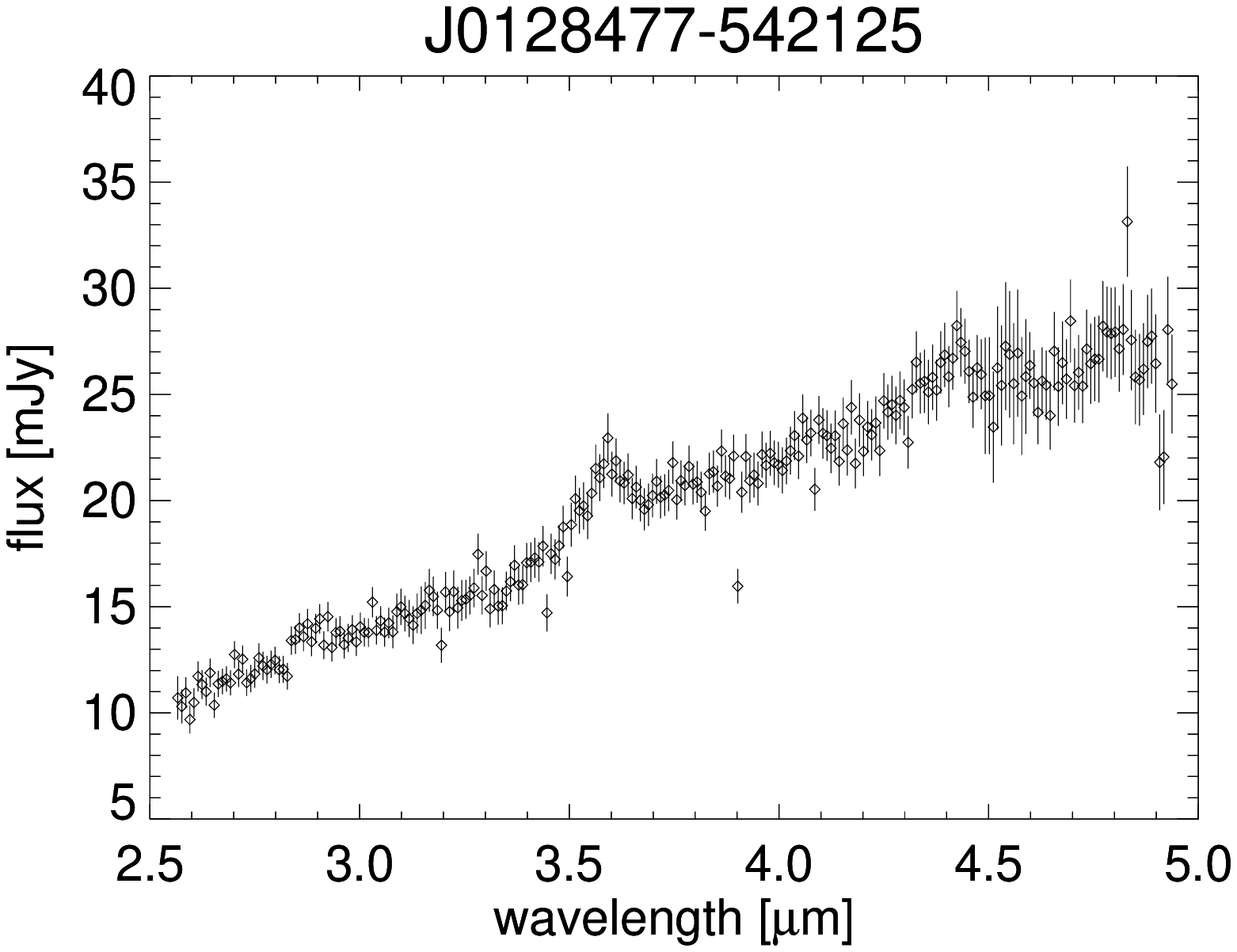}
     \FigureFile(41mm,26mm){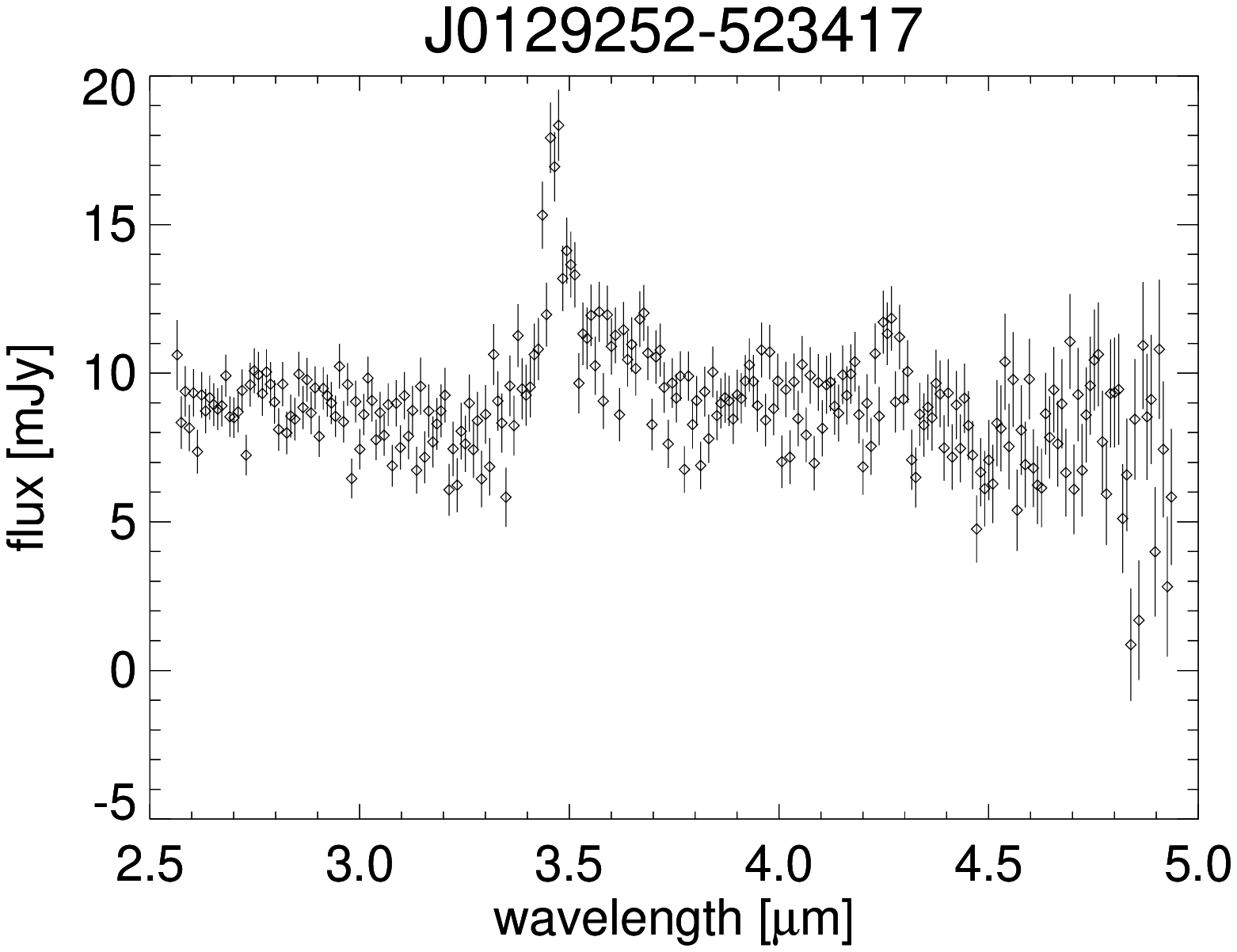}
     \FigureFile(41mm,26mm){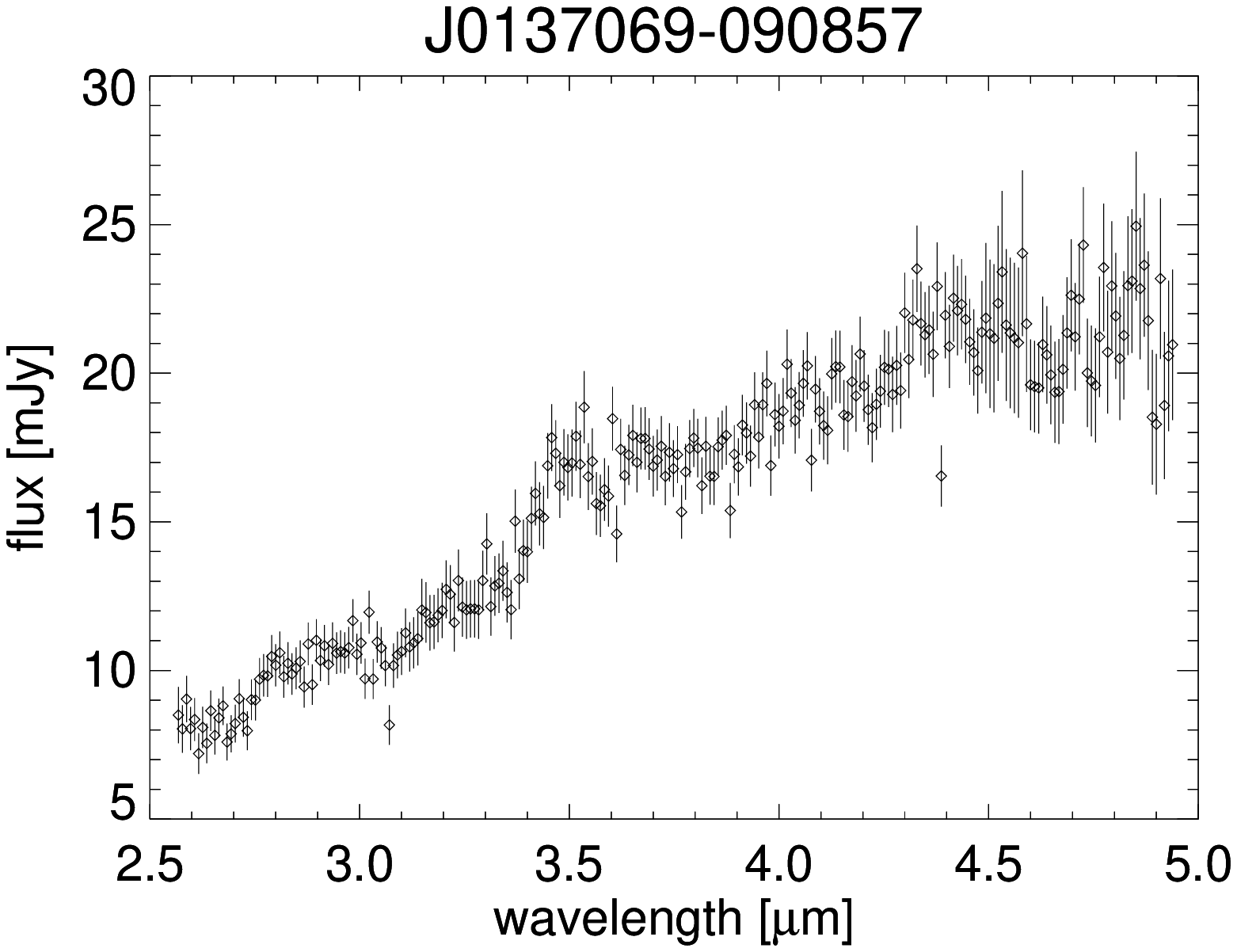}

     \FigureFile(41mm,26mm){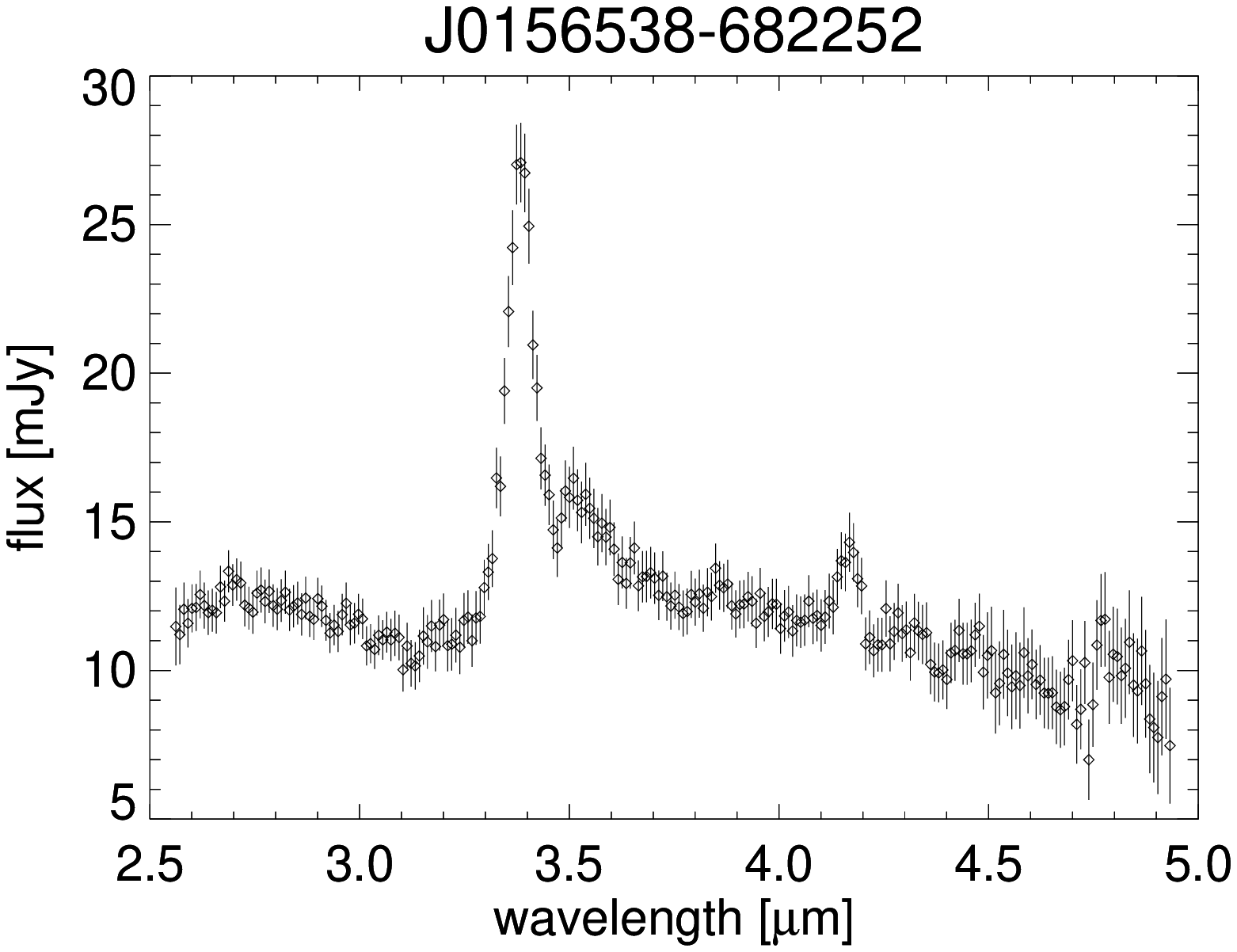}
     \FigureFile(41mm,26mm){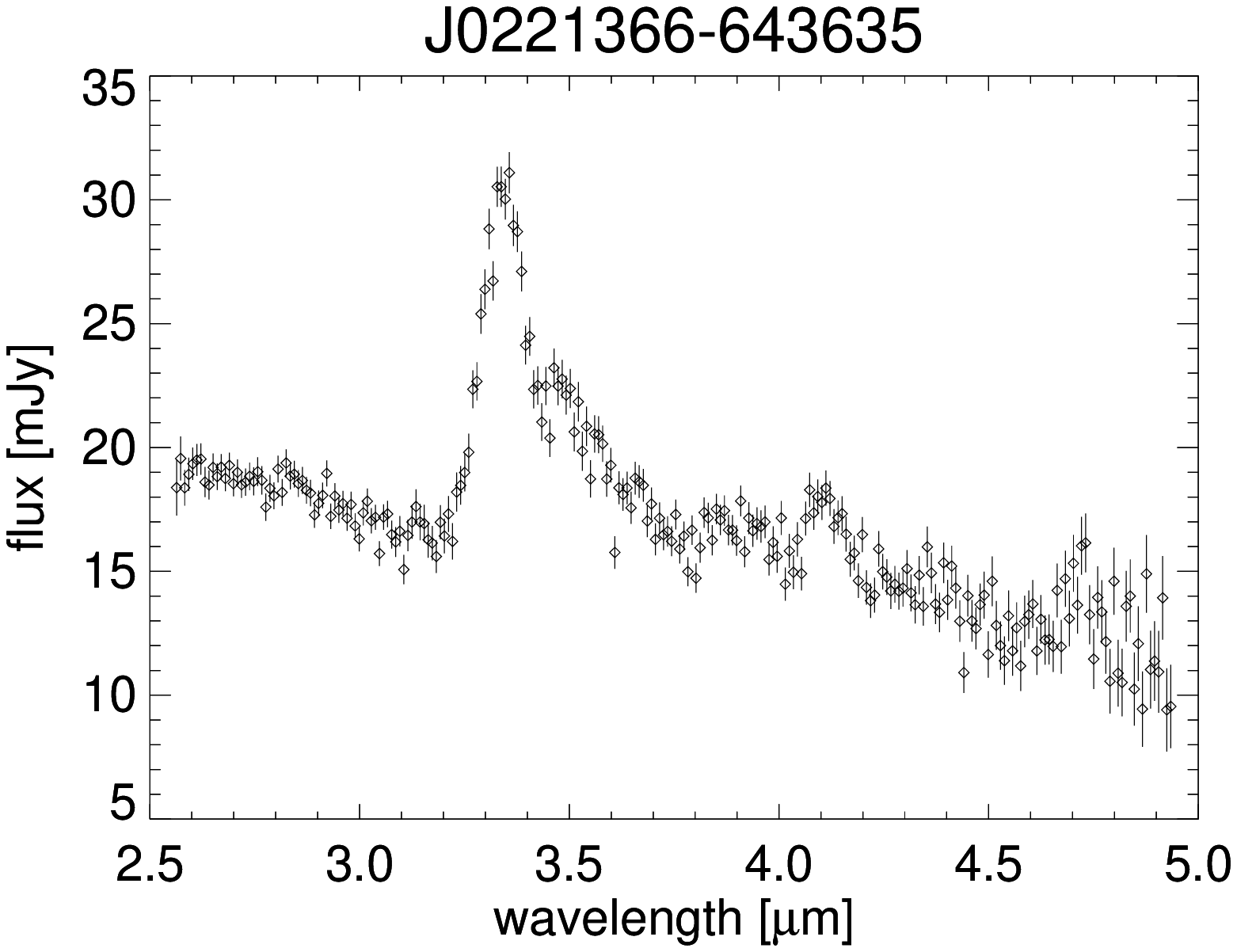}
     \FigureFile(41mm,26mm){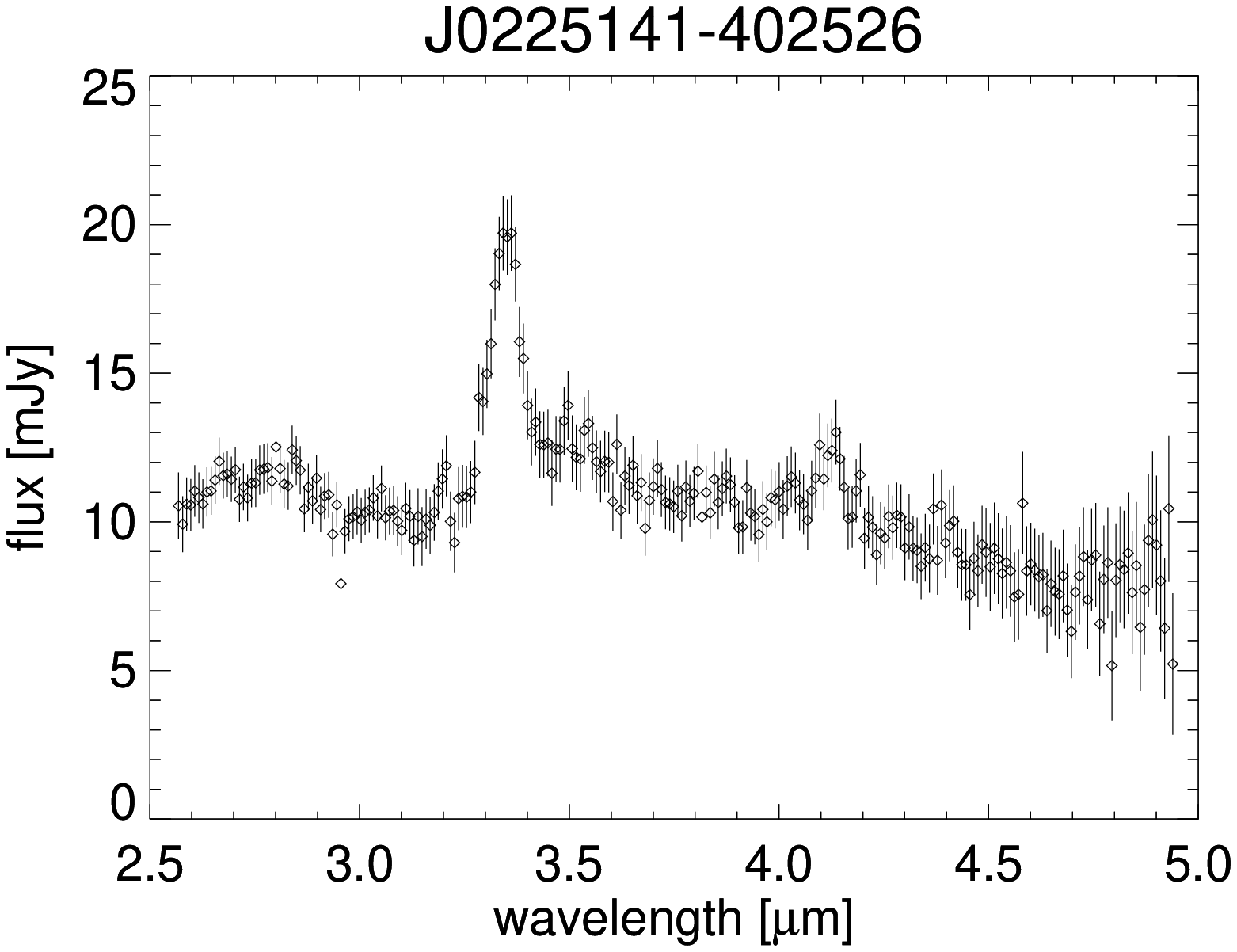}
     \FigureFile(41mm,26mm){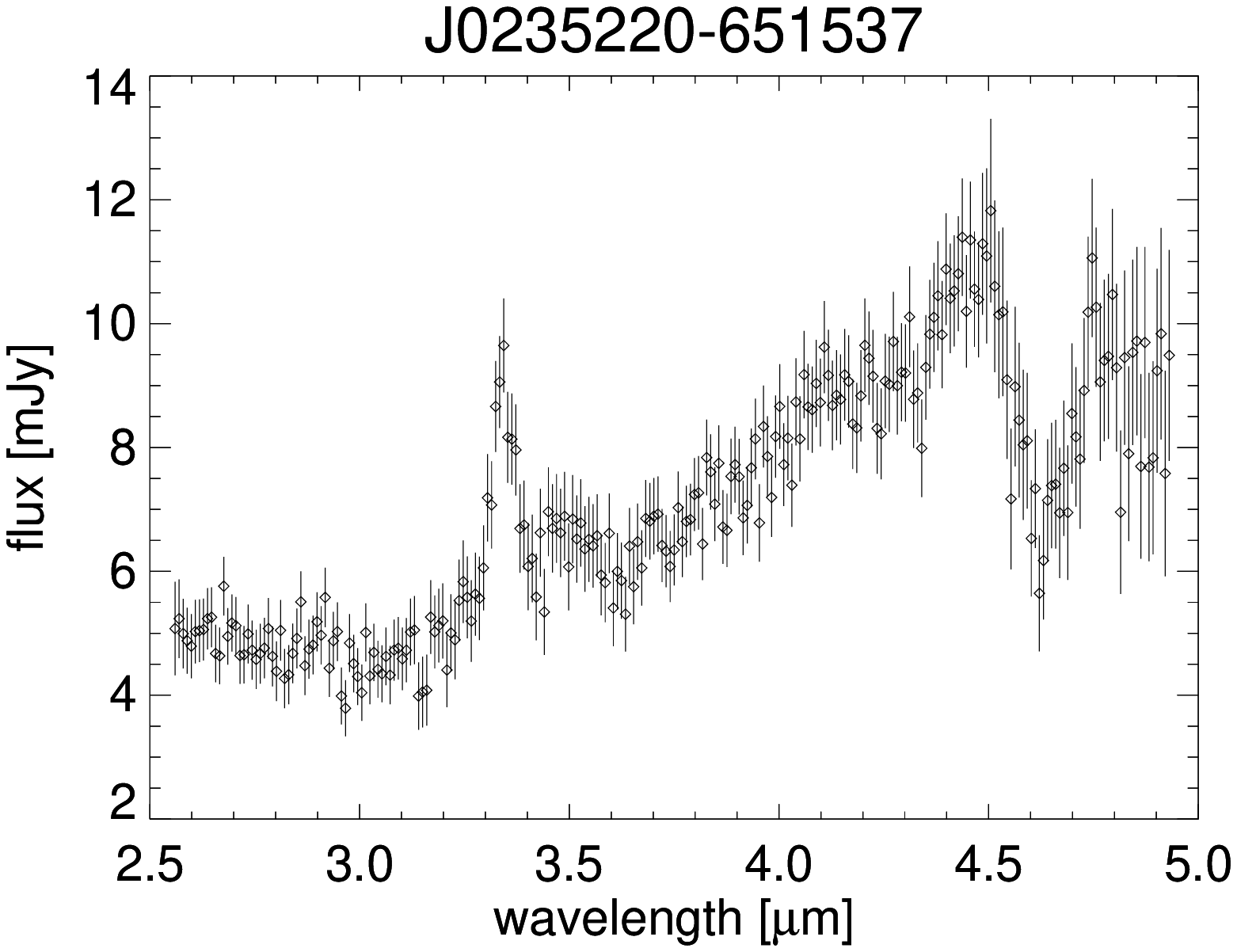}

     \FigureFile(41mm,26mm){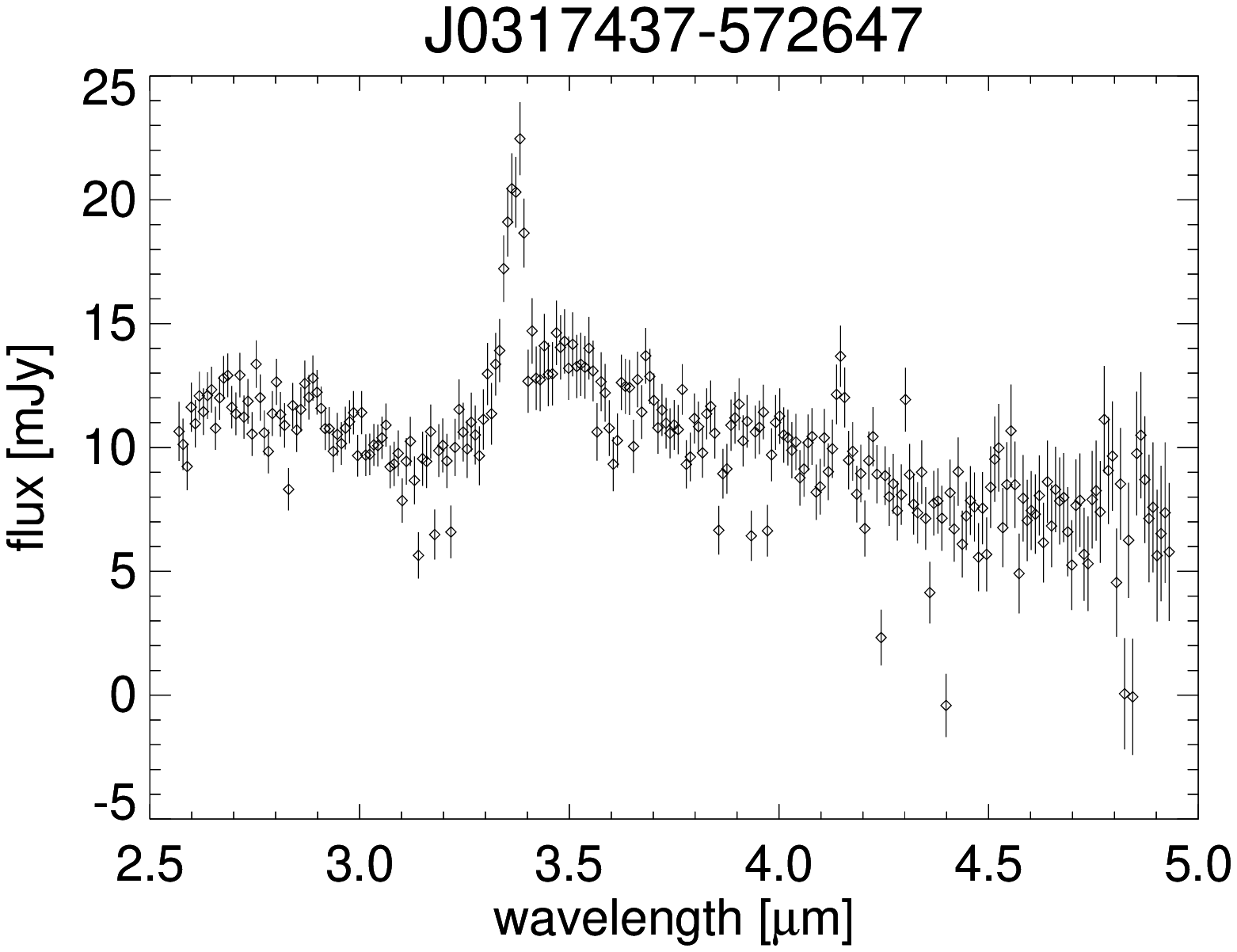}
     \FigureFile(41mm,26mm){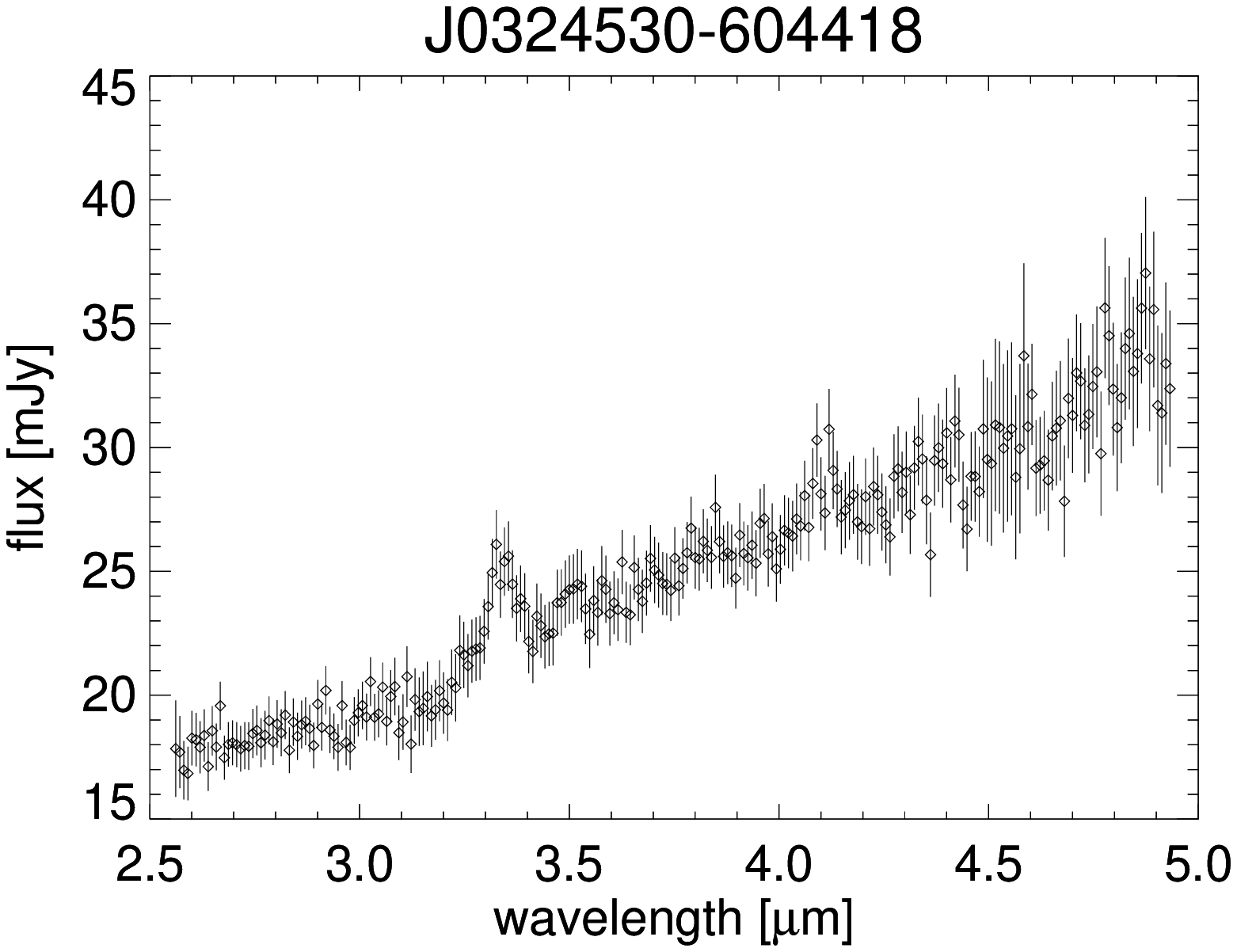}
     \FigureFile(41mm,26mm){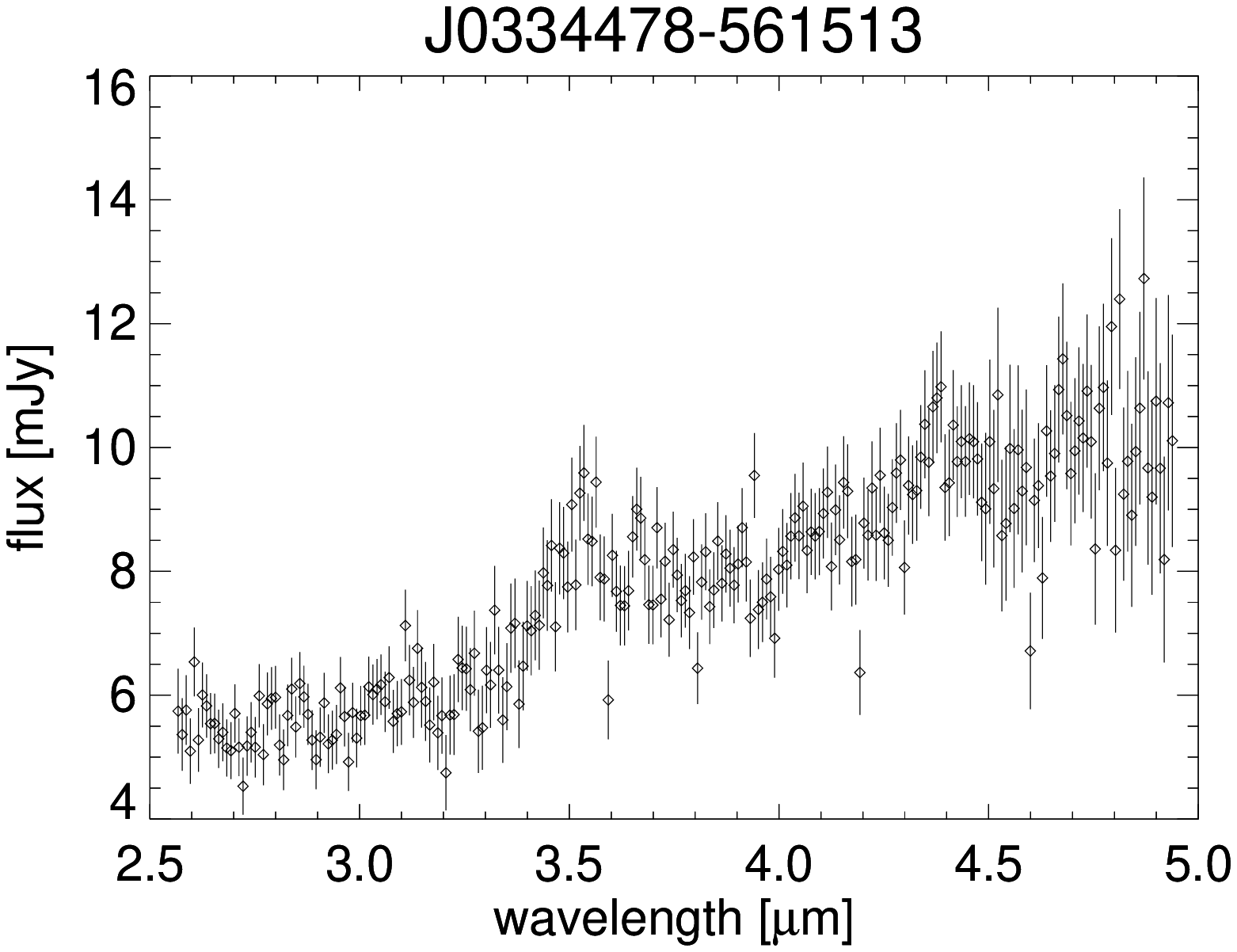}
     \FigureFile(41mm,26mm){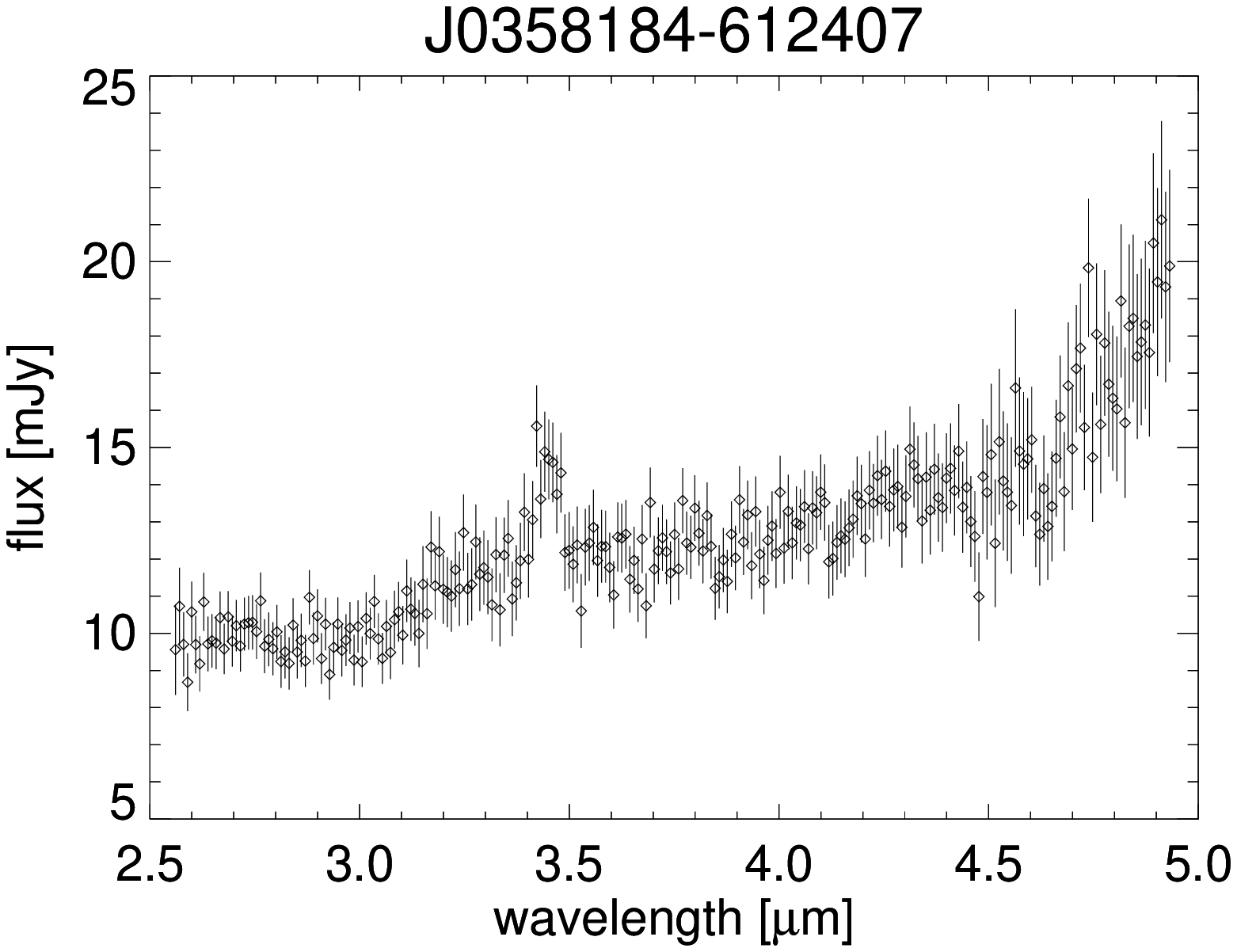}

     \FigureFile(41mm,26mm){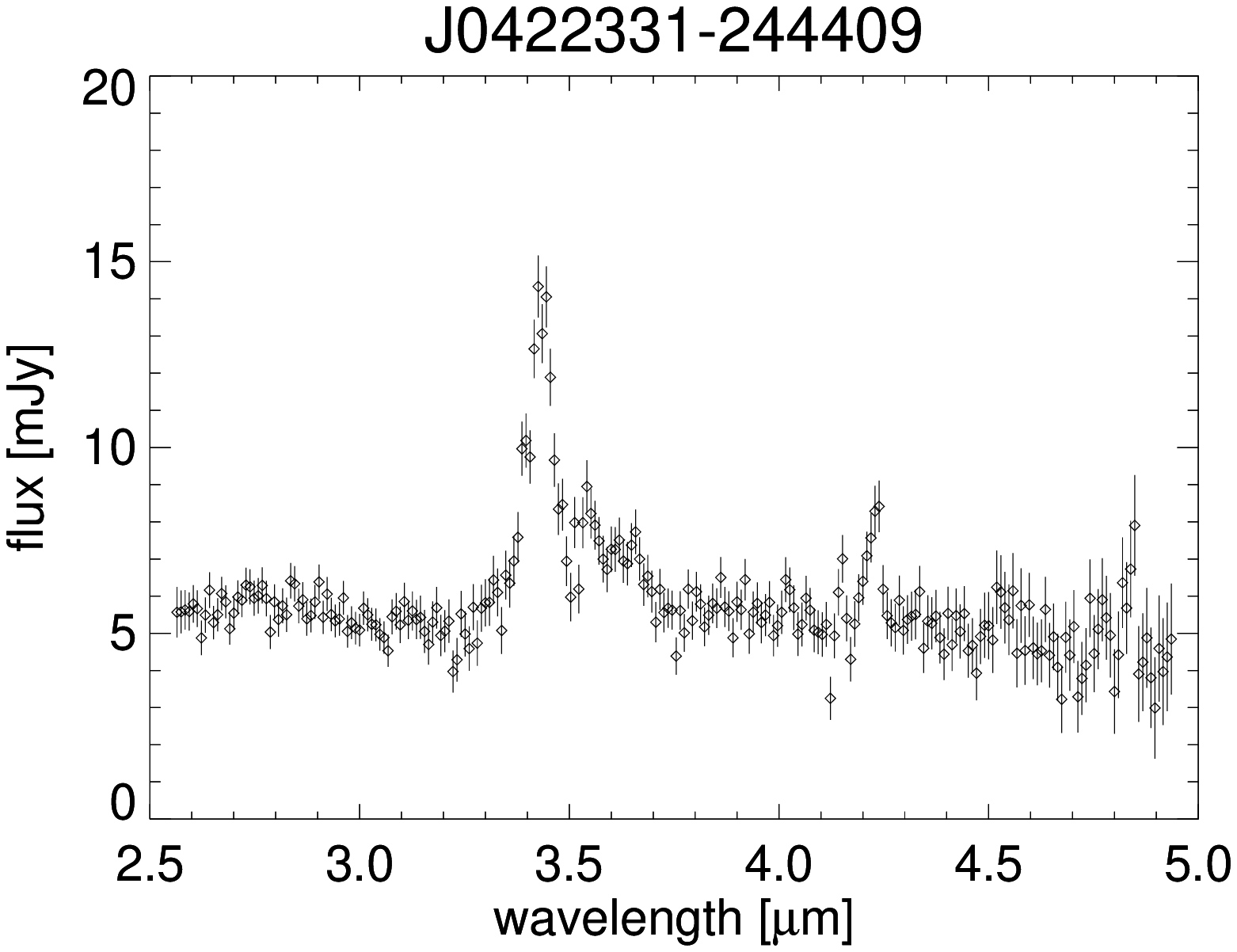}
     \FigureFile(41mm,26mm){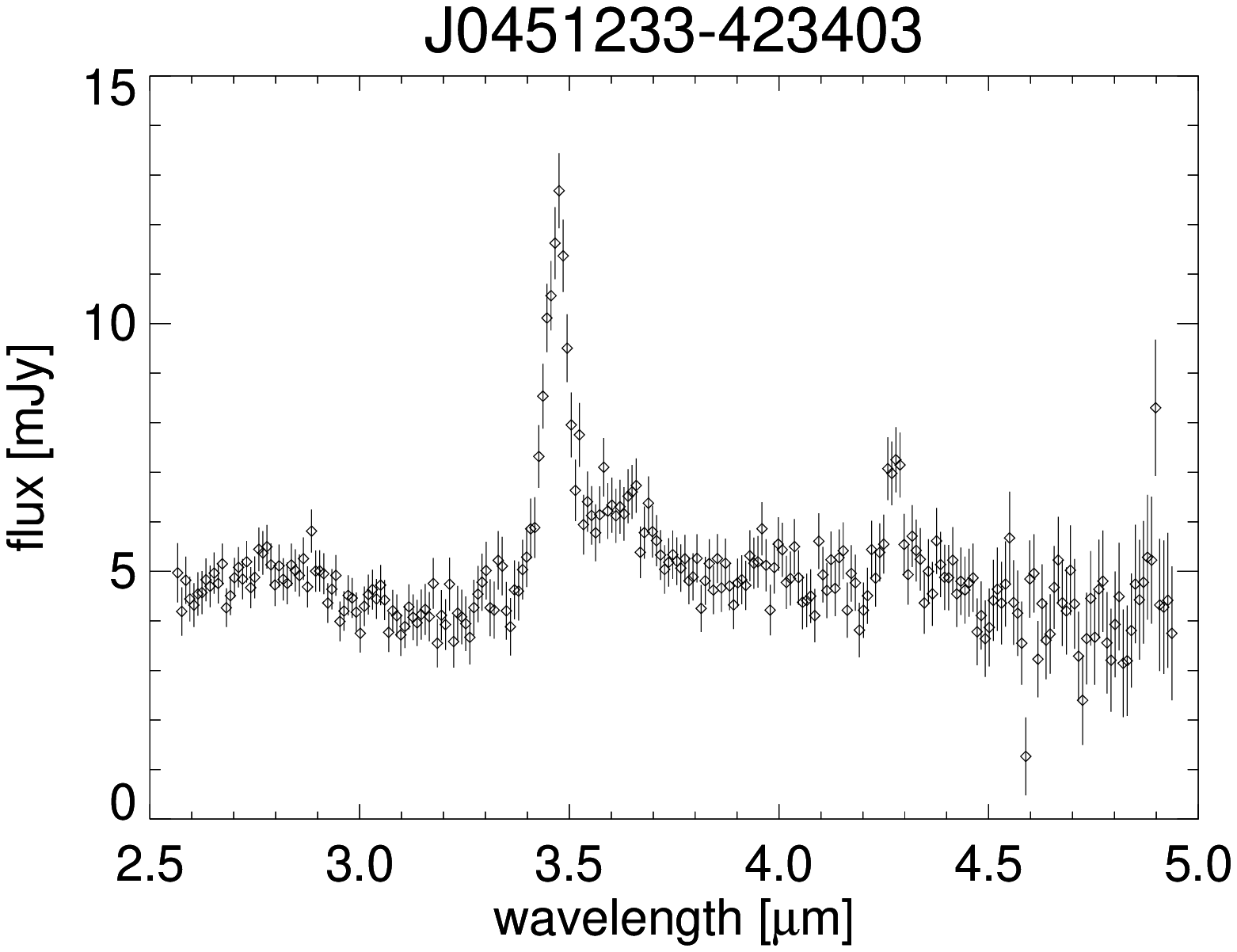}
     \FigureFile(41mm,26mm){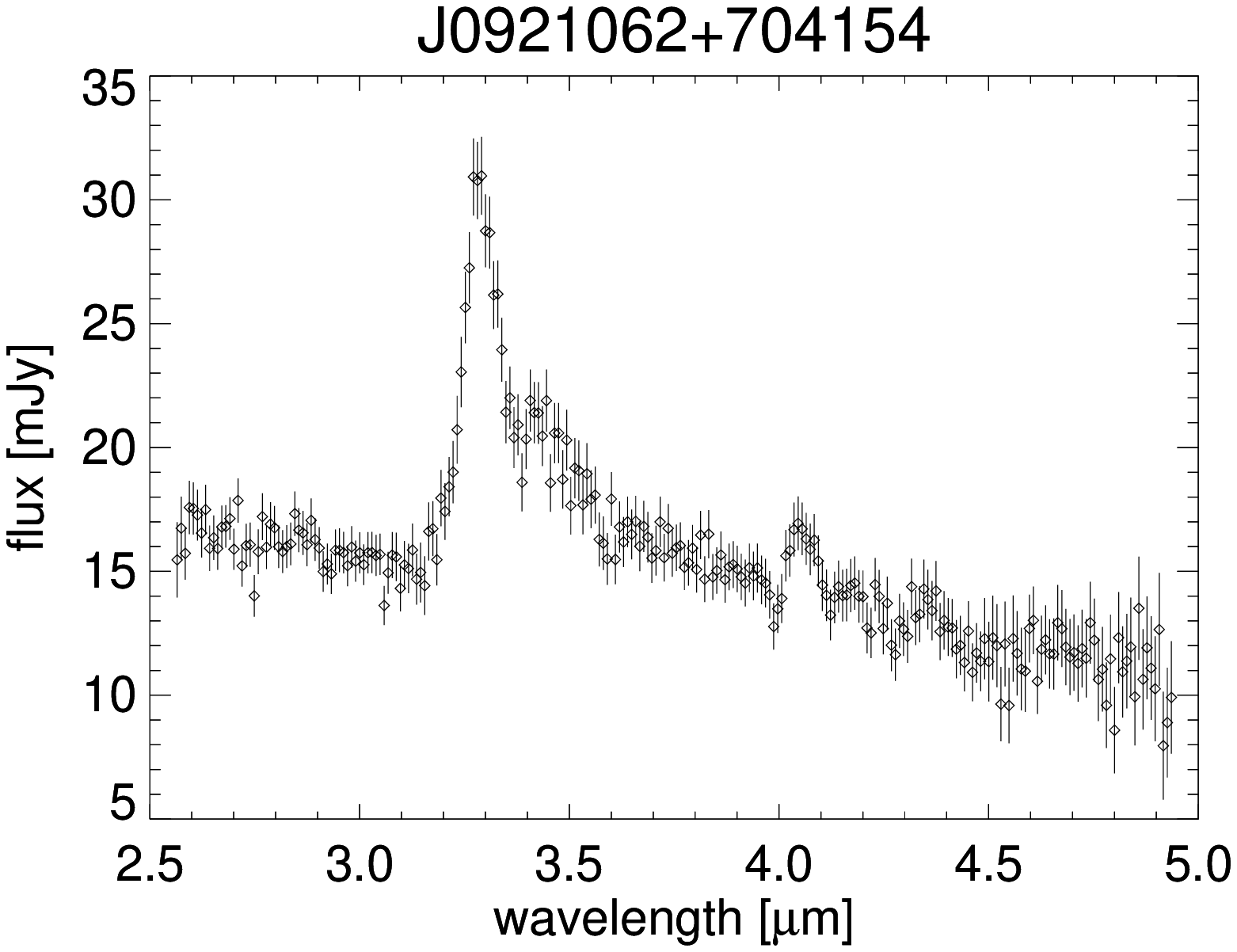}
     \FigureFile(41mm,26mm){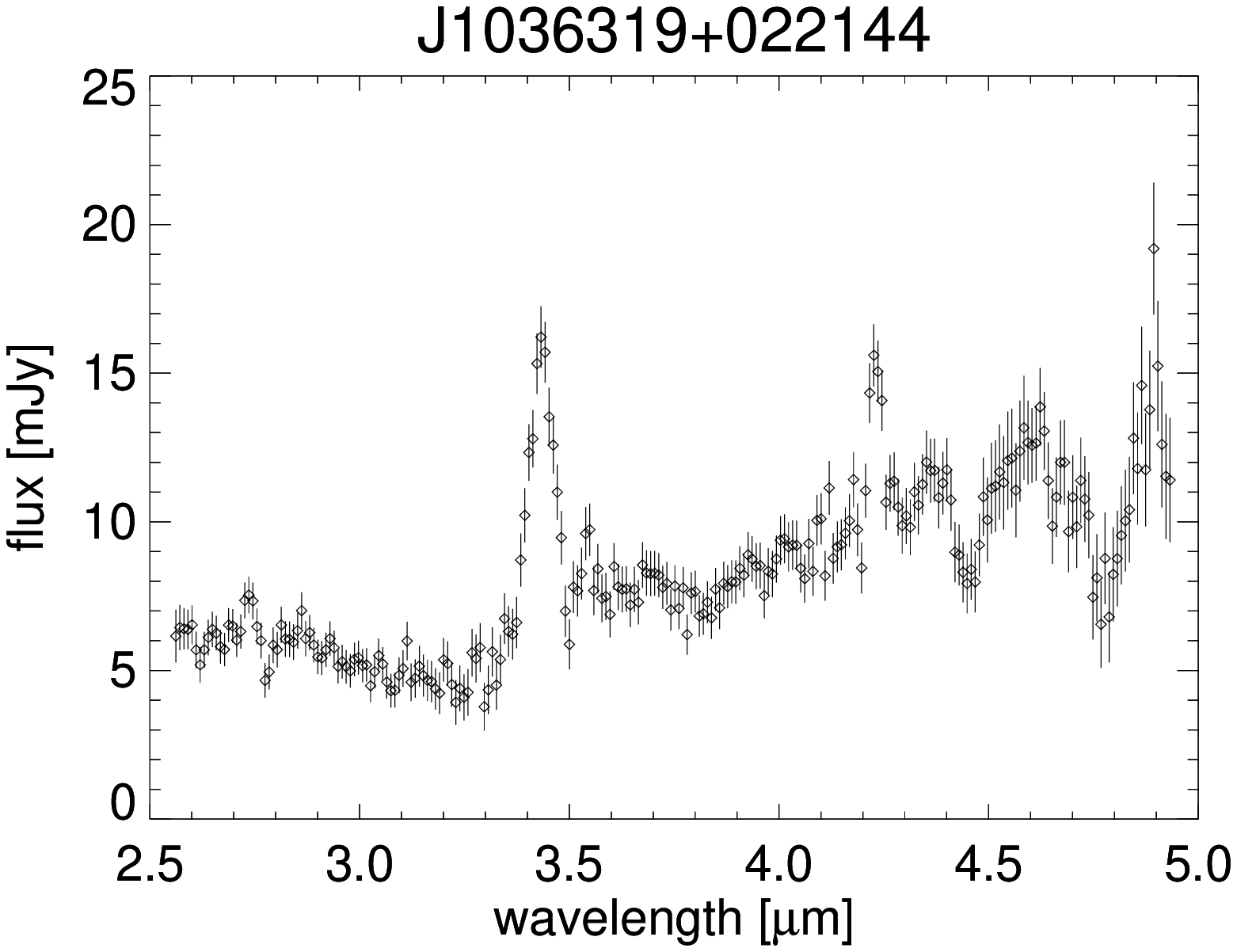}

     \FigureFile(41mm,26mm){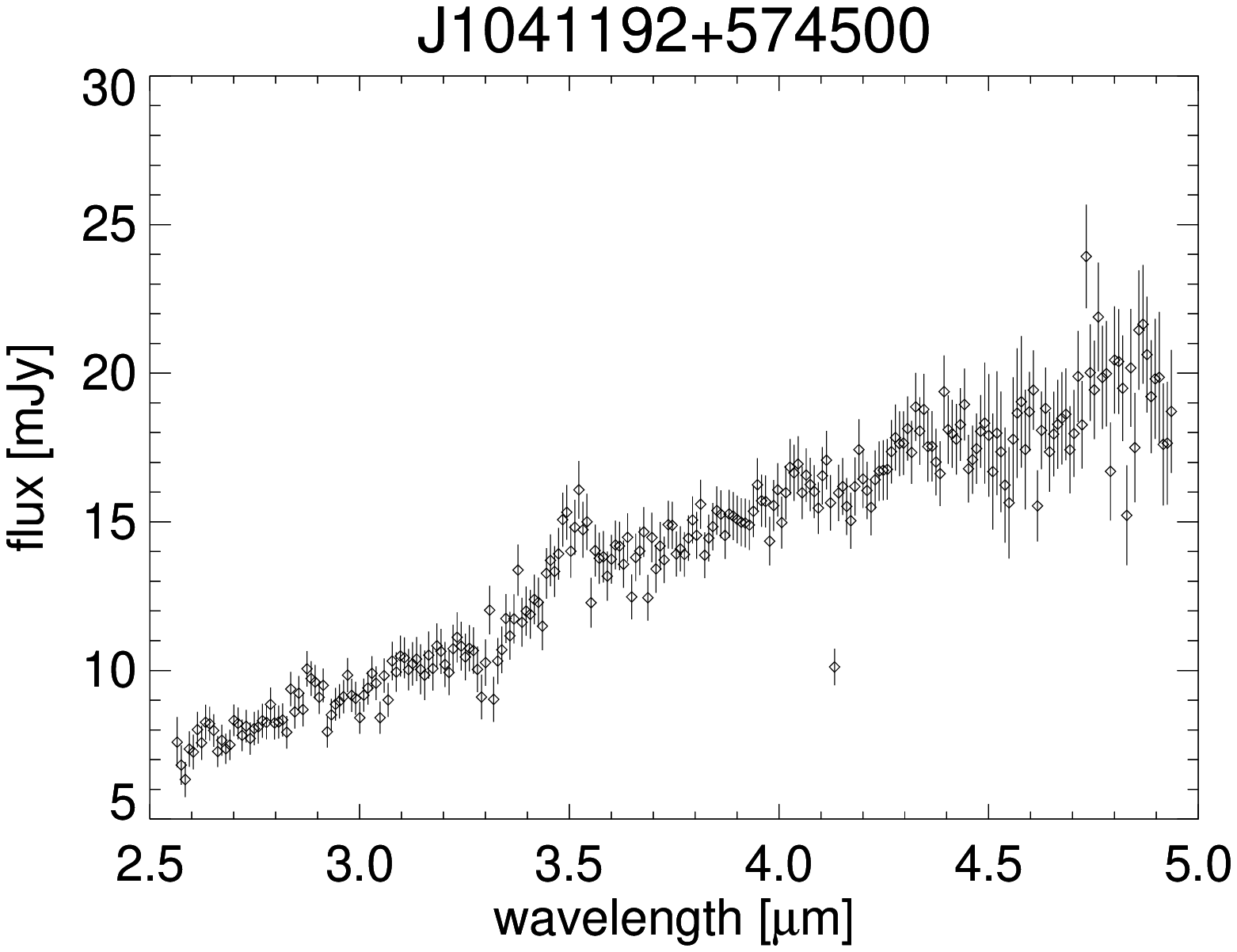}
     \FigureFile(41mm,26mm){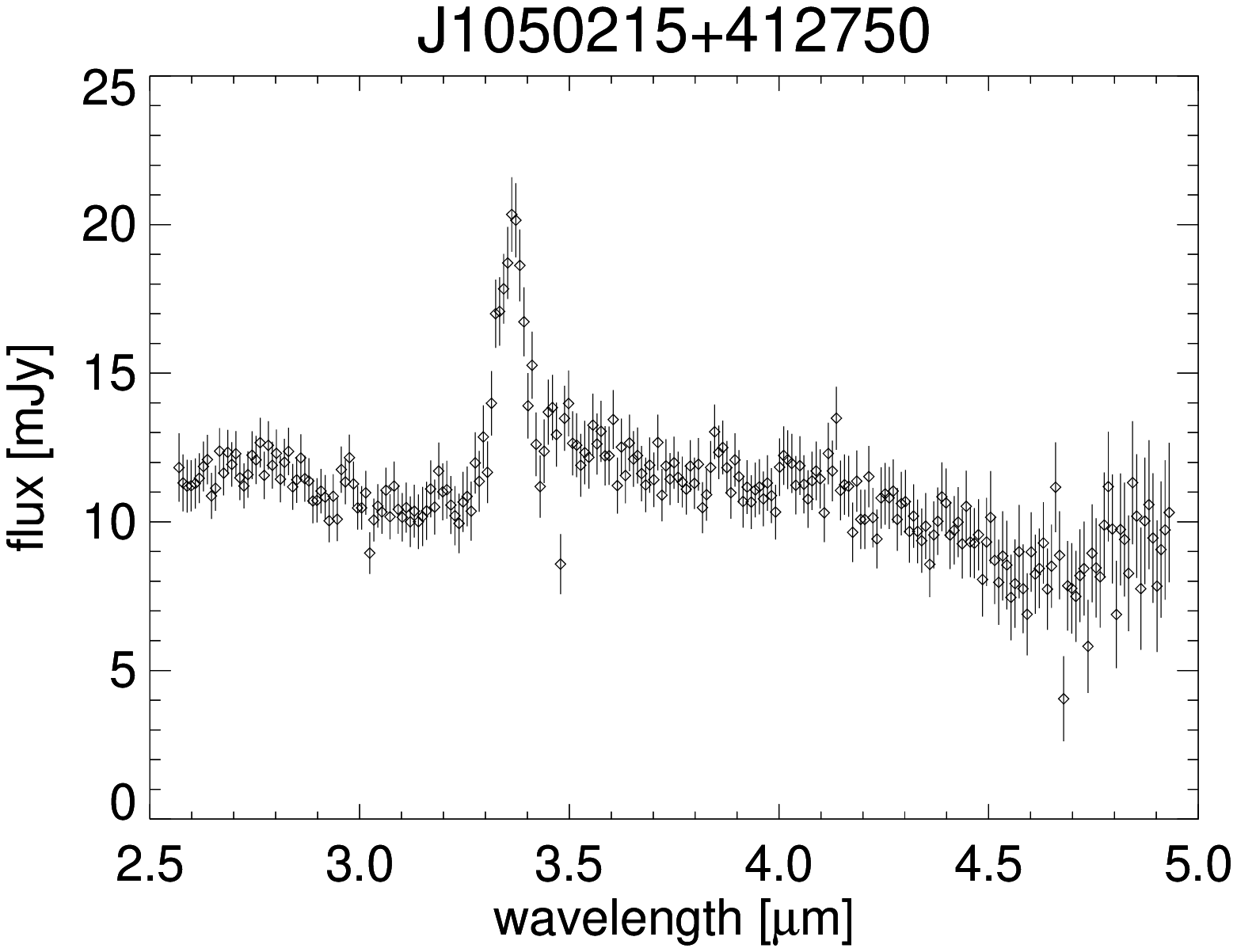}
     \FigureFile(41mm,26mm){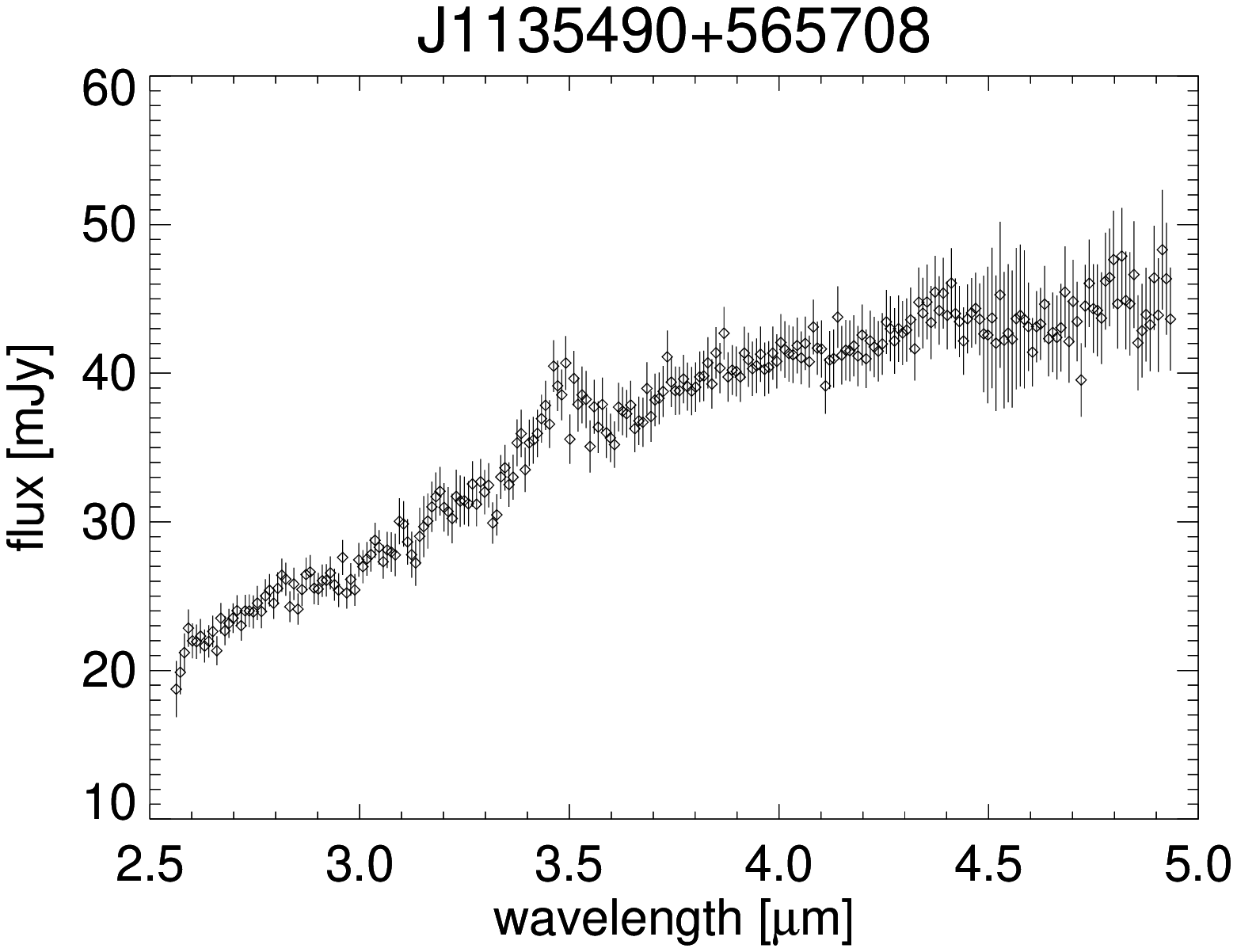}
     \FigureFile(41mm,26mm){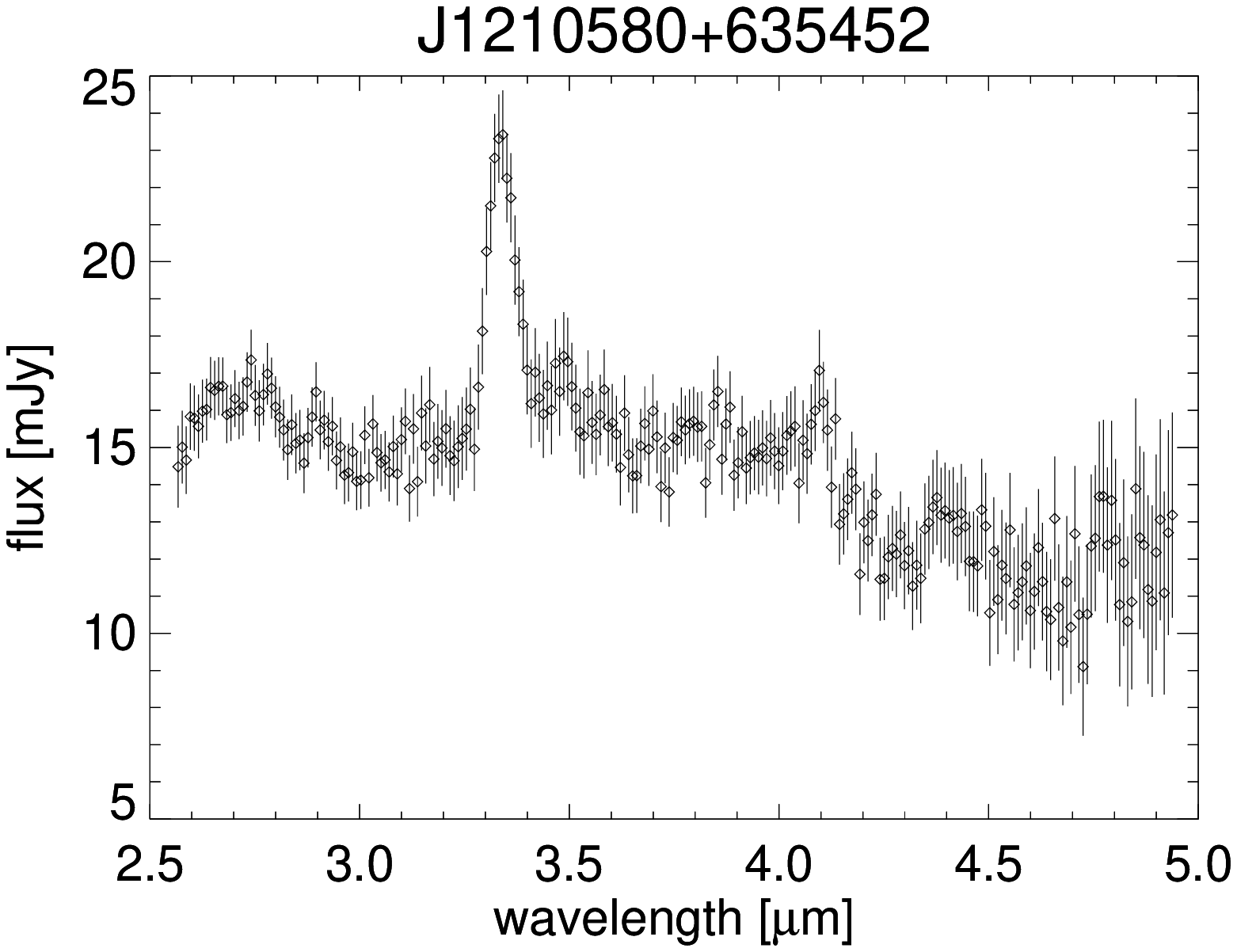}

     \FigureFile(41mm,26mm){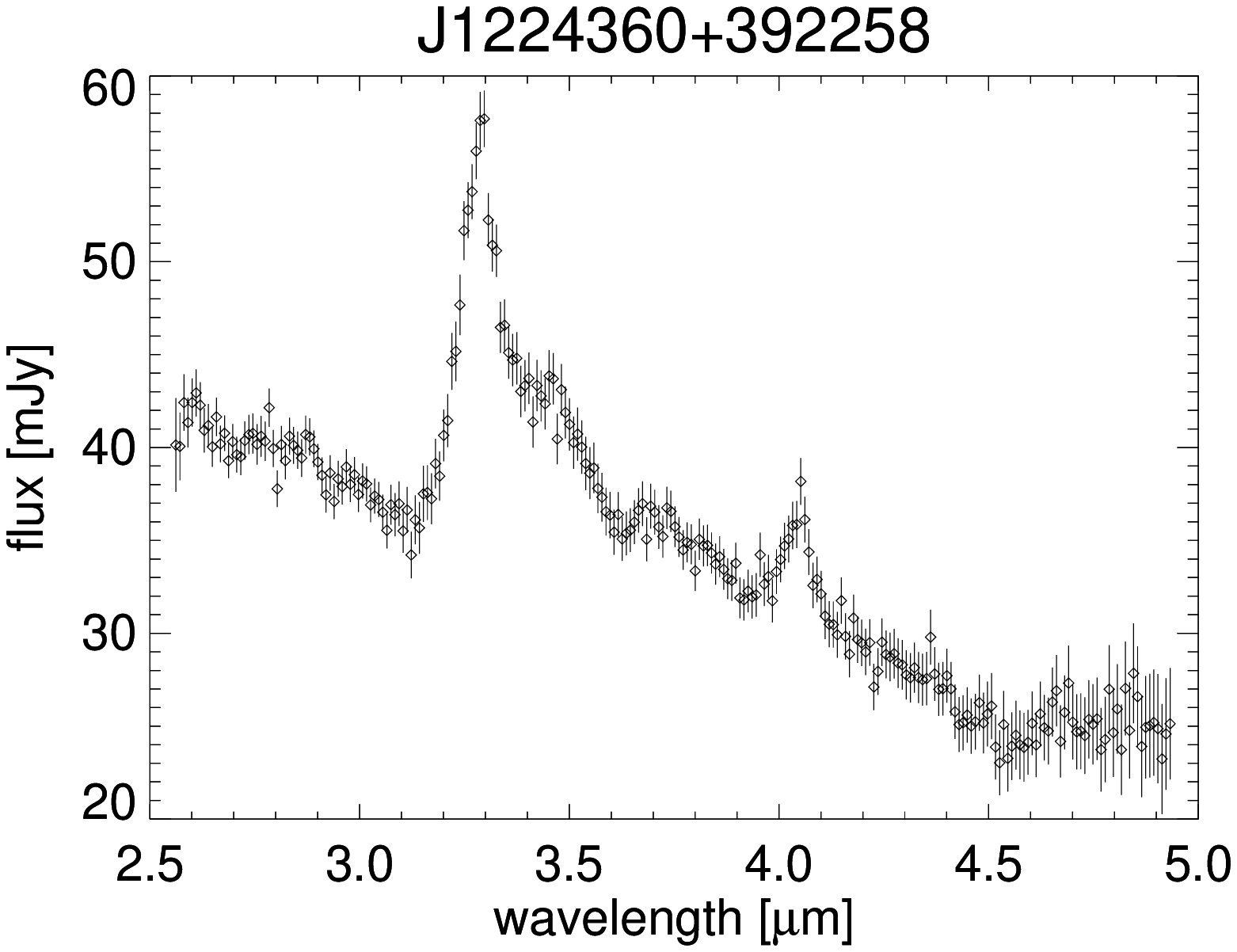}
     \FigureFile(41mm,26mm){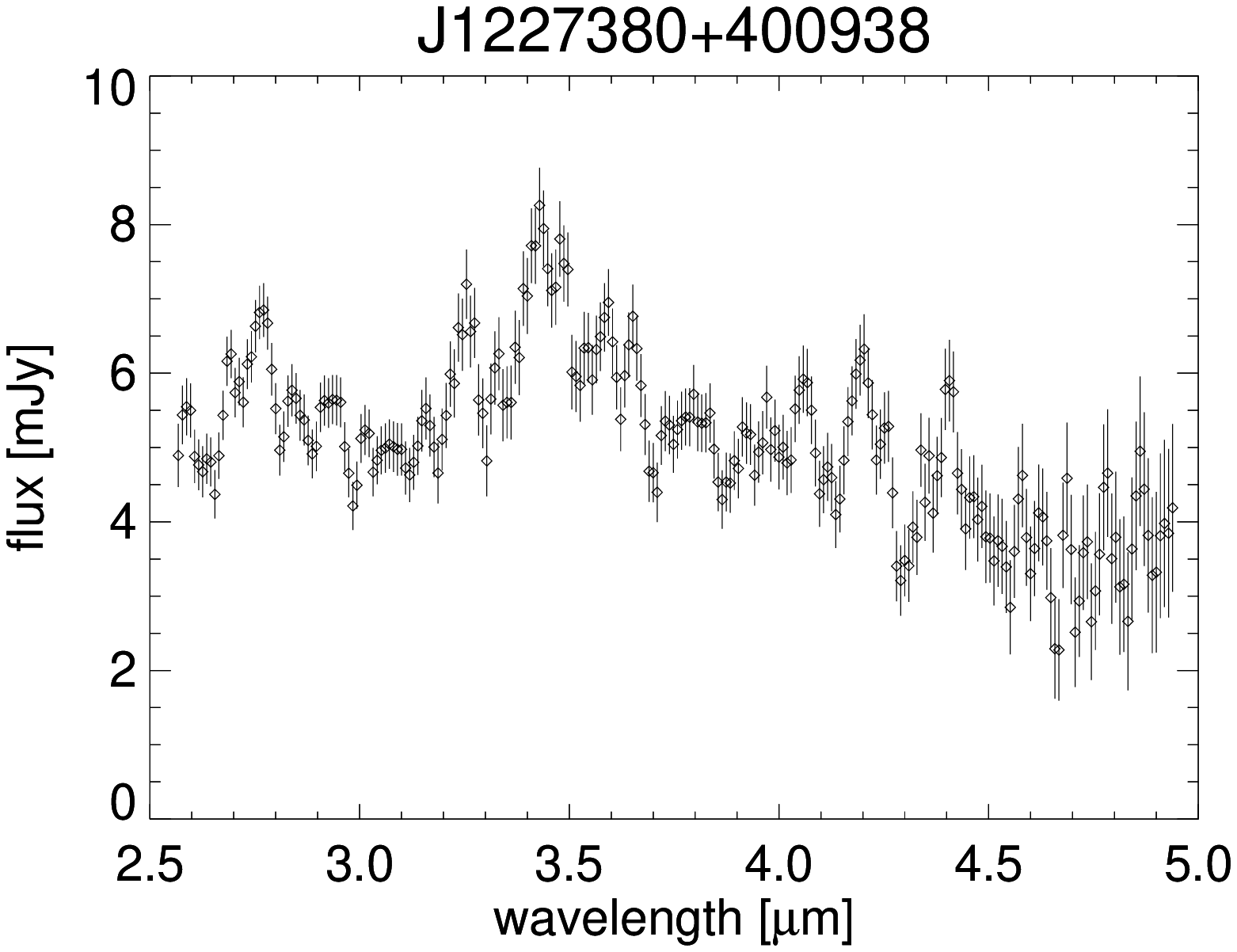}
     \FigureFile(41mm,26mm){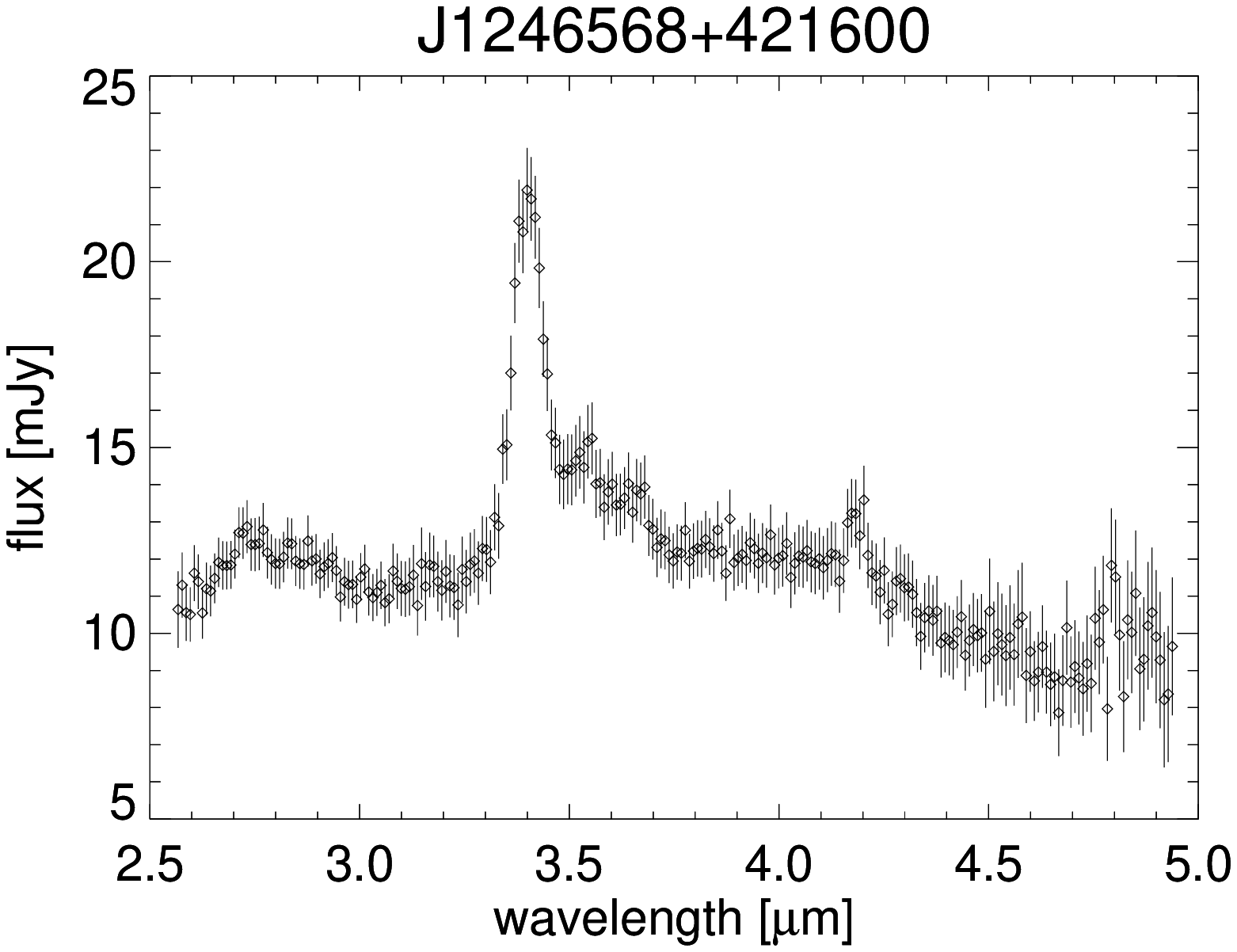}
     \FigureFile(41mm,26mm){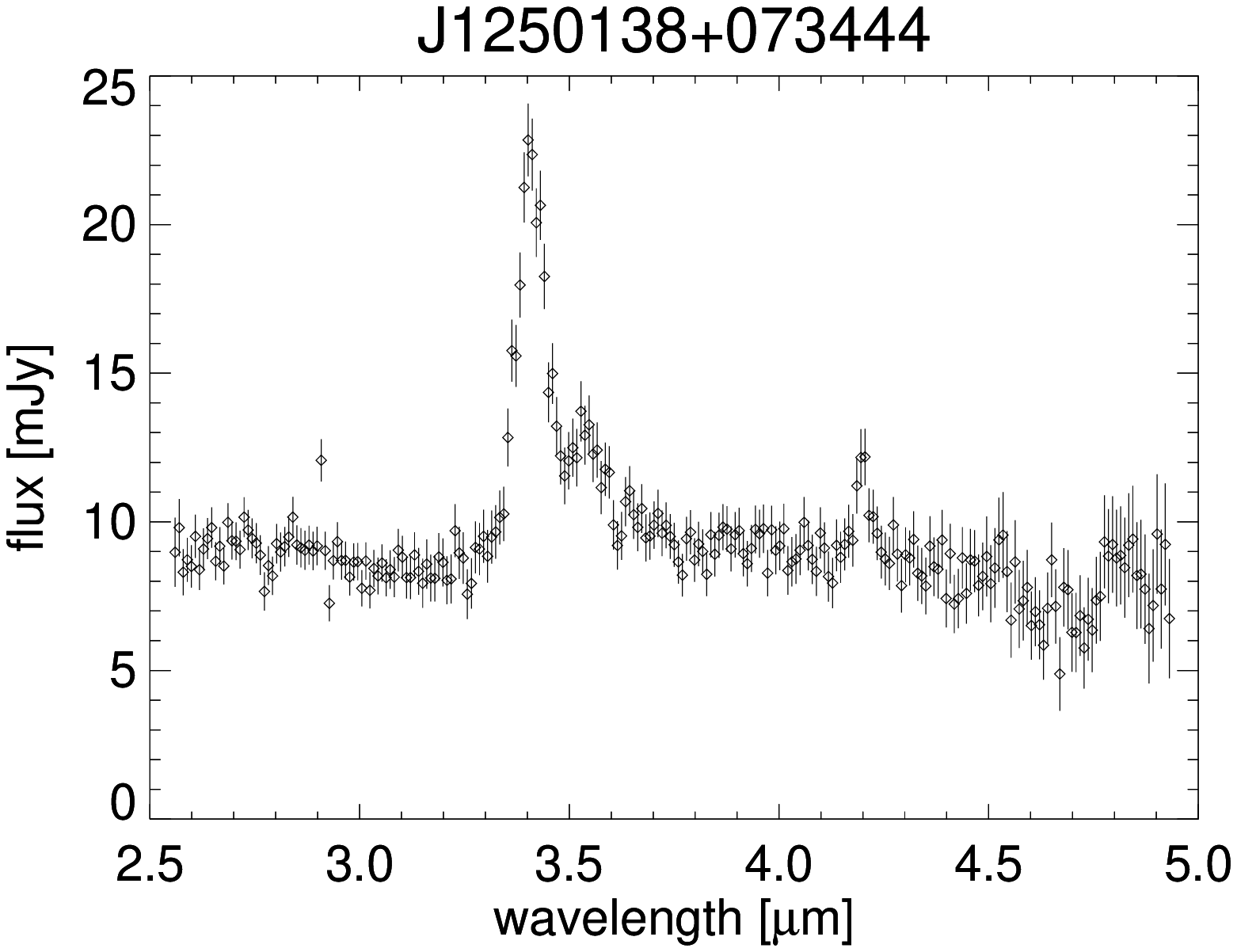}

 \end{center}
\end{figure}
\begin{figure}
 \begin{center}

     \FigureFile(41mm,26mm){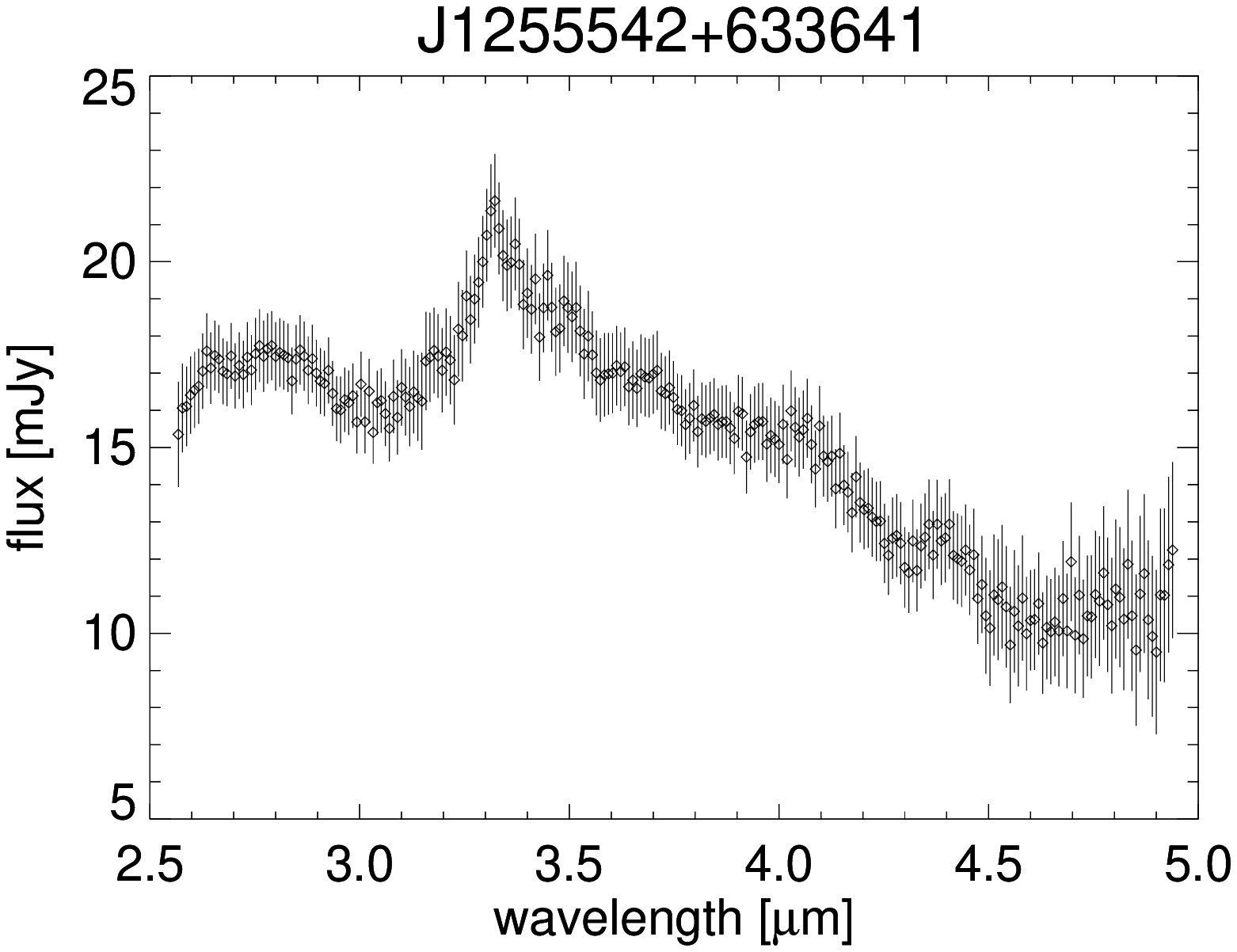}
     \FigureFile(41mm,26mm){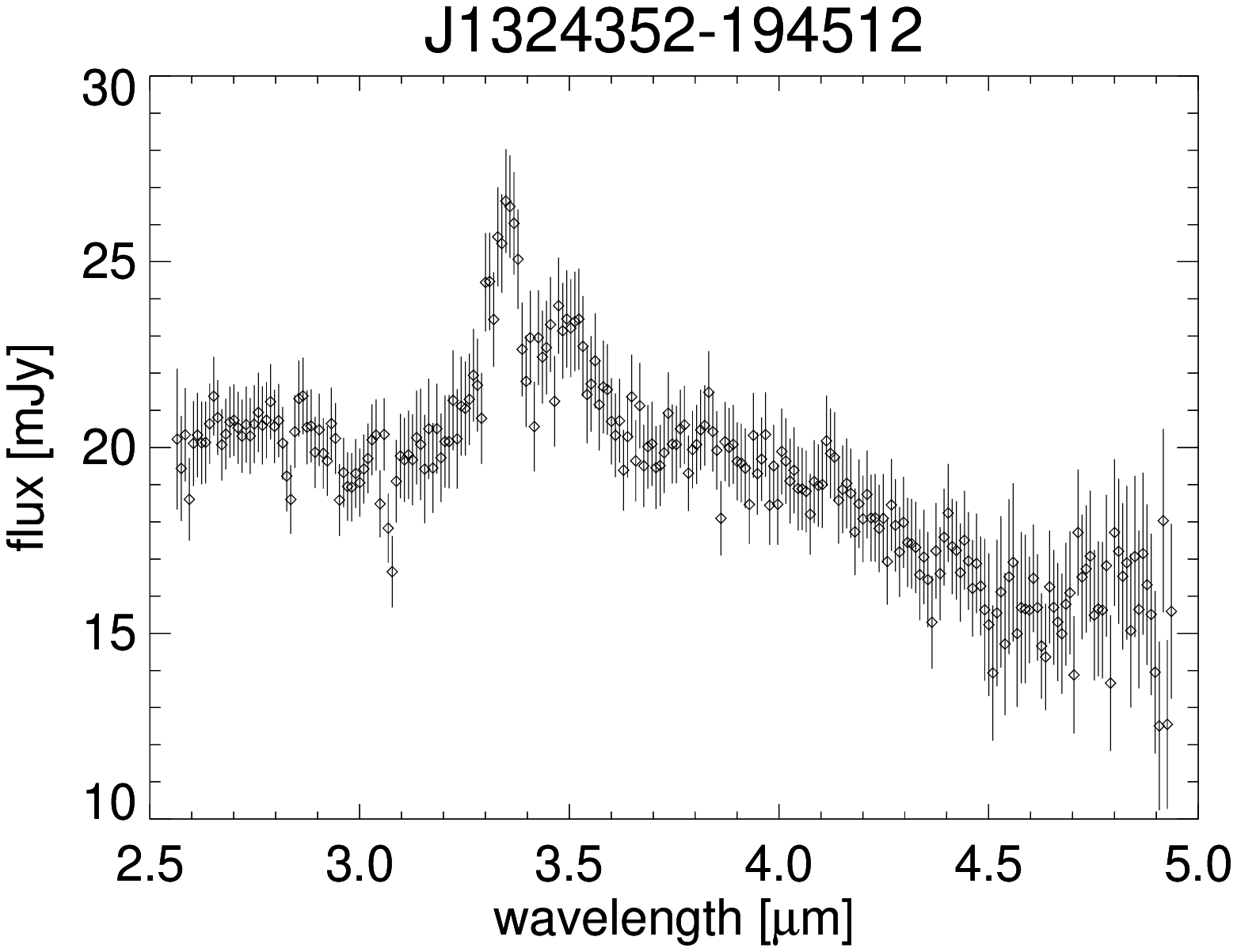}
     \FigureFile(41mm,26mm){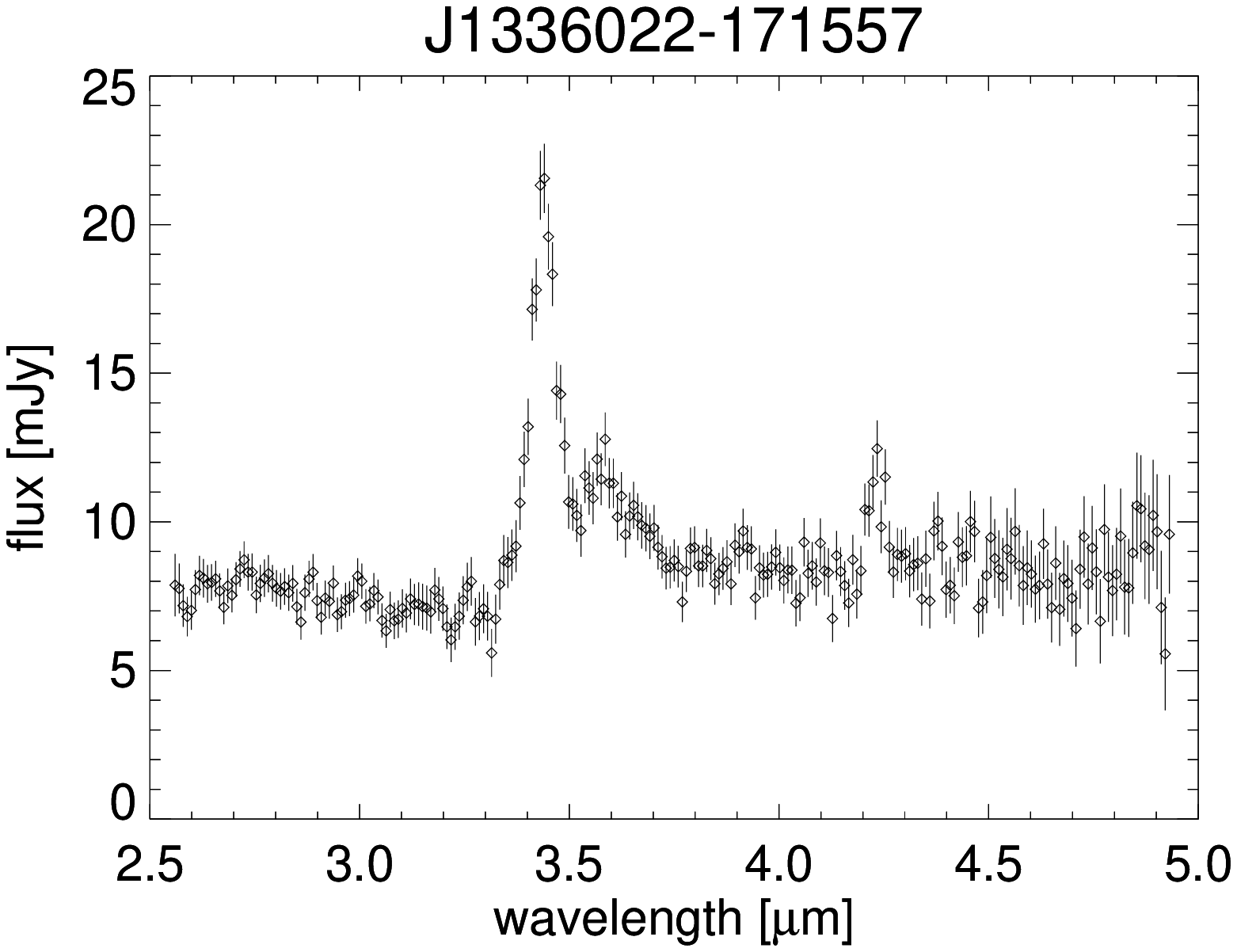}
     \FigureFile(41mm,26mm){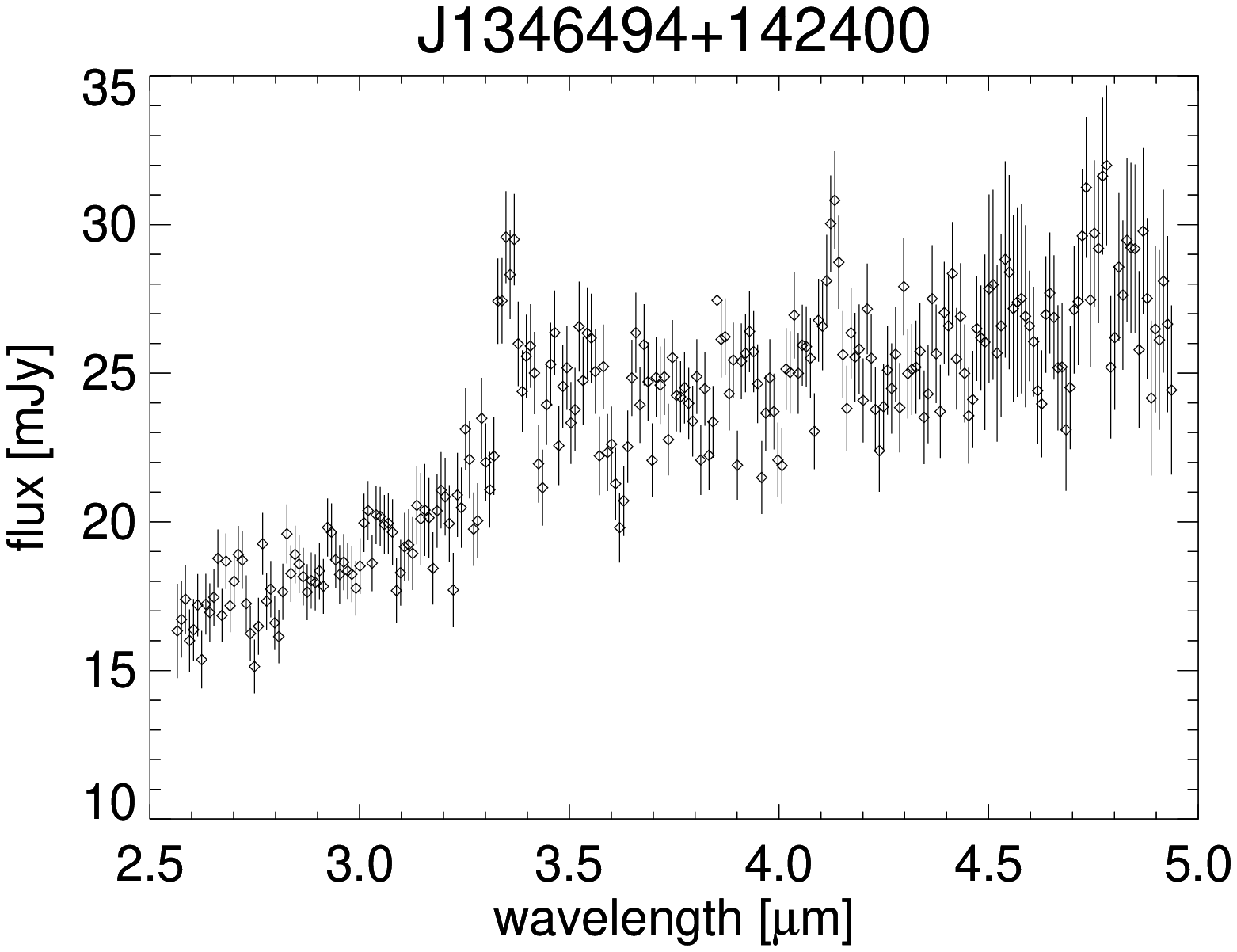}

     \FigureFile(41mm,26mm){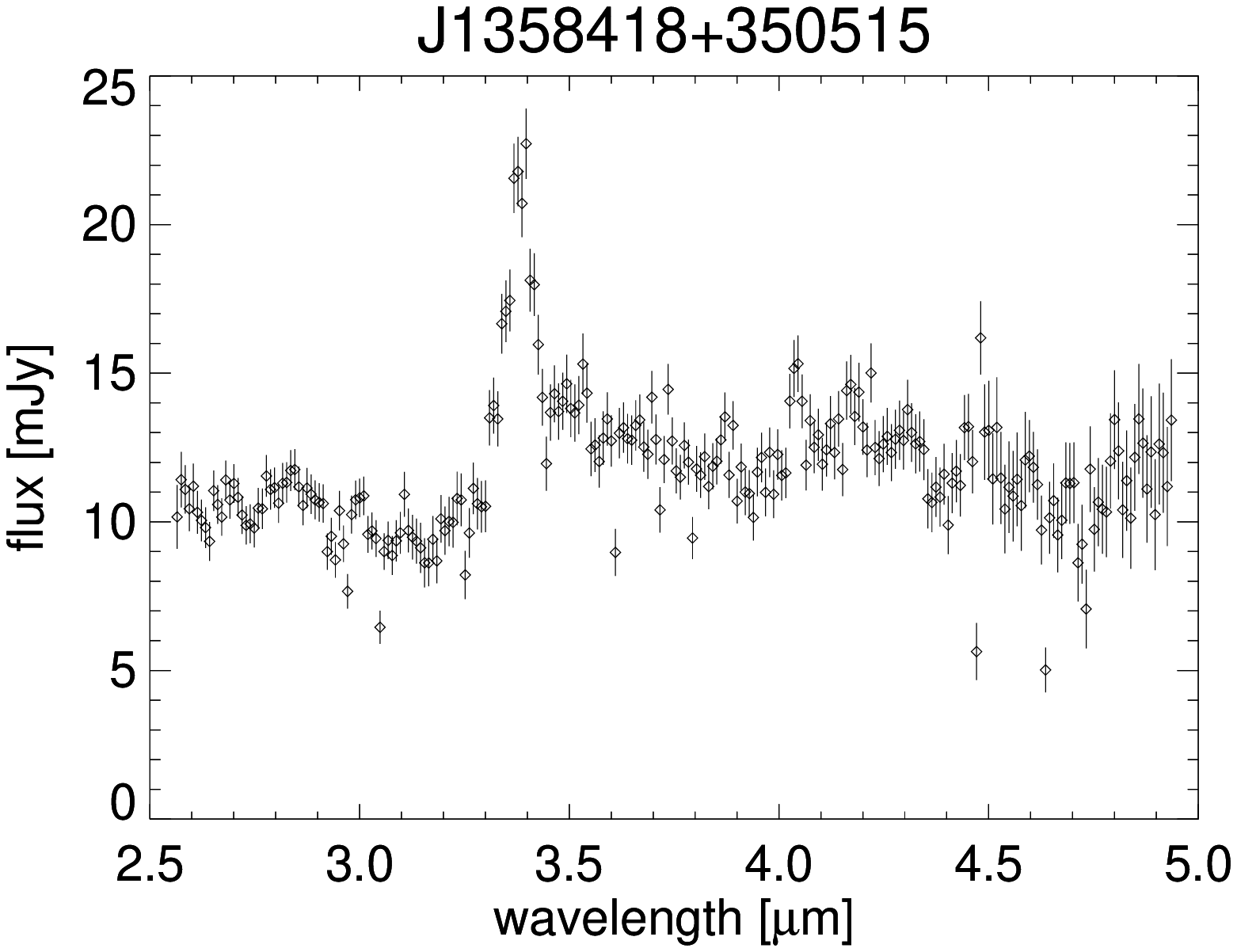}
     \FigureFile(41mm,26mm){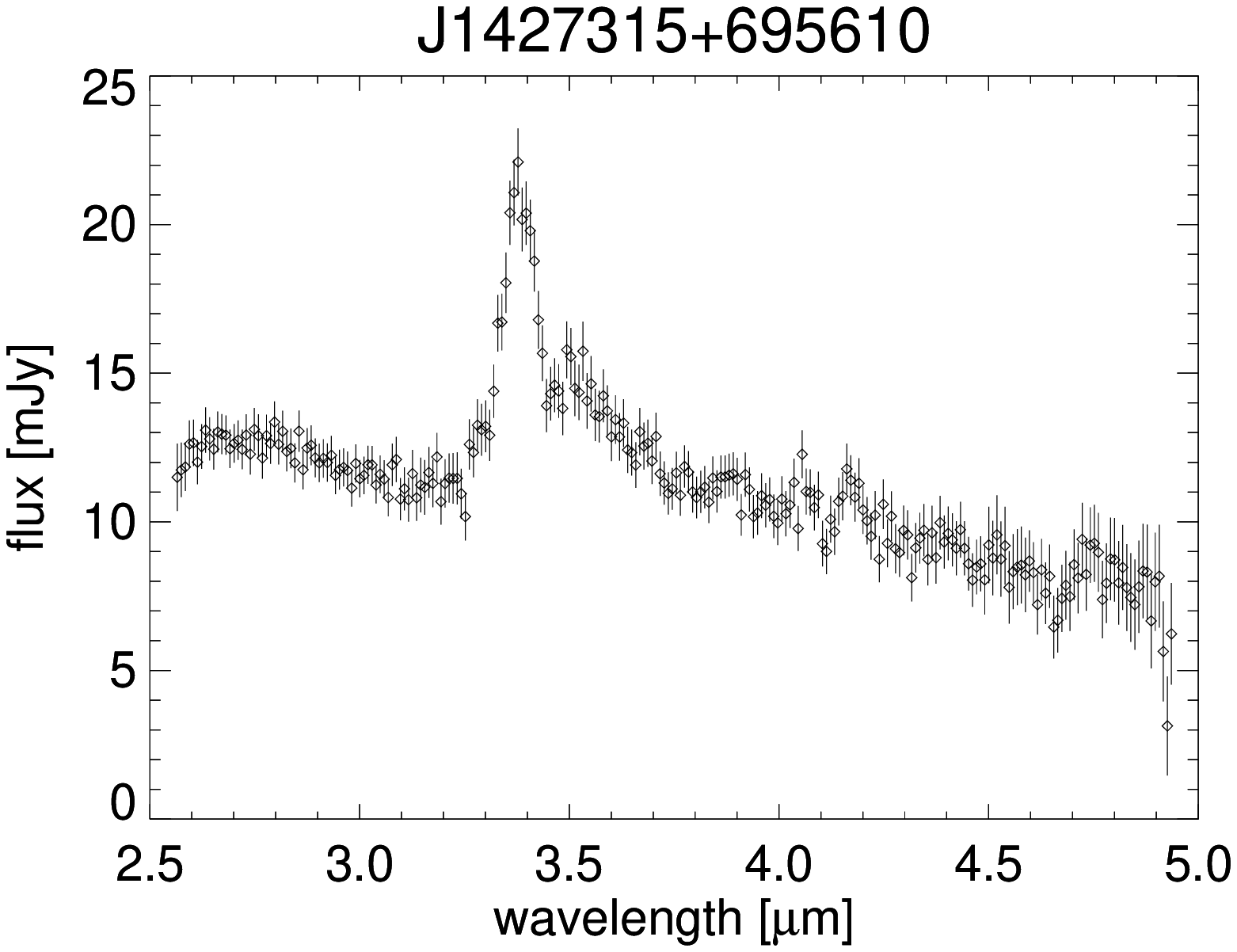}
     \FigureFile(41mm,26mm){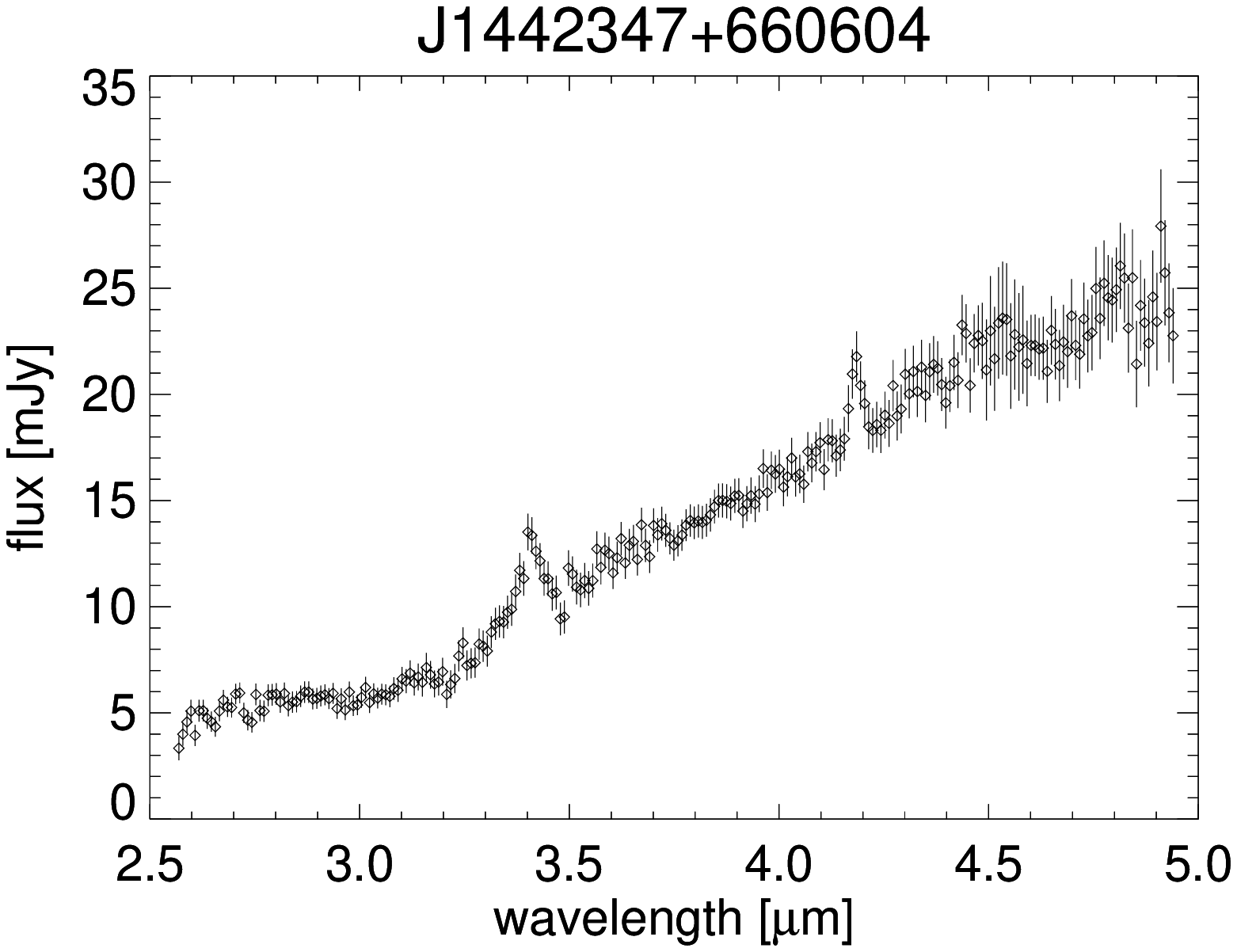}
     \FigureFile(41mm,26mm){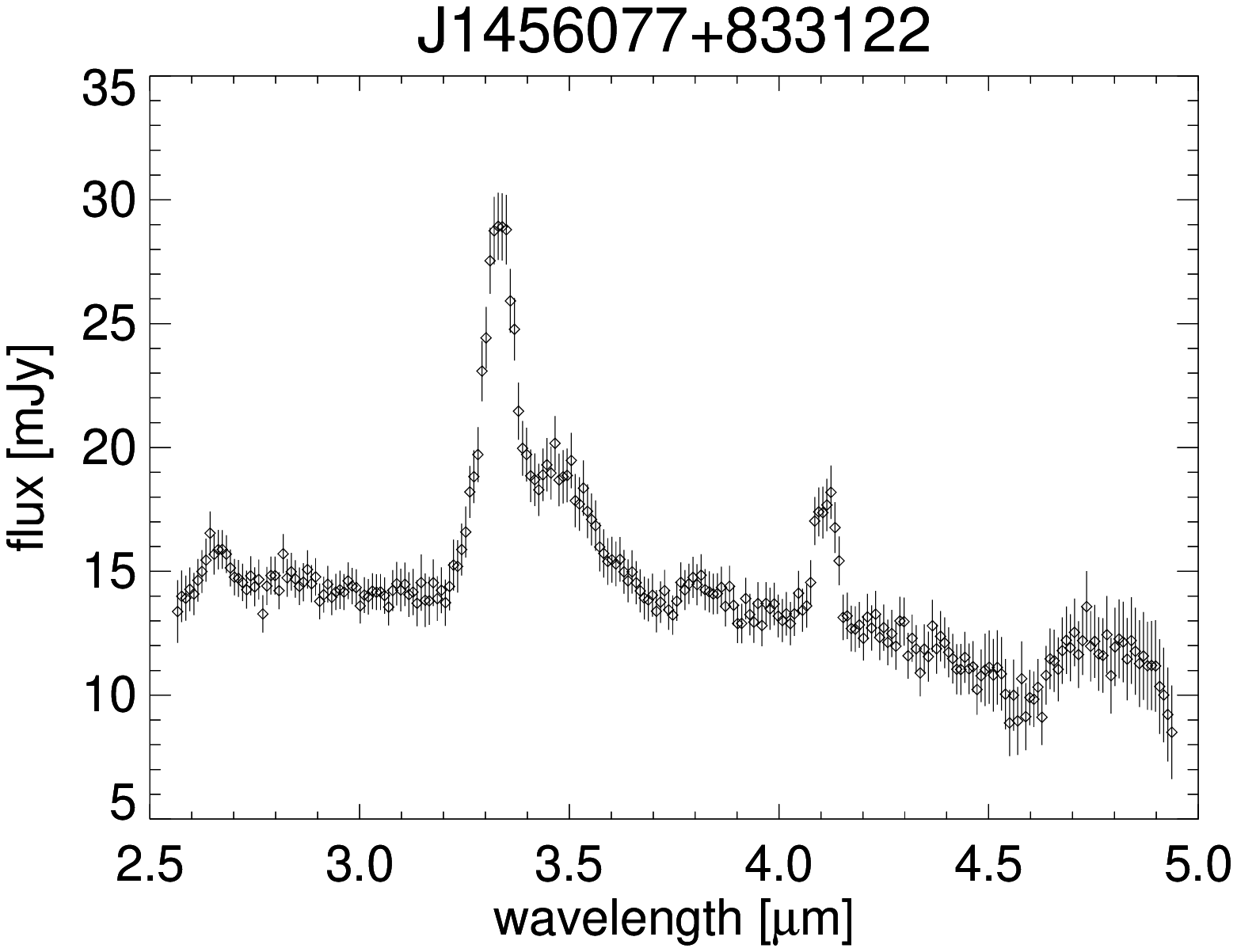}
 
     \FigureFile(41mm,26mm){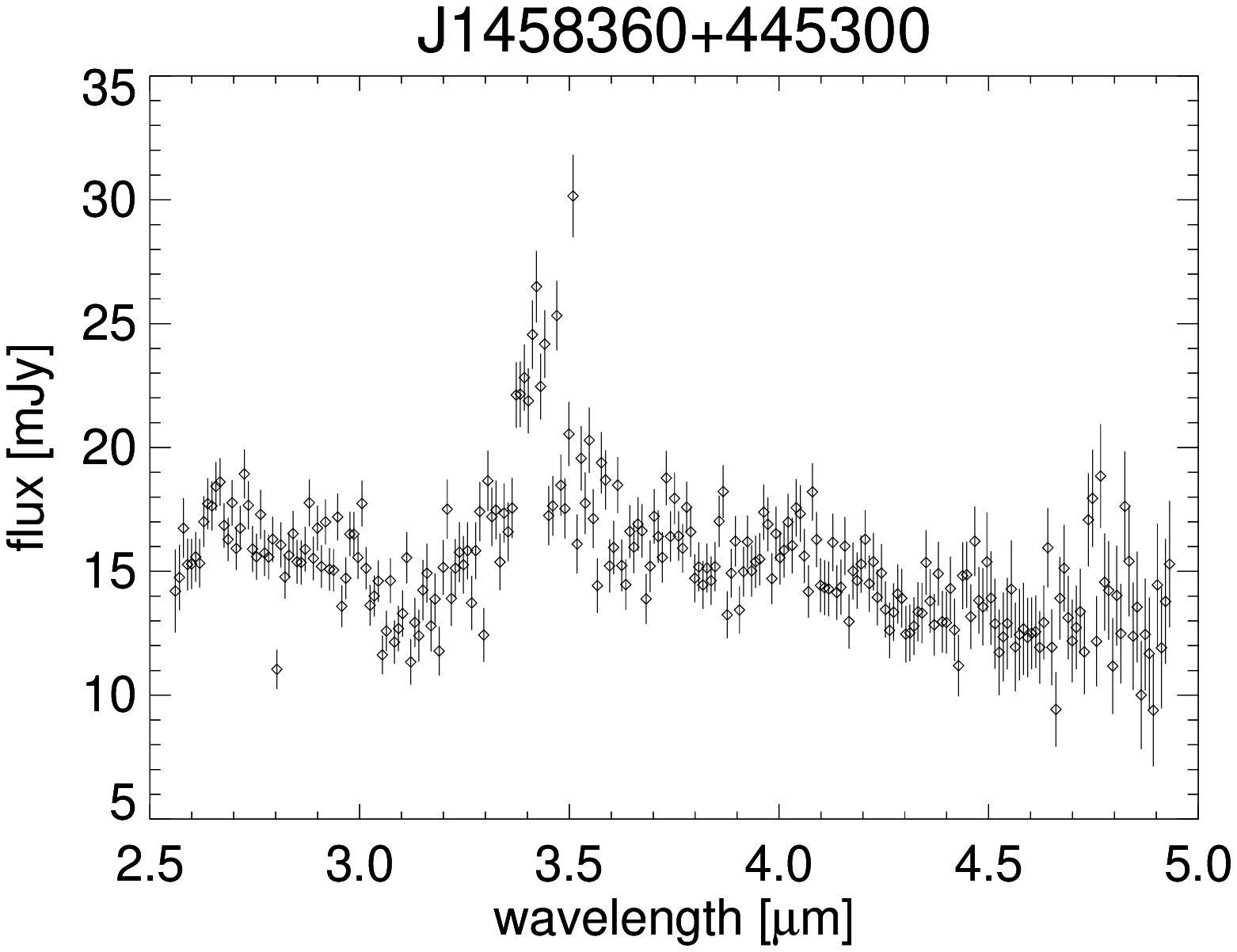}
     \FigureFile(41mm,26mm){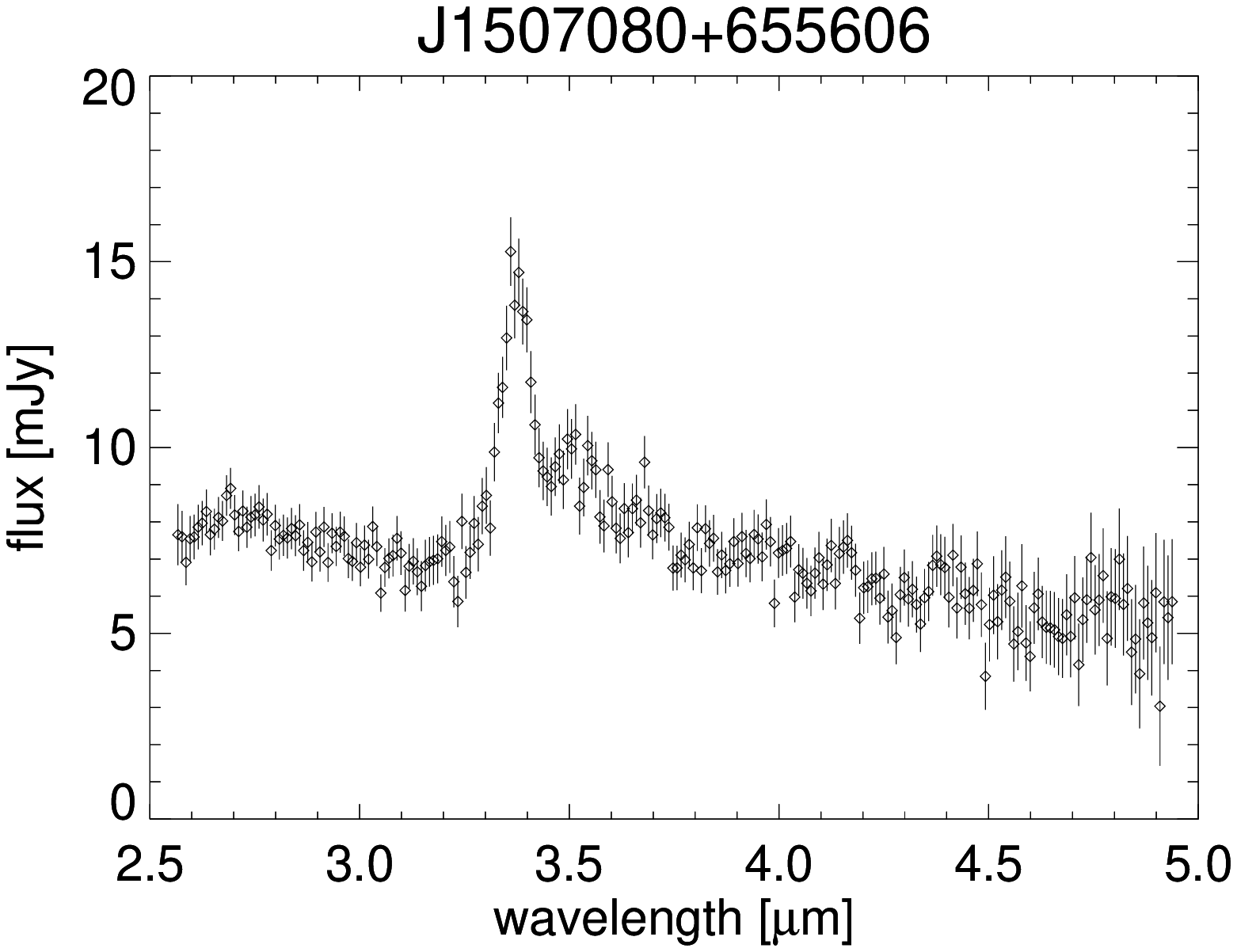}
     \FigureFile(41mm,26mm){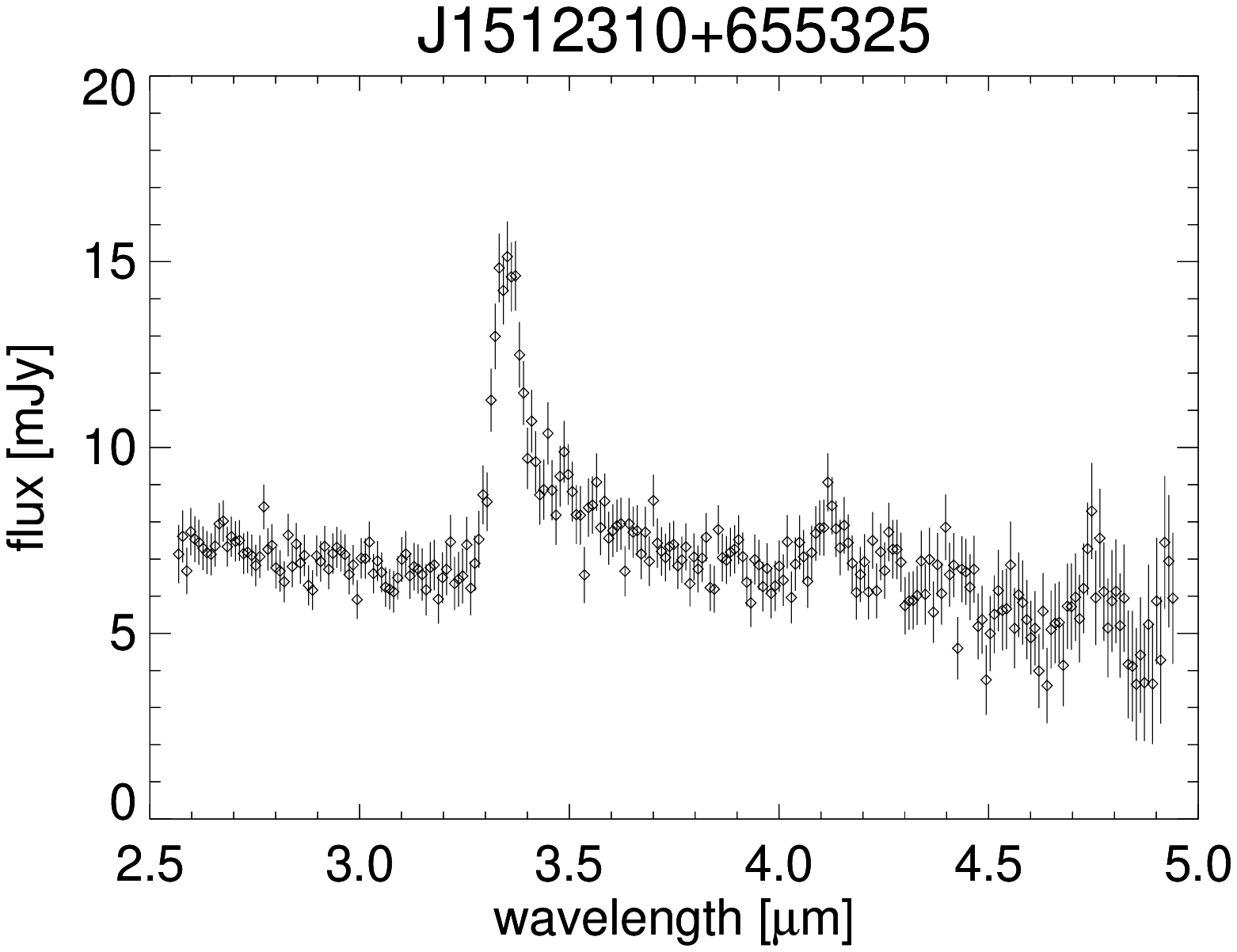}
     \FigureFile(41mm,26mm){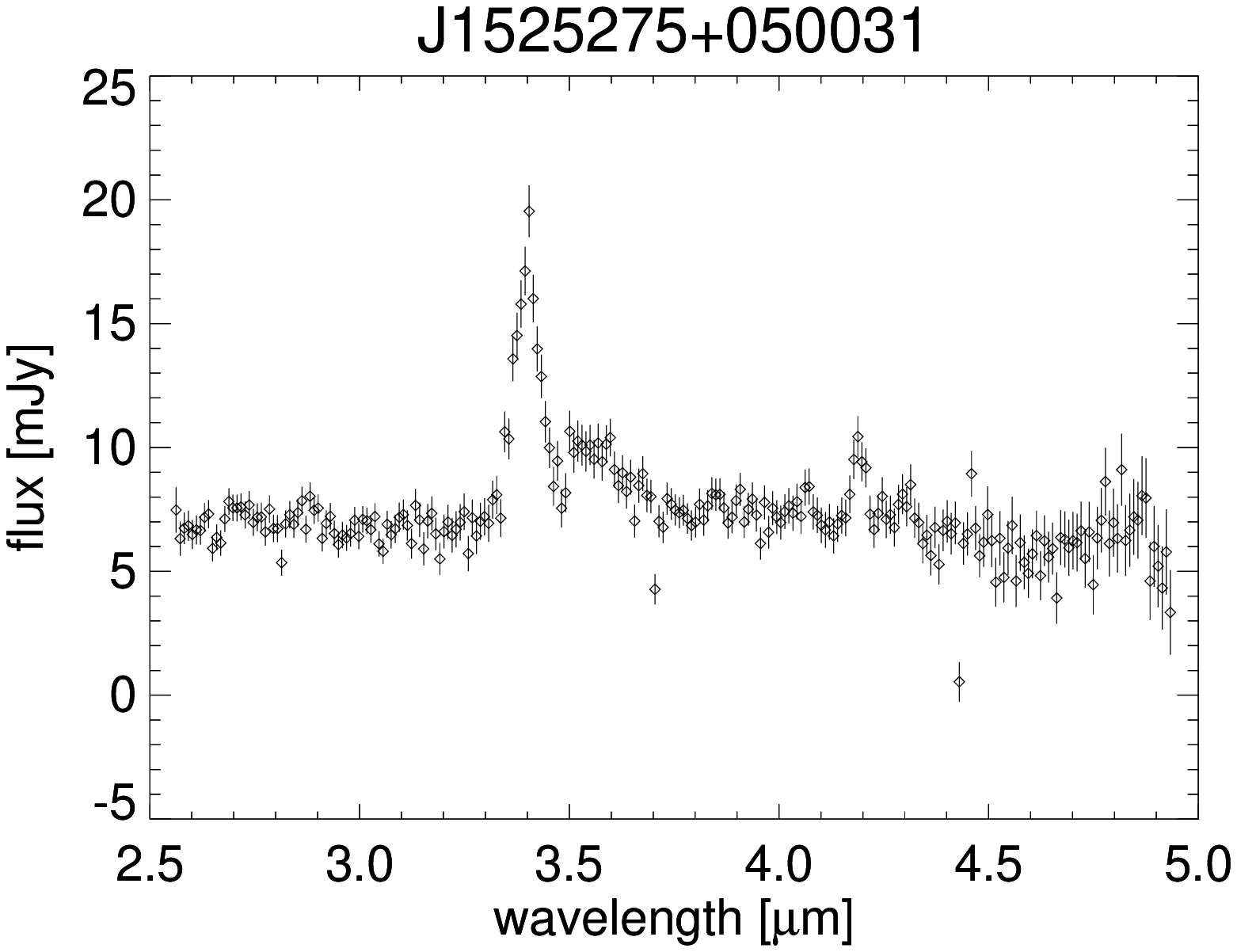}

     \FigureFile(41mm,26mm){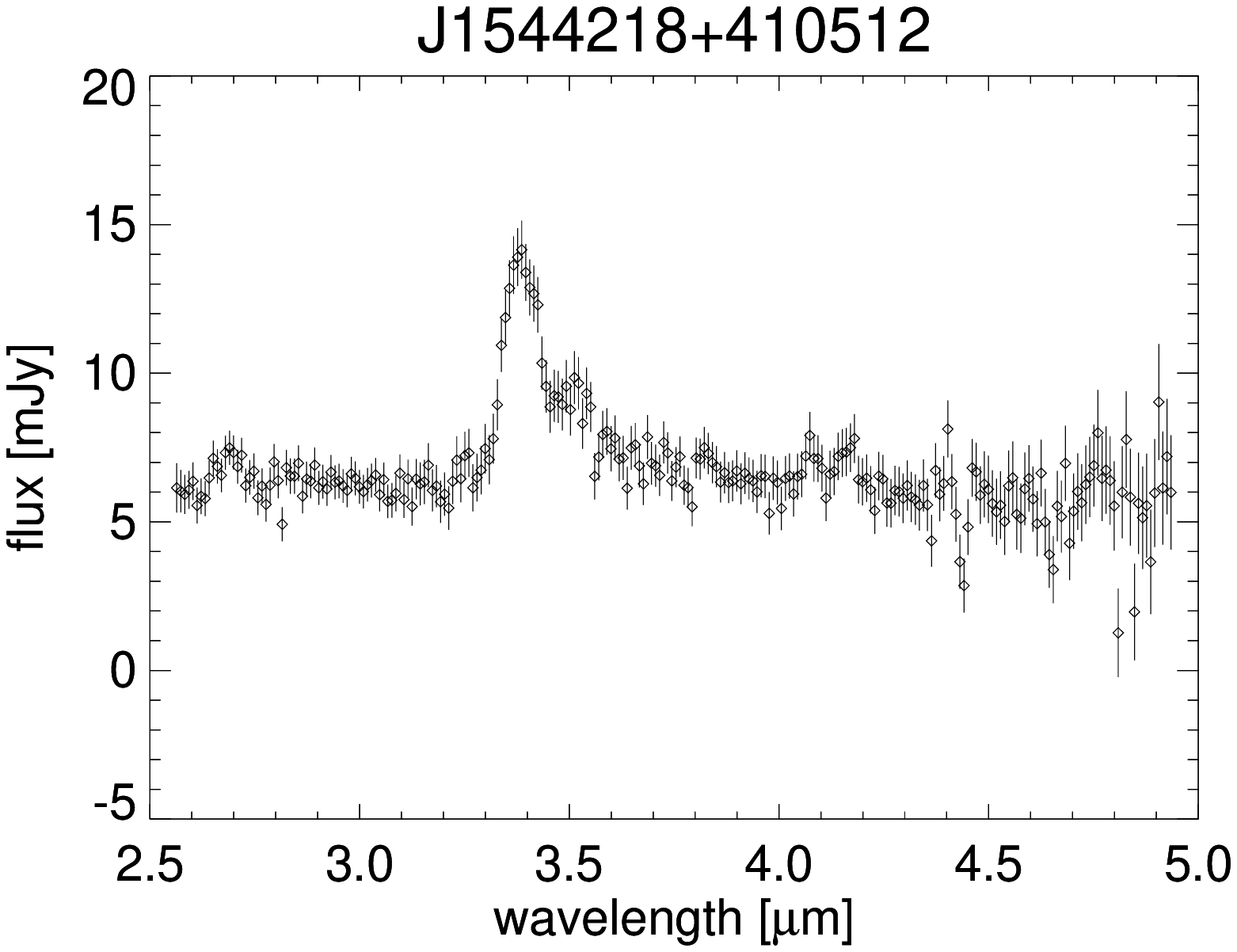}
     \FigureFile(41mm,26mm){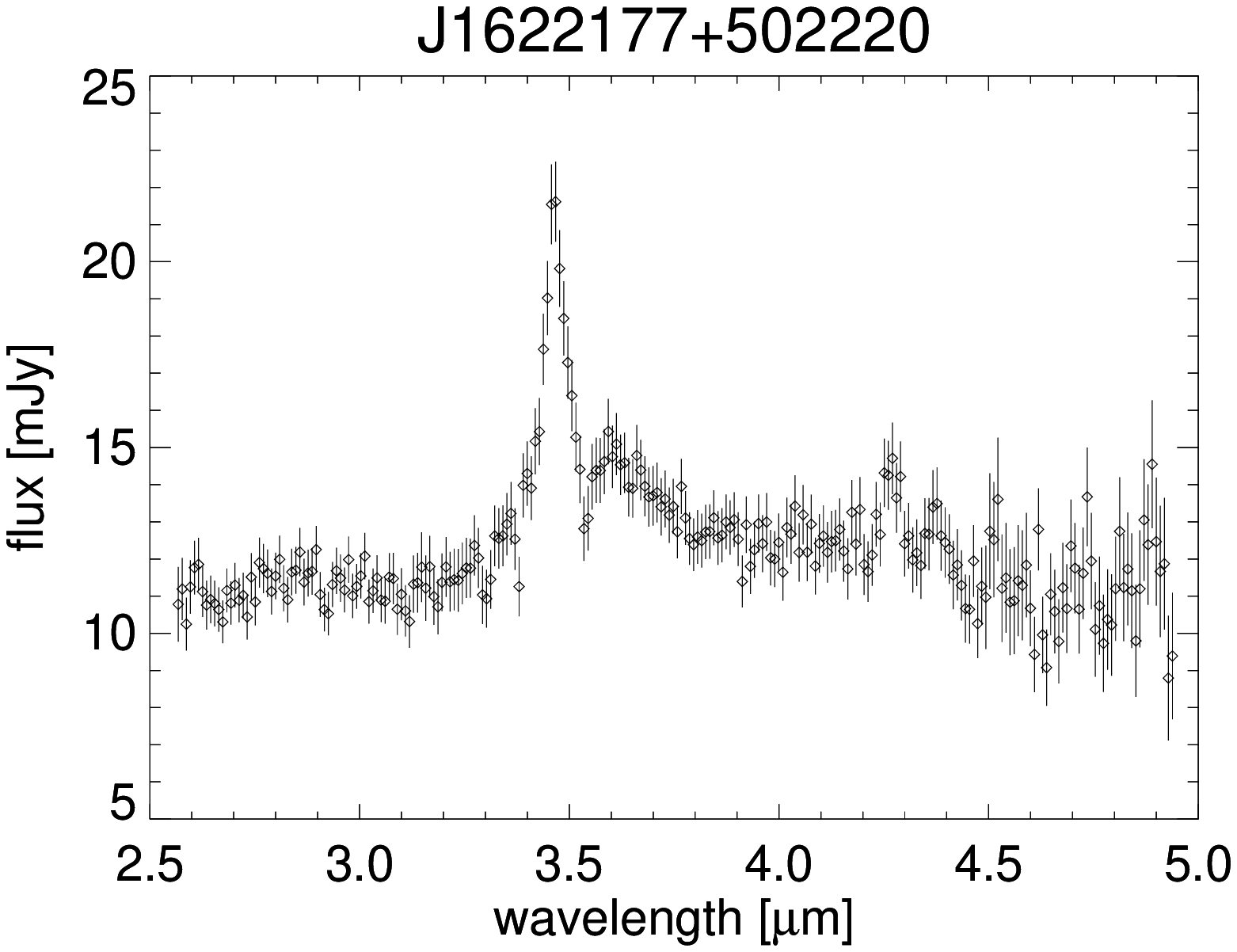}
     \FigureFile(41mm,26mm){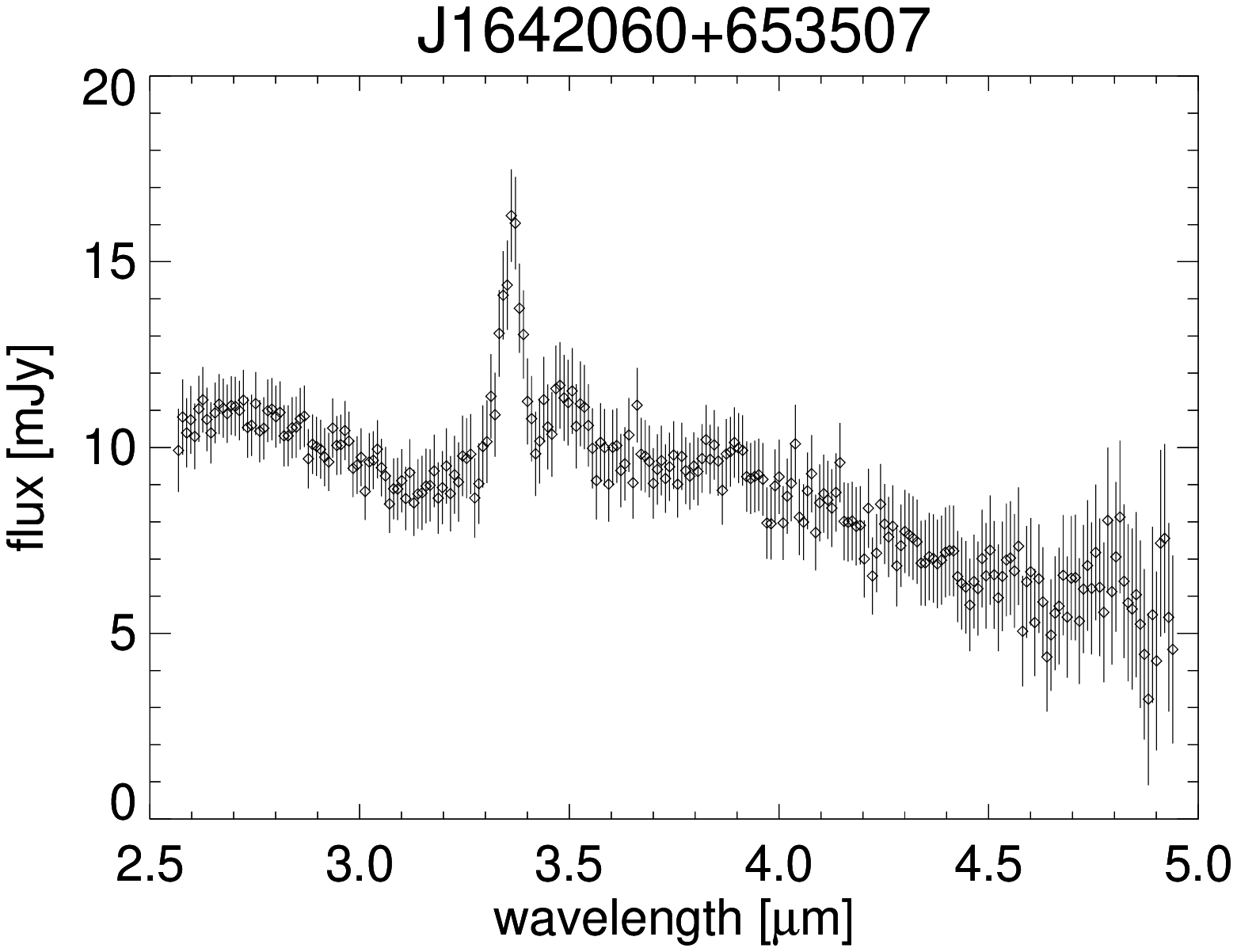}
     \FigureFile(41mm,26mm){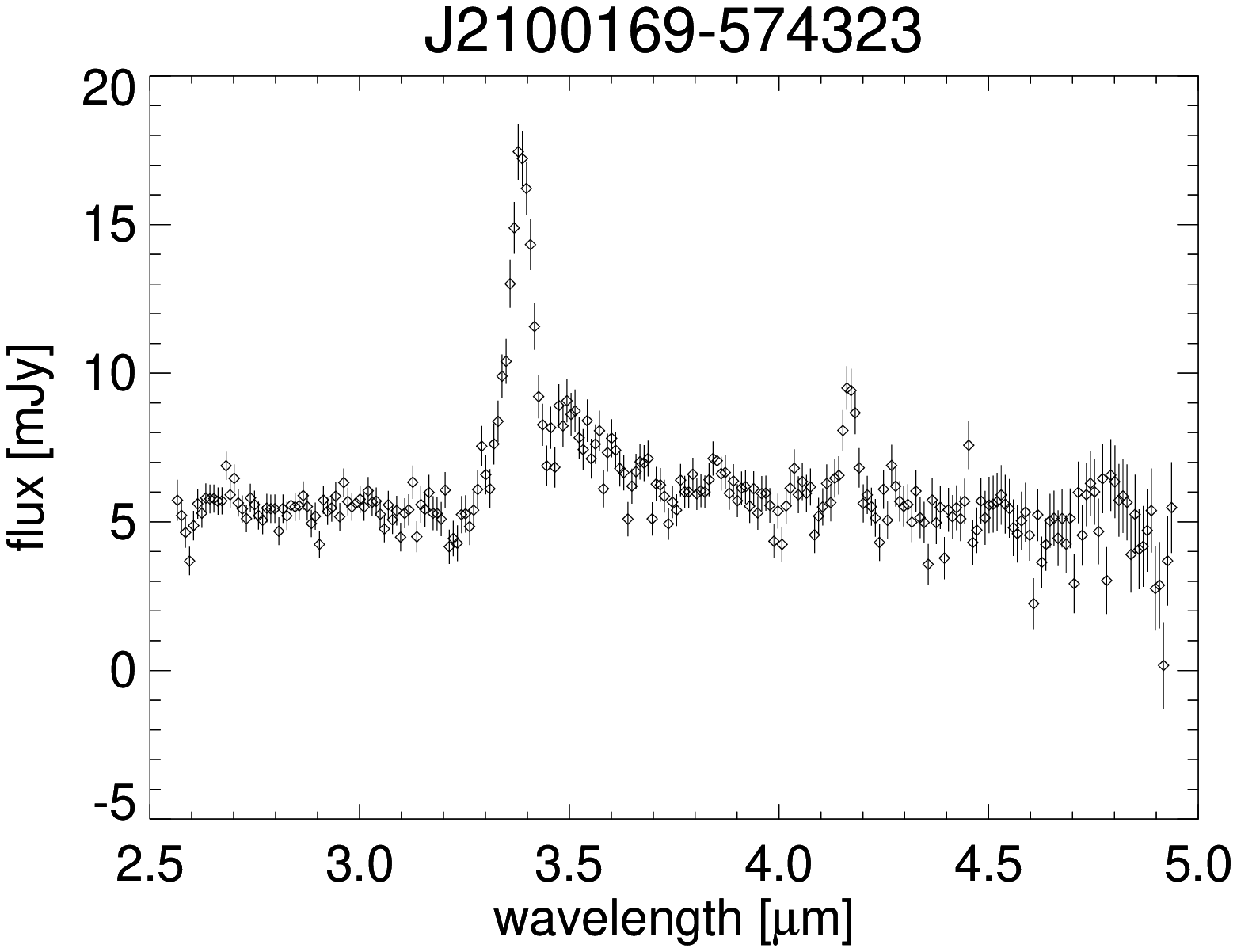}

     \FigureFile(41mm,26mm){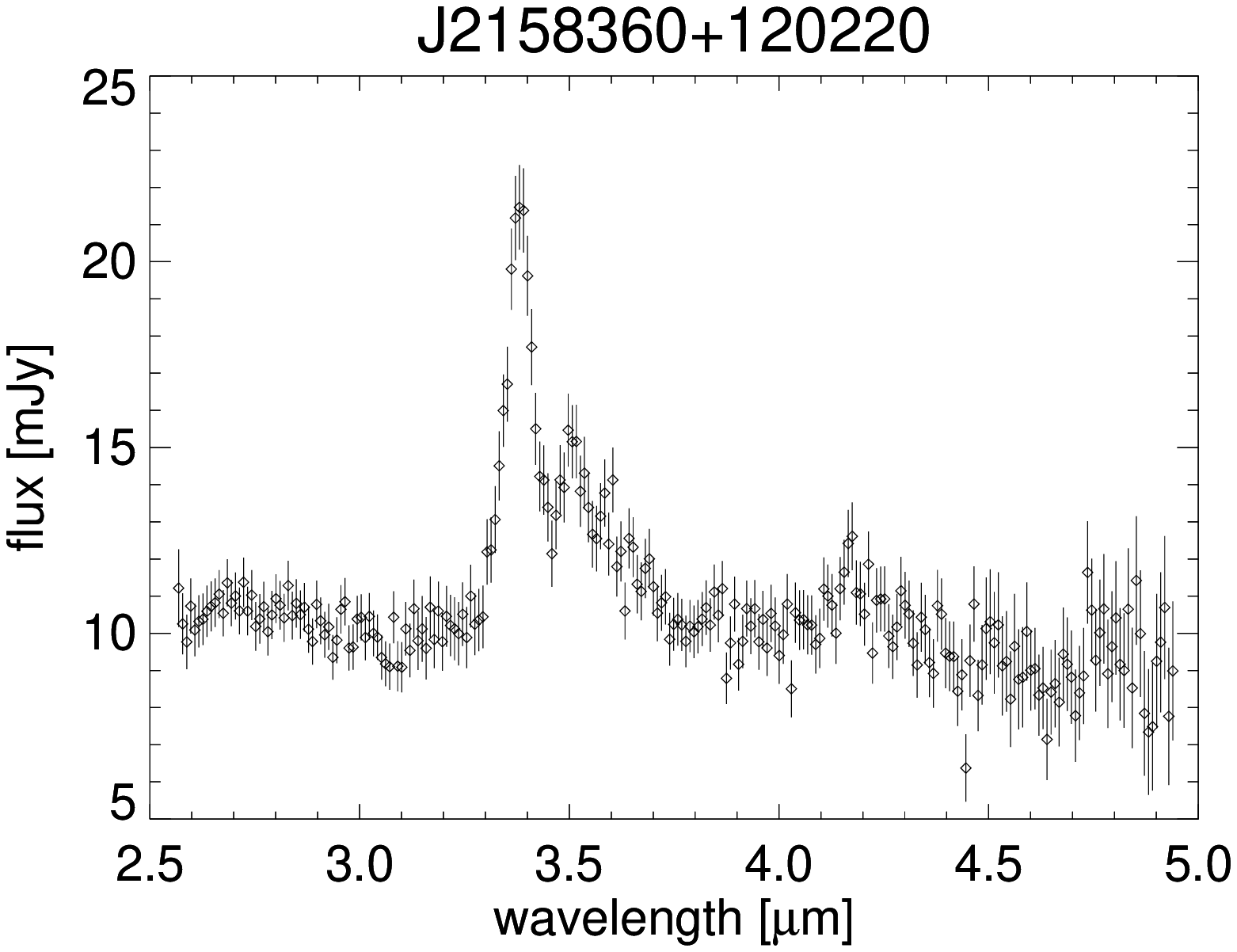}
     \FigureFile(41mm,26mm){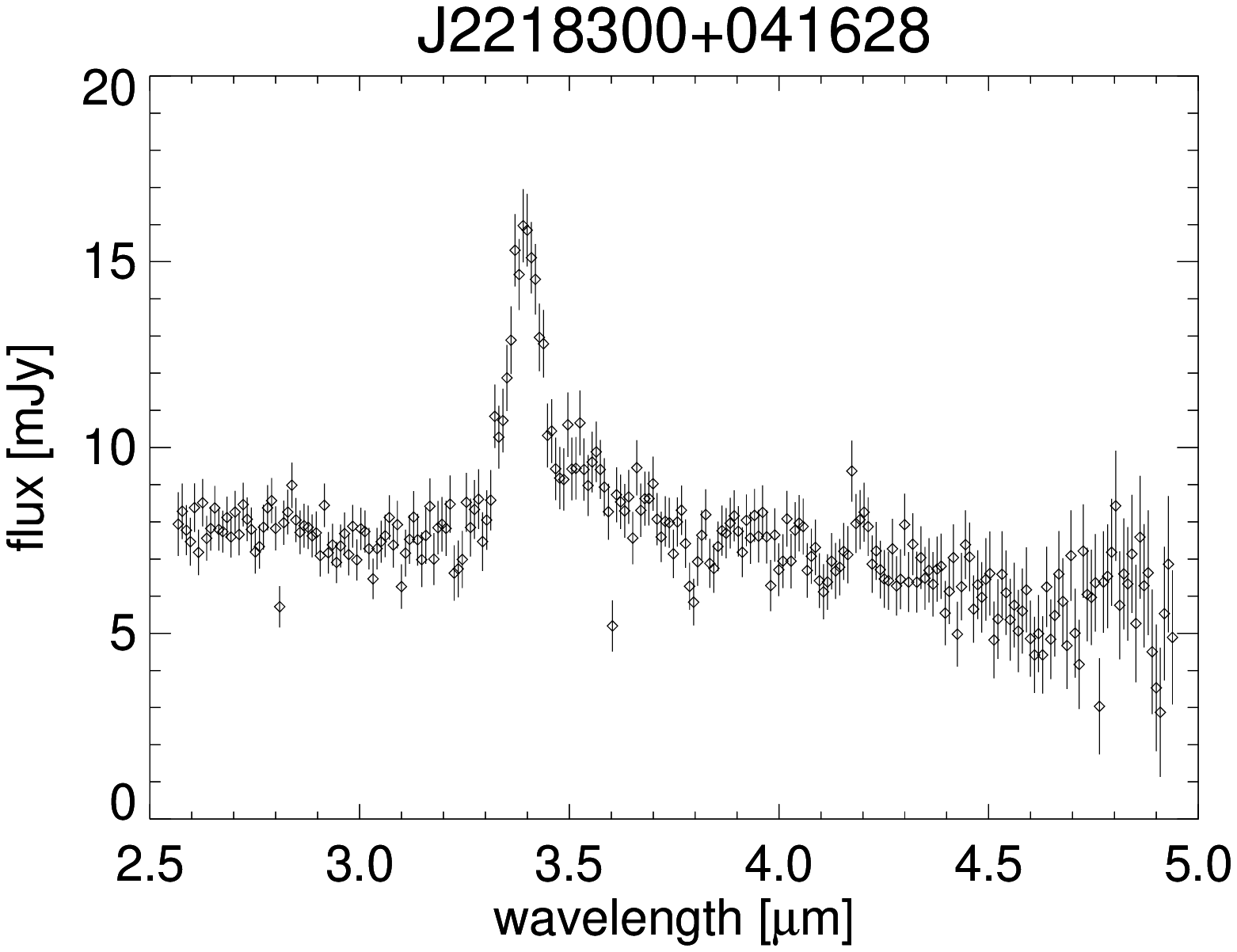}
     \FigureFile(41mm,26mm){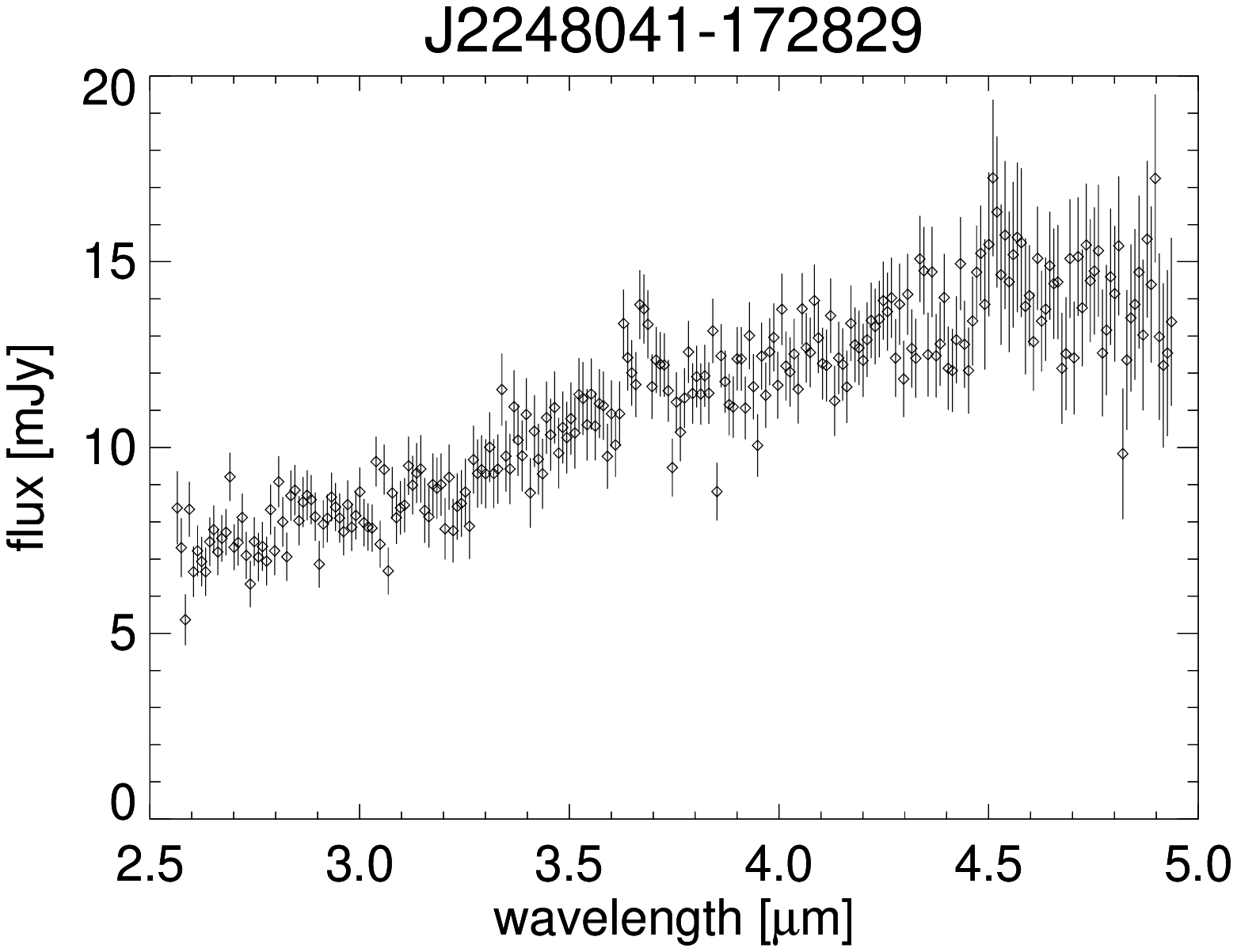}

 \end{center}
    \caption{AKARI IRC 2.5-5 $\mu$m spectra of the galaxies with a 5 $\sigma$ detection of the PAH 3.3 $\mu$m emission feature.} \label{1}
\end{figure}

\begin{figure}
 \begin{center}
     \FigureFile(41mm,26mm){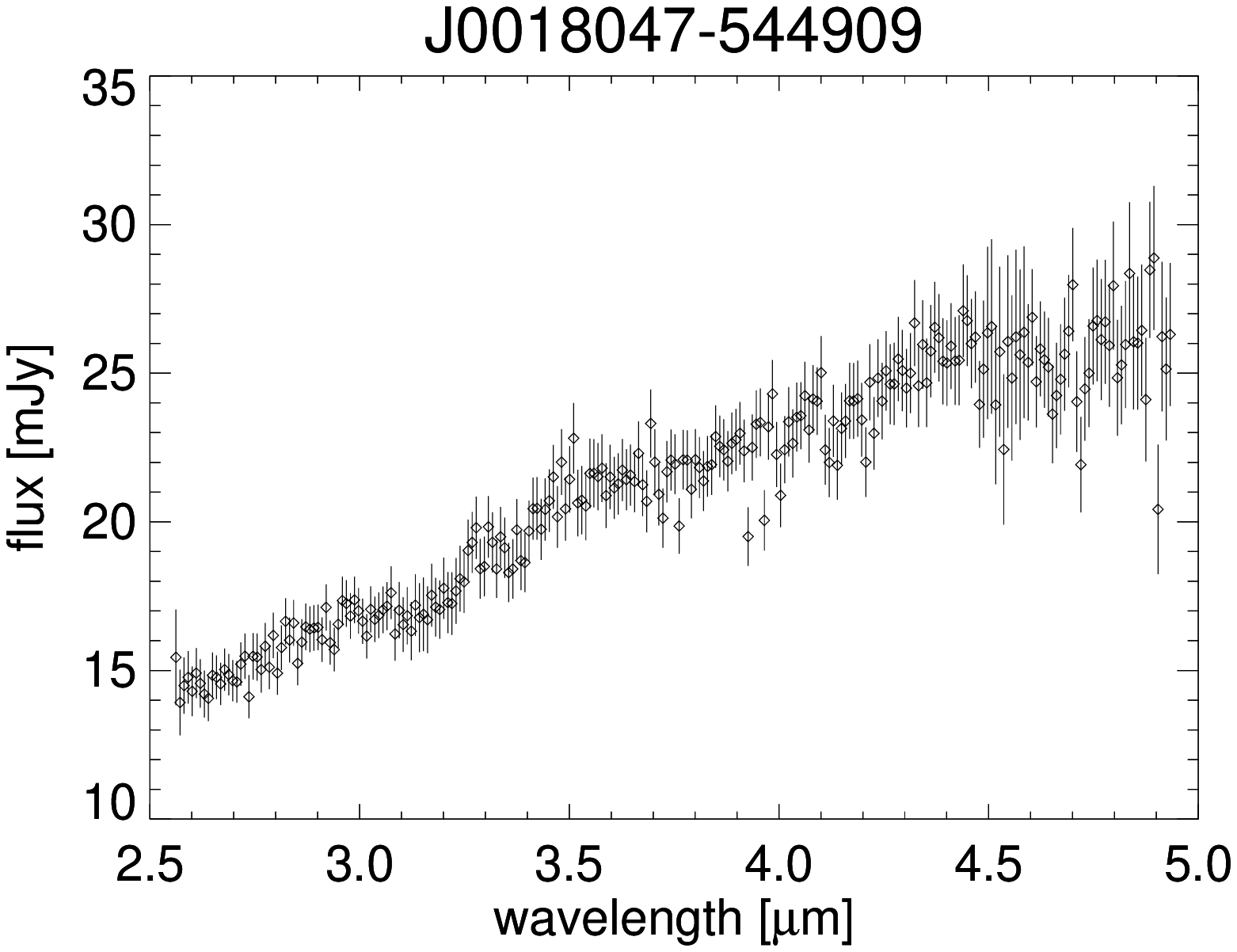}
     \FigureFile(41mm,26mm){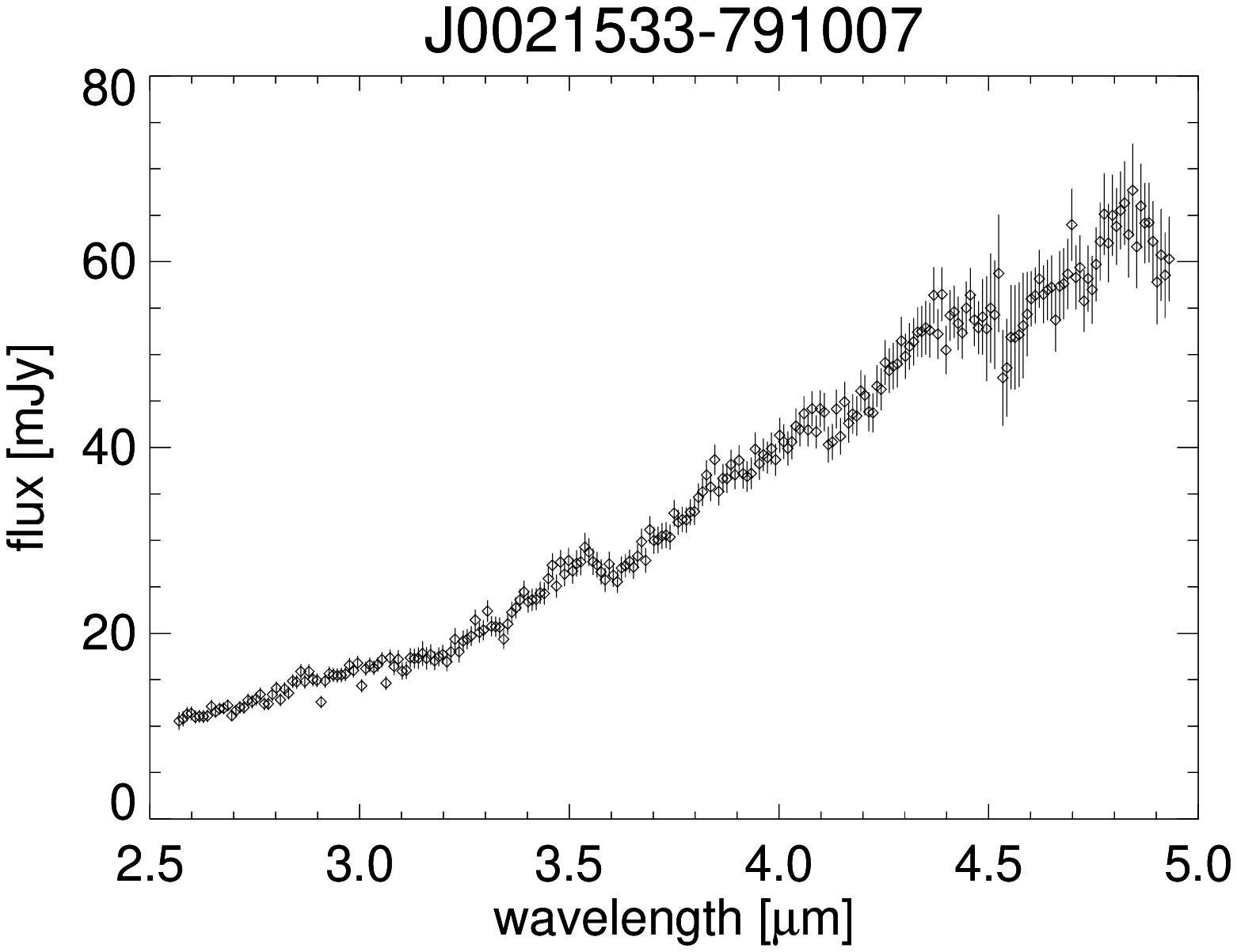}
     \FigureFile(41mm,26mm){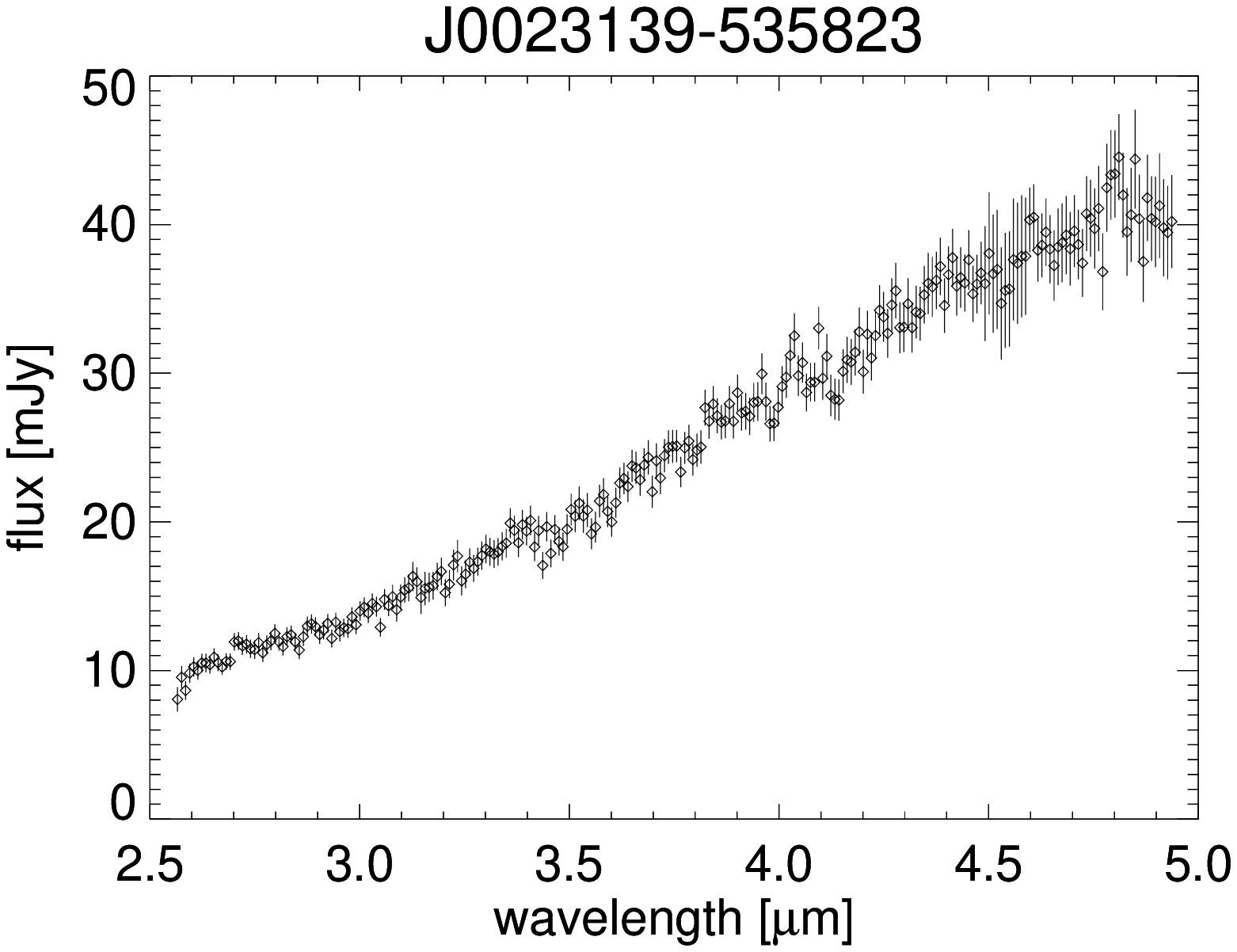}
     \FigureFile(41mm,26mm){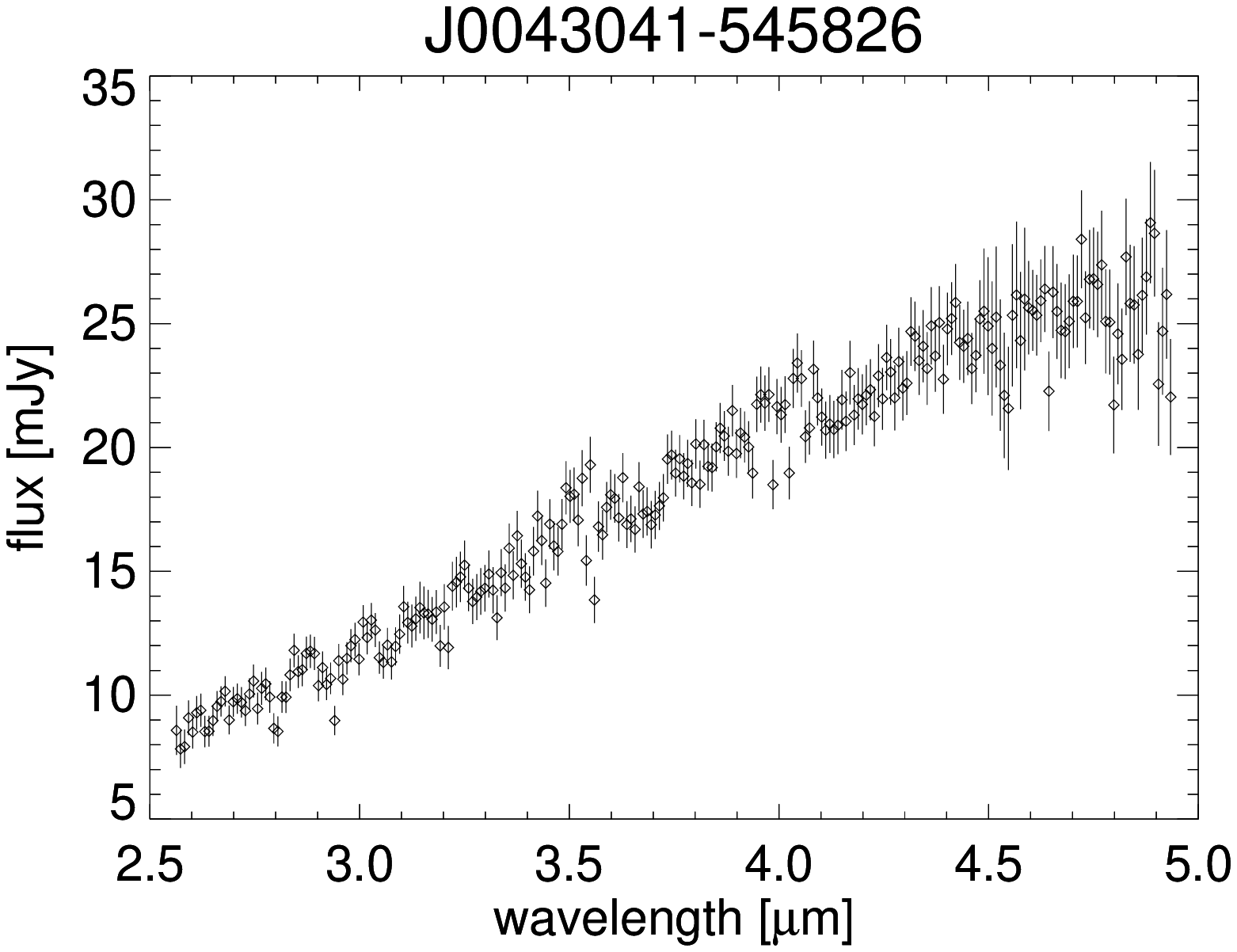}

     \FigureFile(41mm,26mm){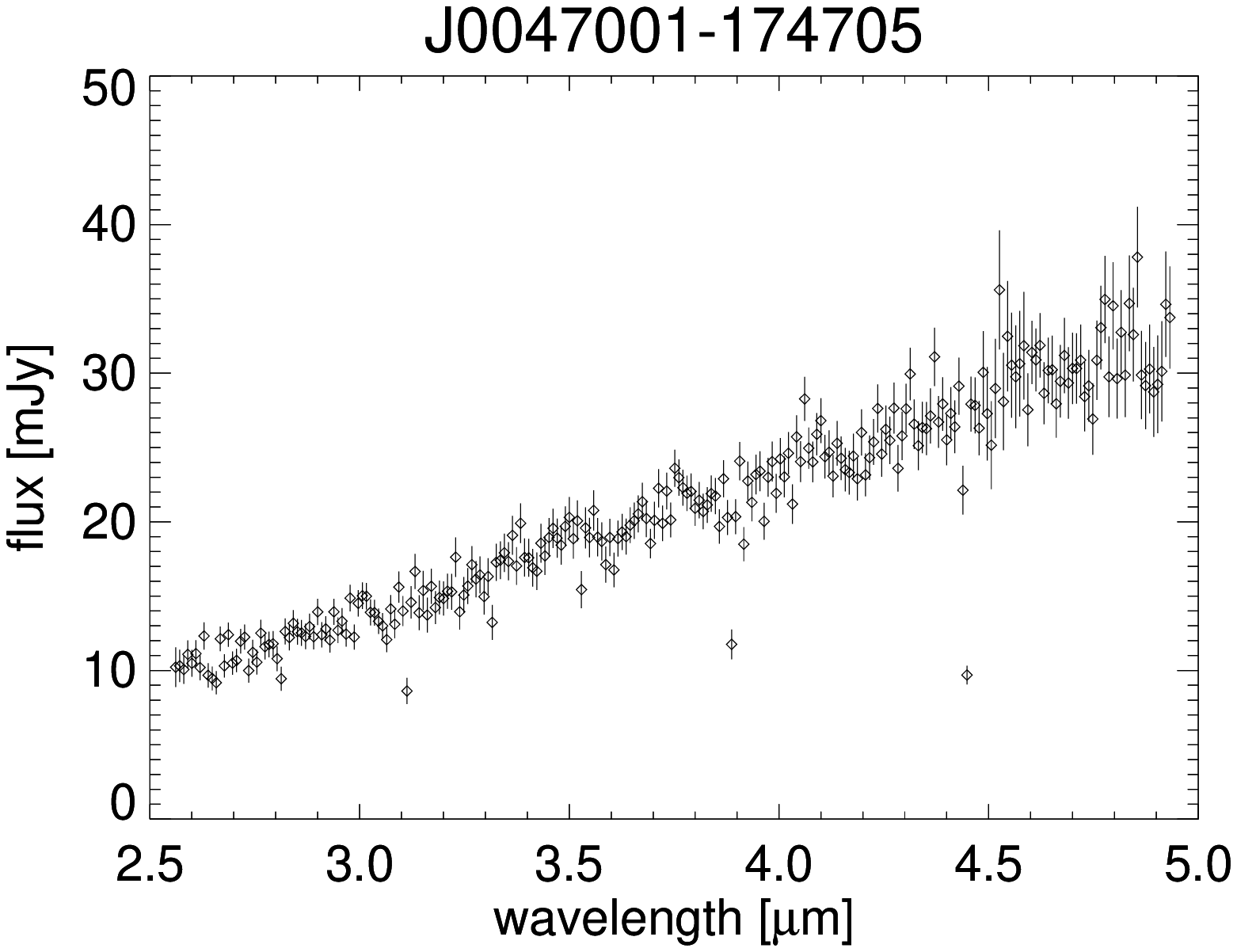}
     \FigureFile(41mm,26mm){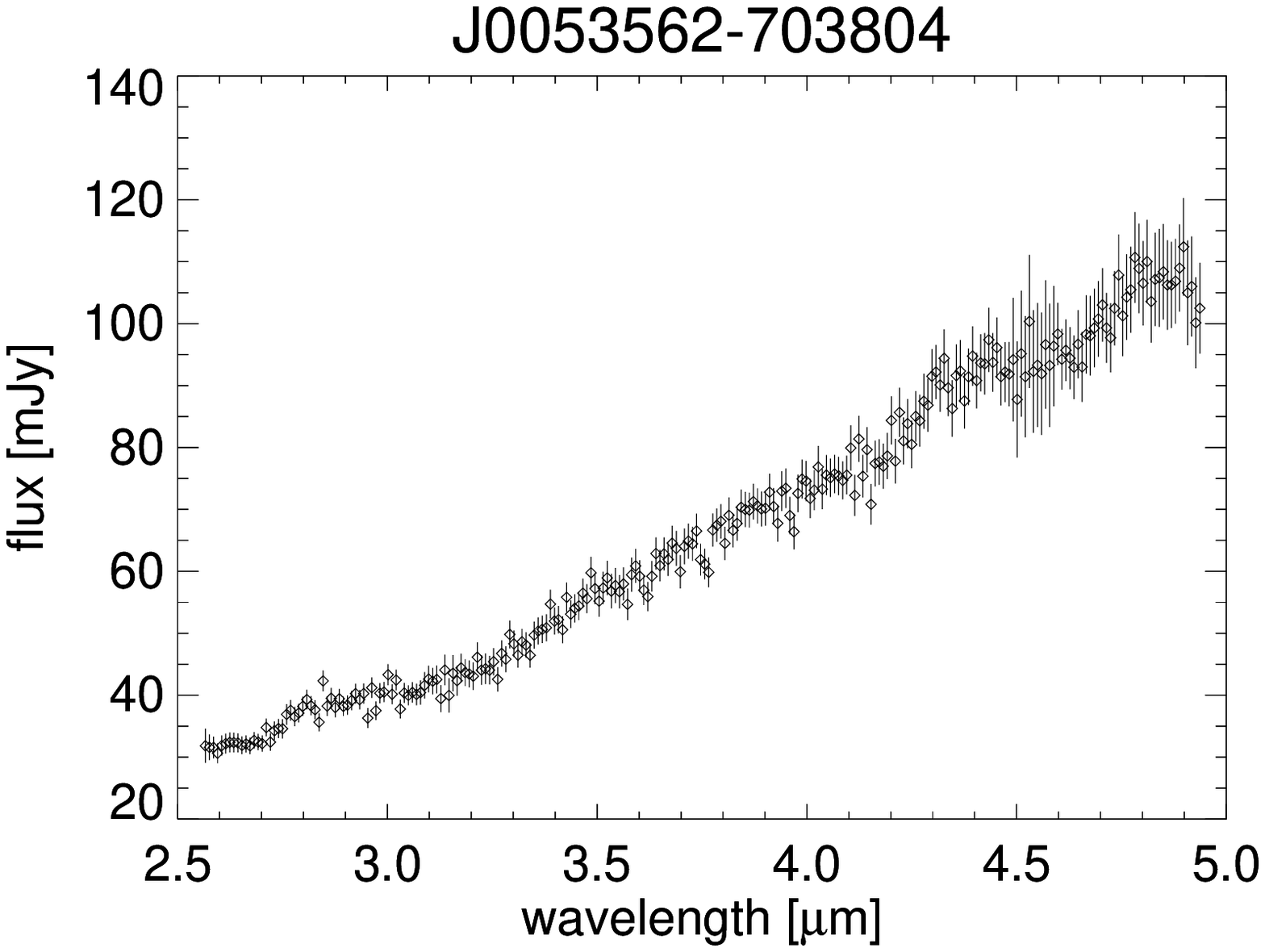}
     \FigureFile(41mm,26mm){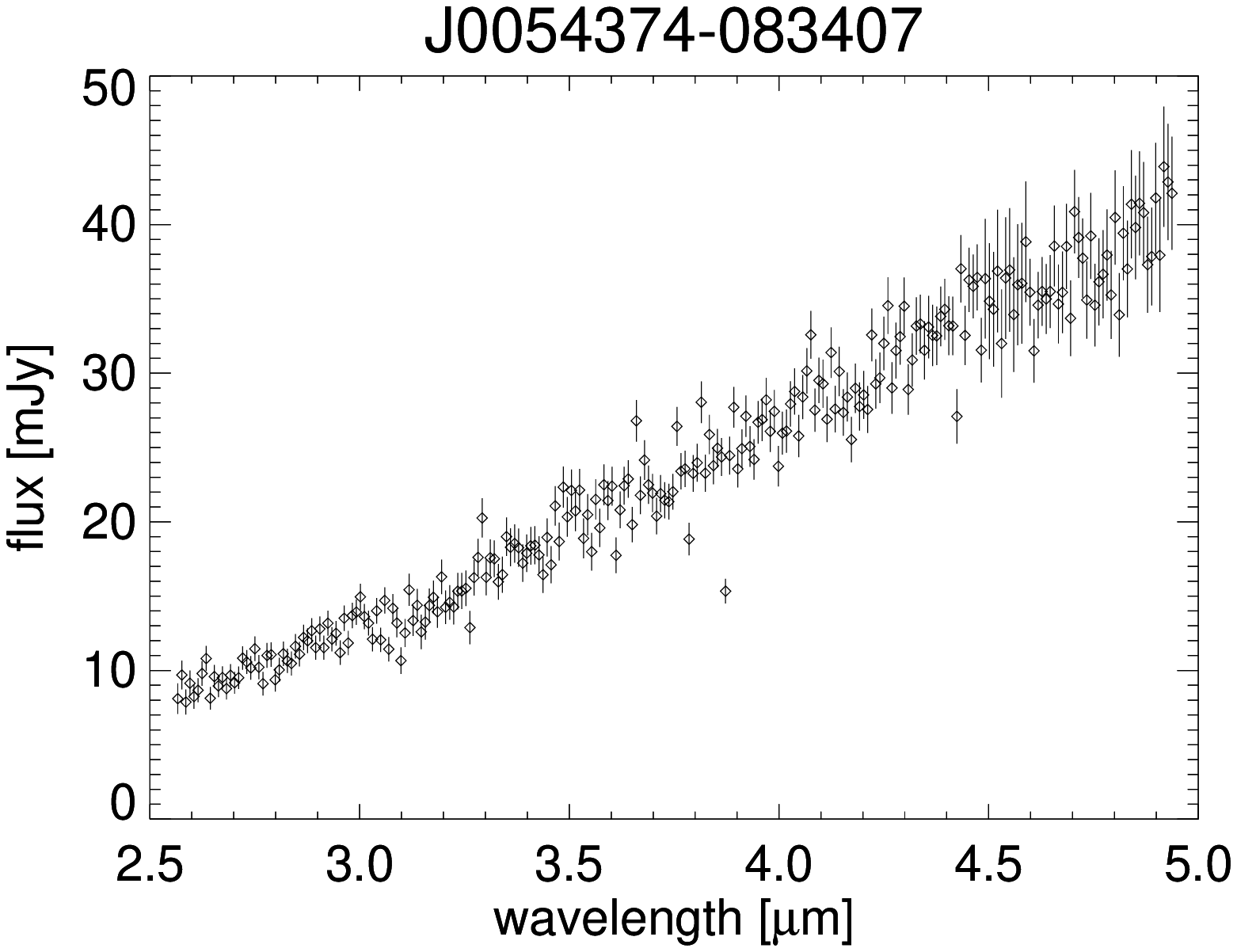}
     \FigureFile(41mm,26mm){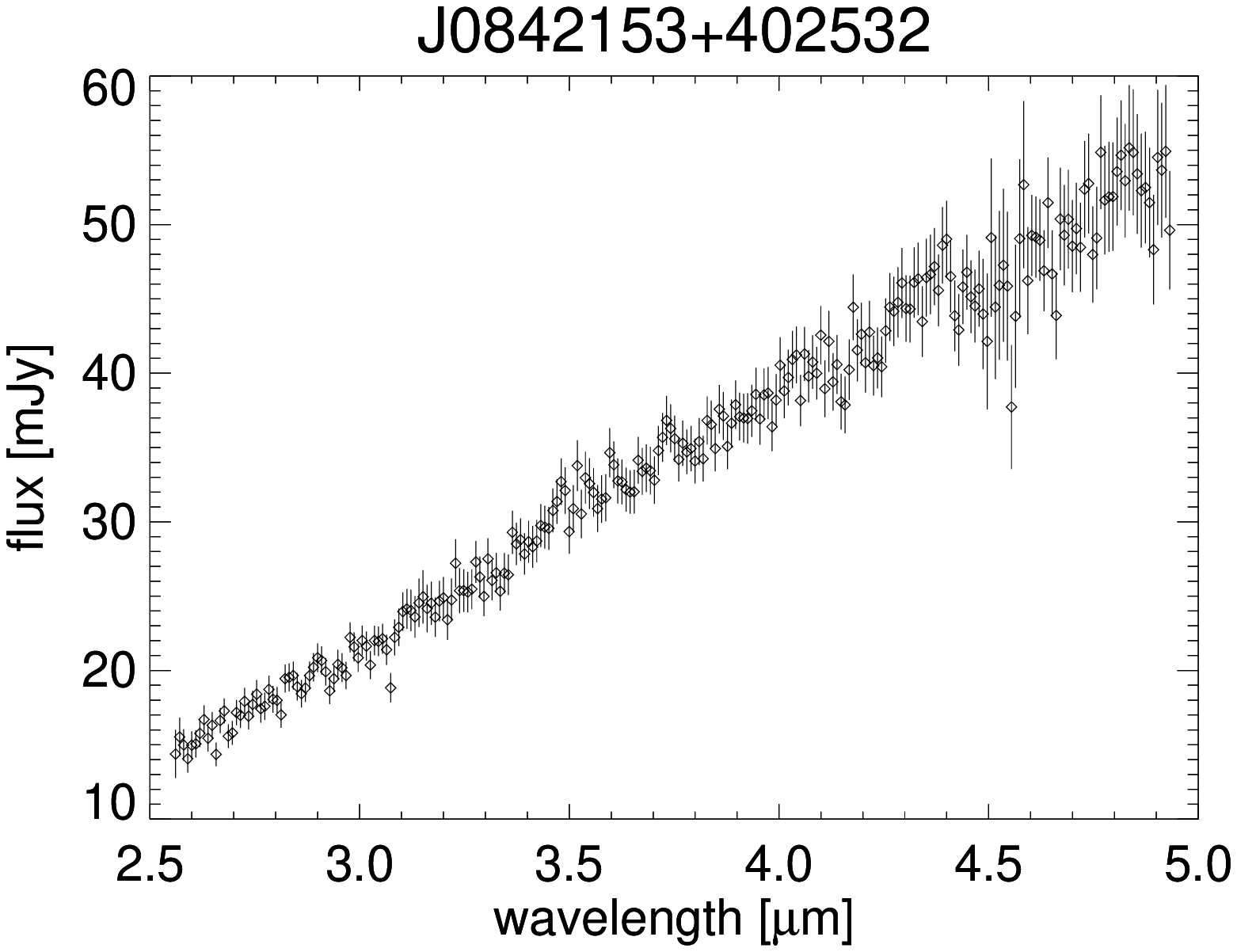}

     \FigureFile(41mm,26mm){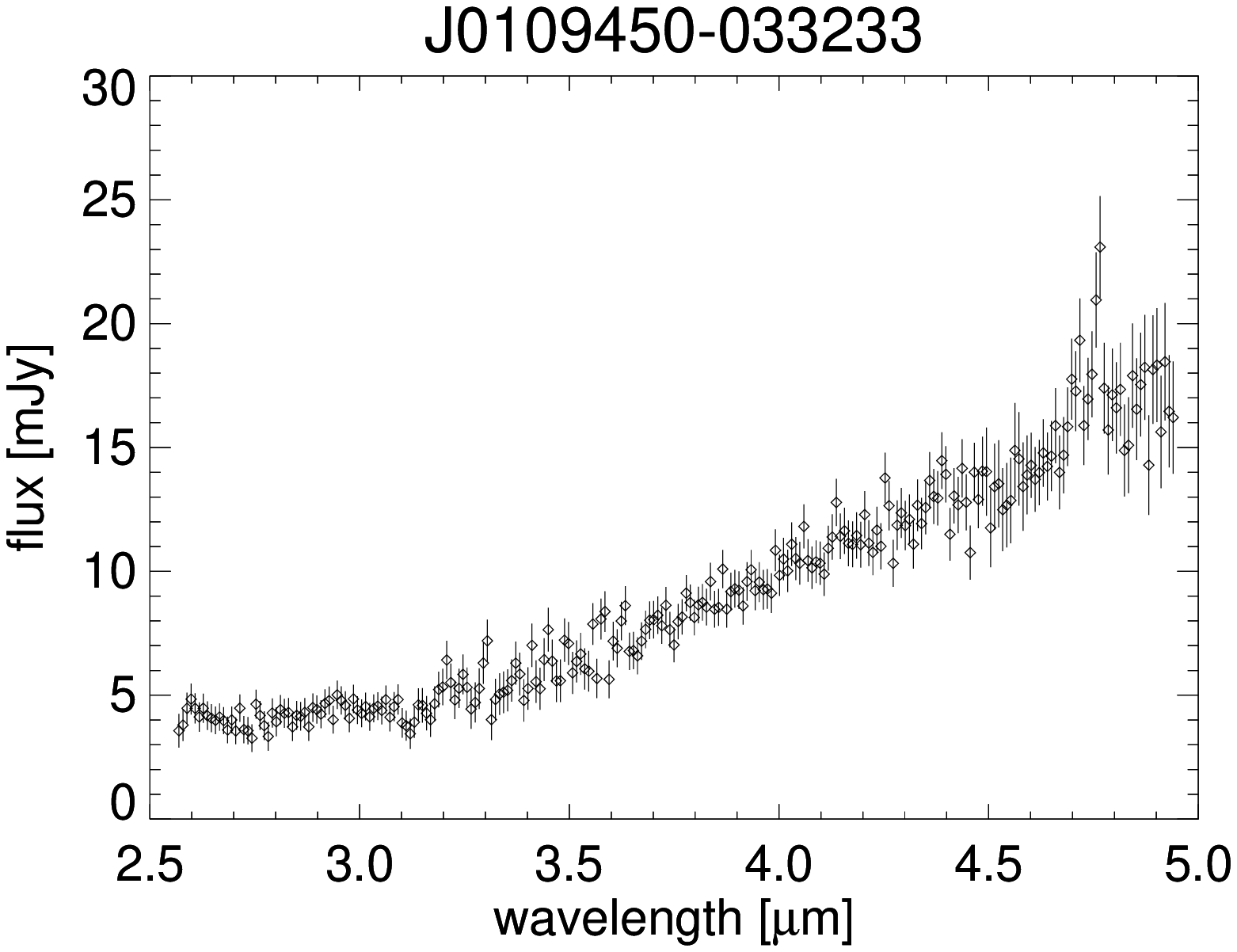}
     \FigureFile(41mm,26mm){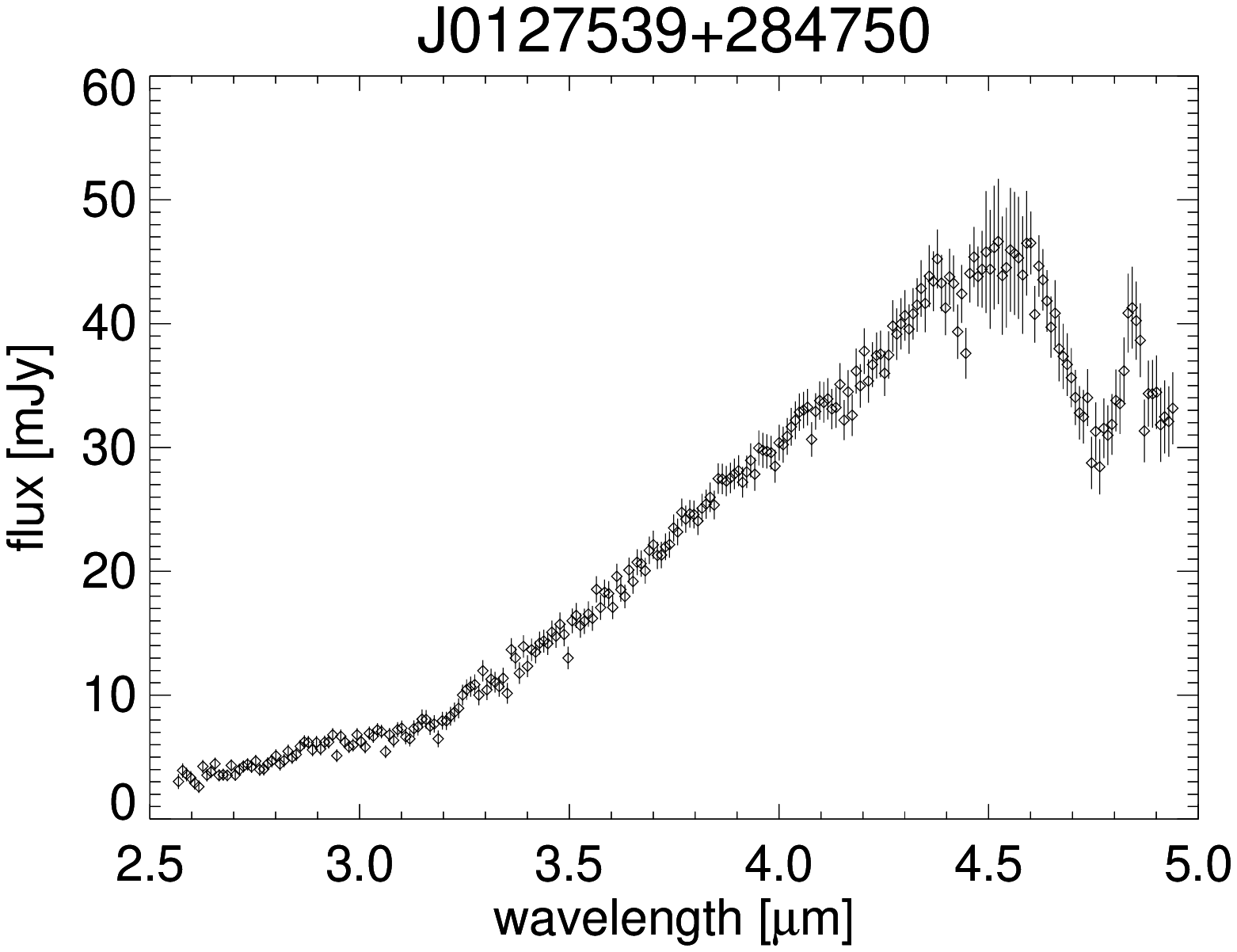}
     \FigureFile(41mm,26mm){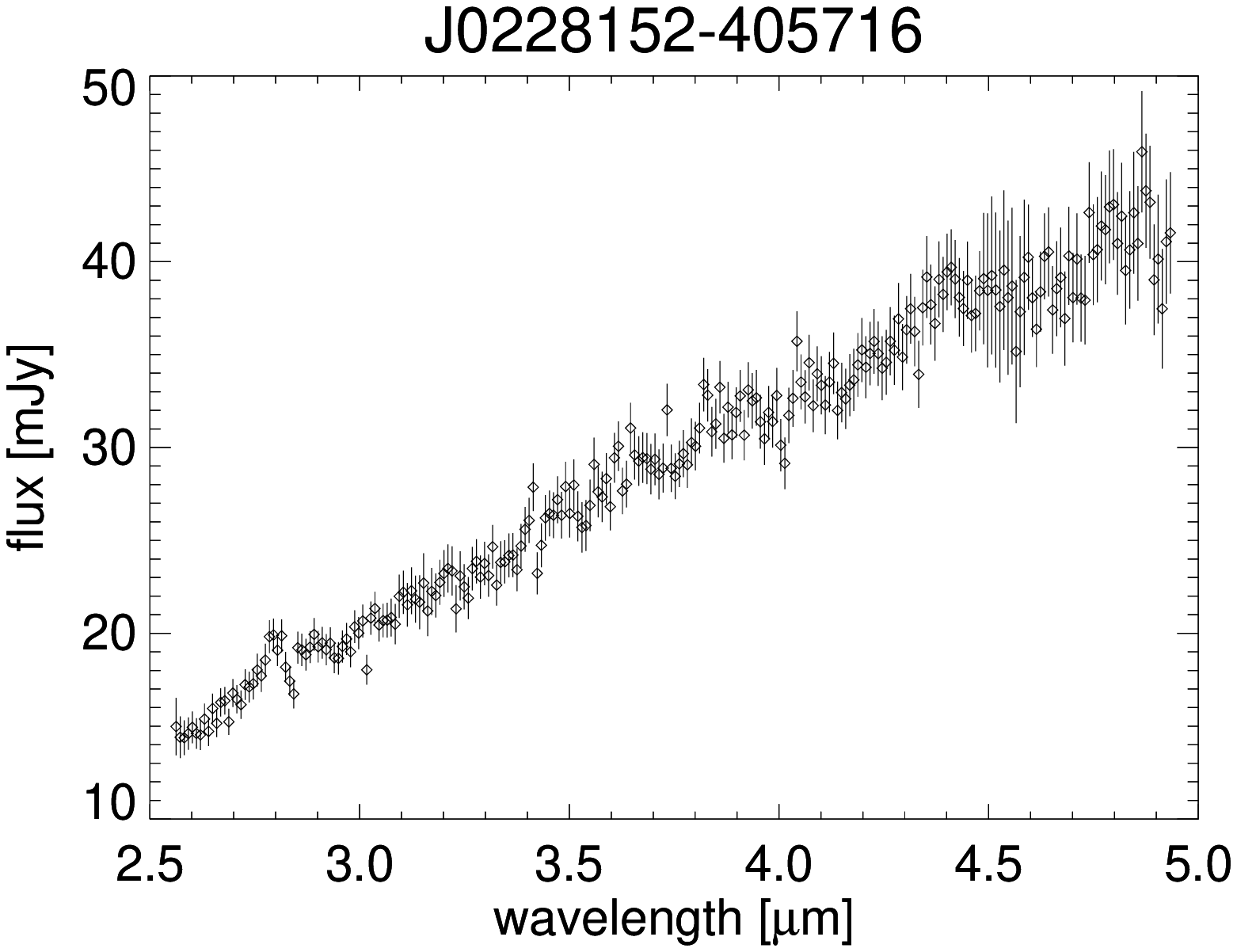}
     \FigureFile(41mm,26mm){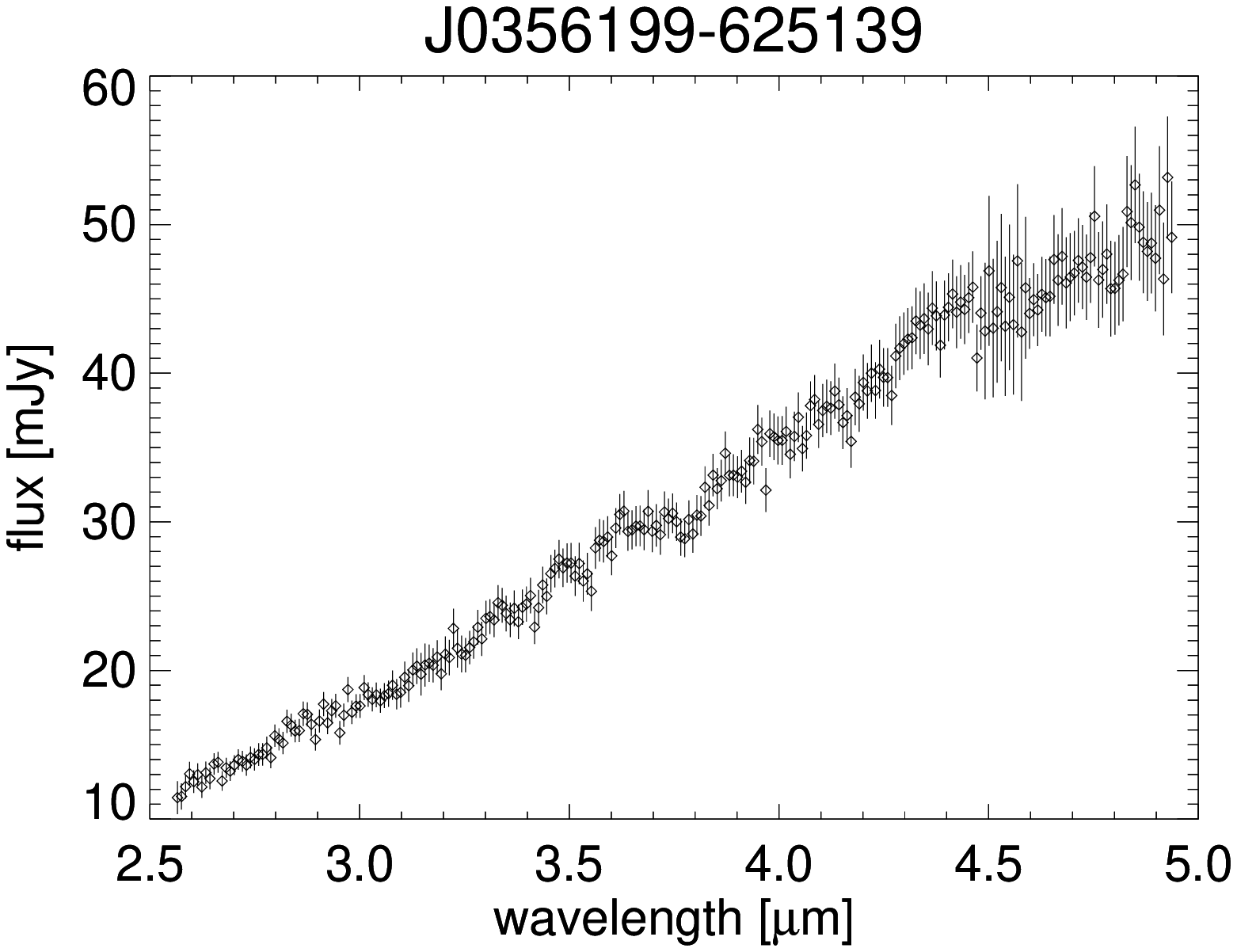}    

     \FigureFile(41mm,26mm){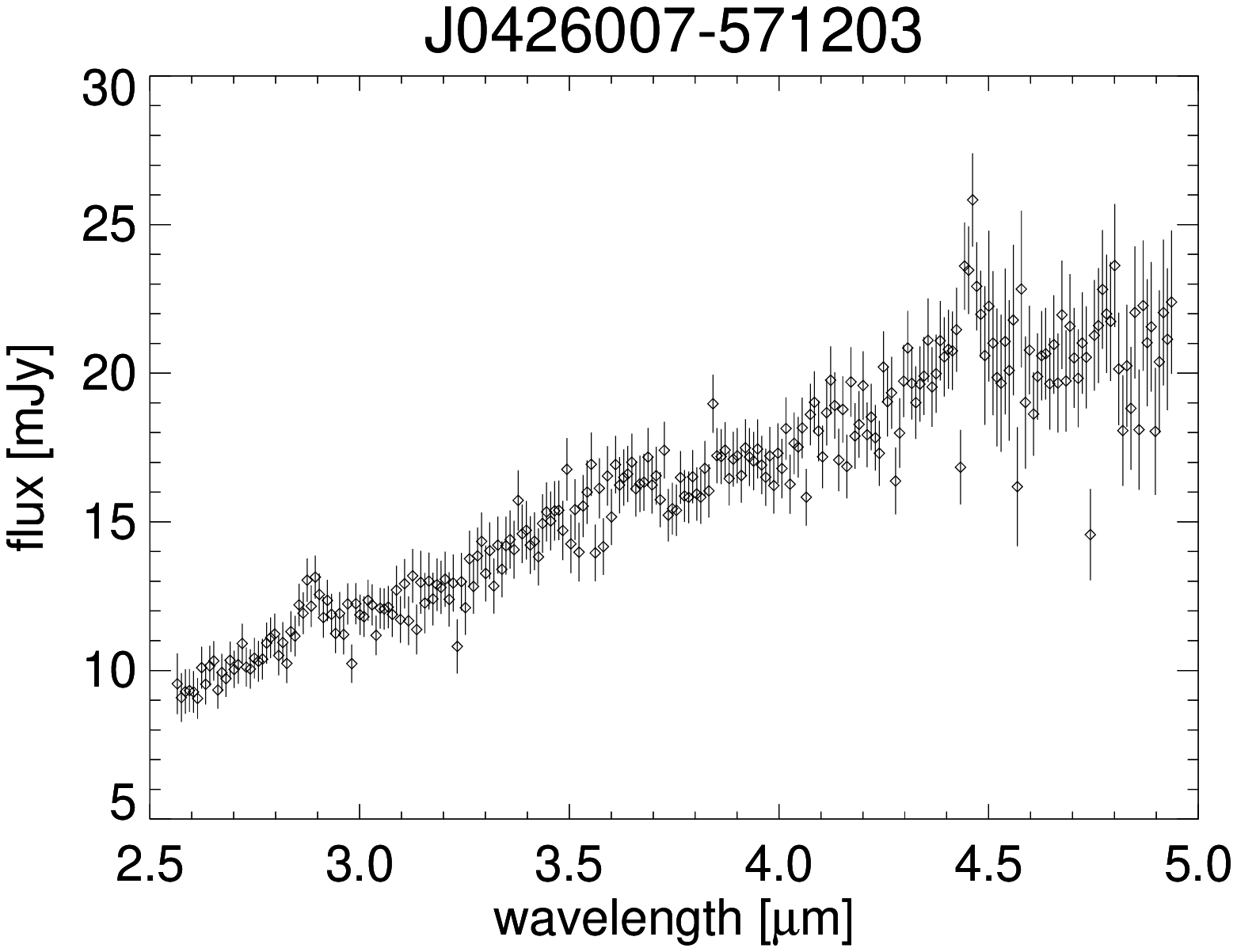}
     \FigureFile(41mm,26mm){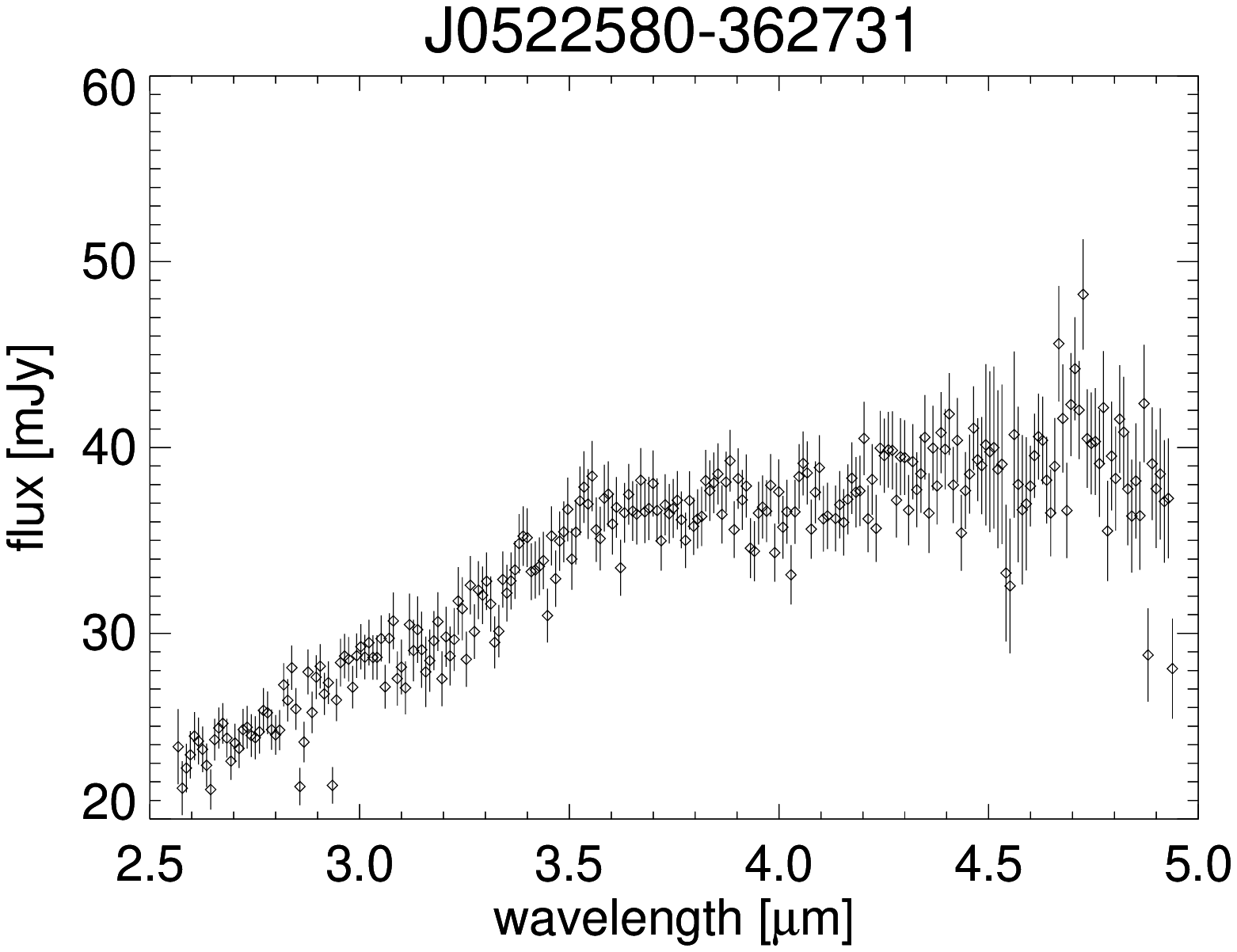}
     \FigureFile(41mm,26mm){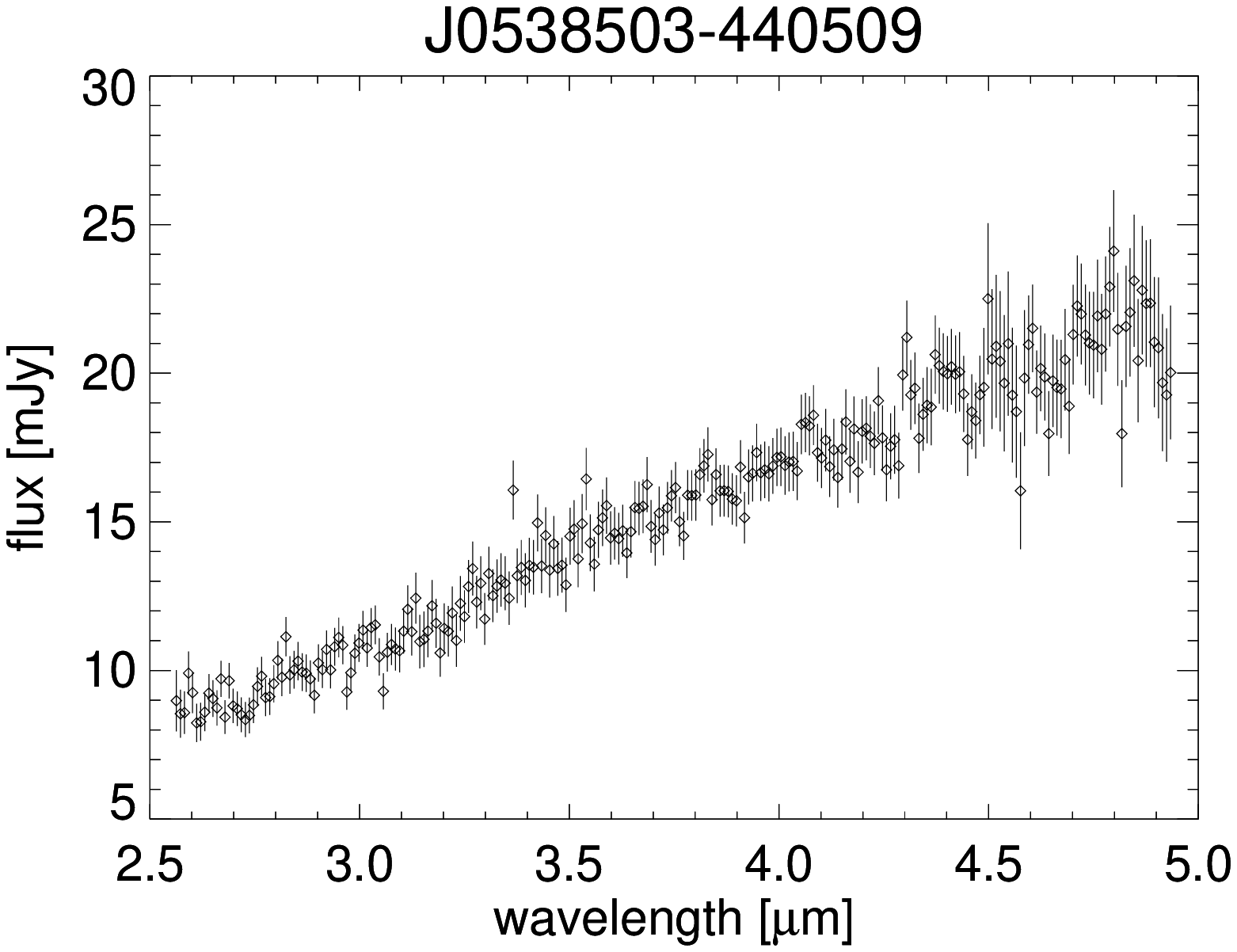}
     \FigureFile(41mm,26mm){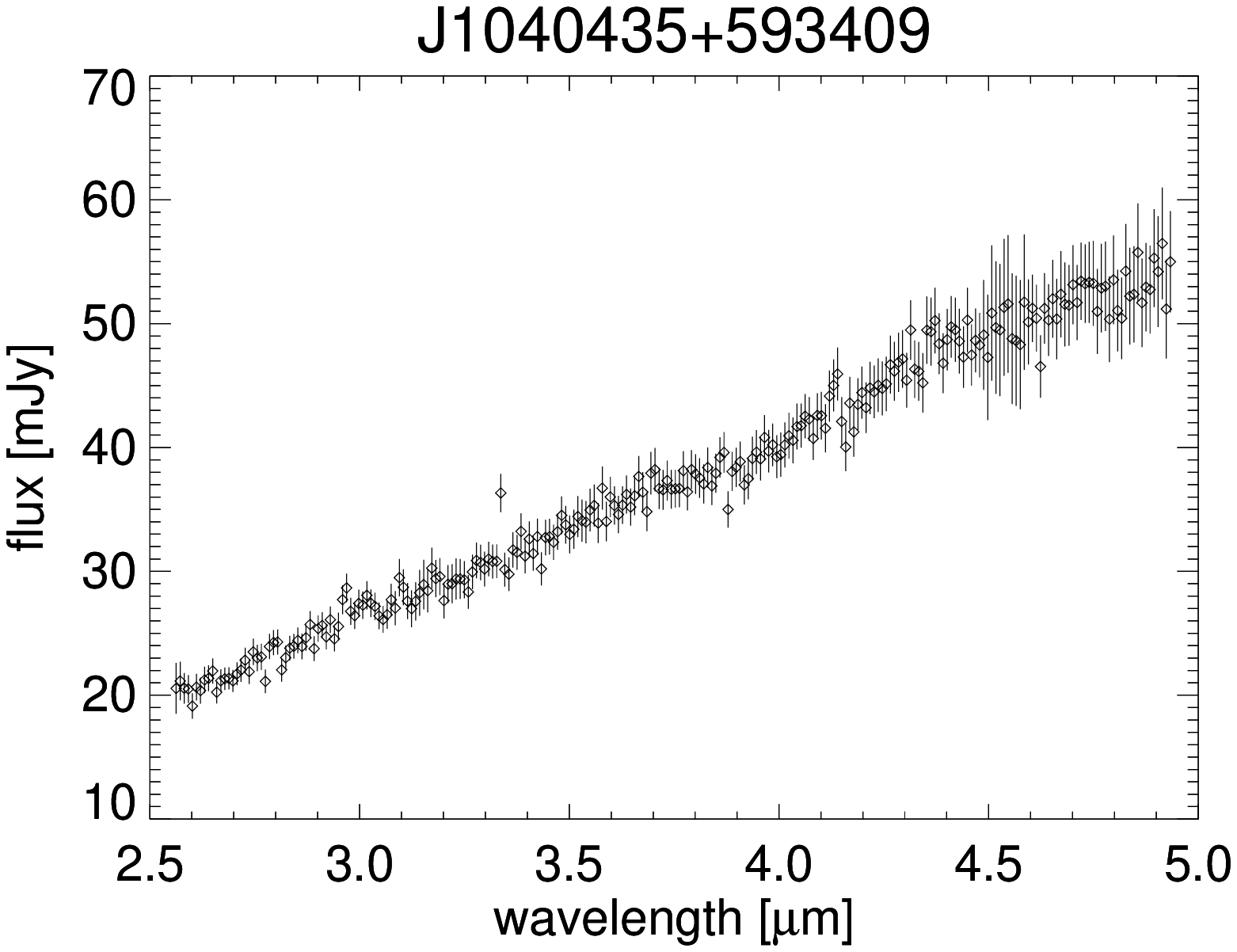}

     \FigureFile(41mm,26mm){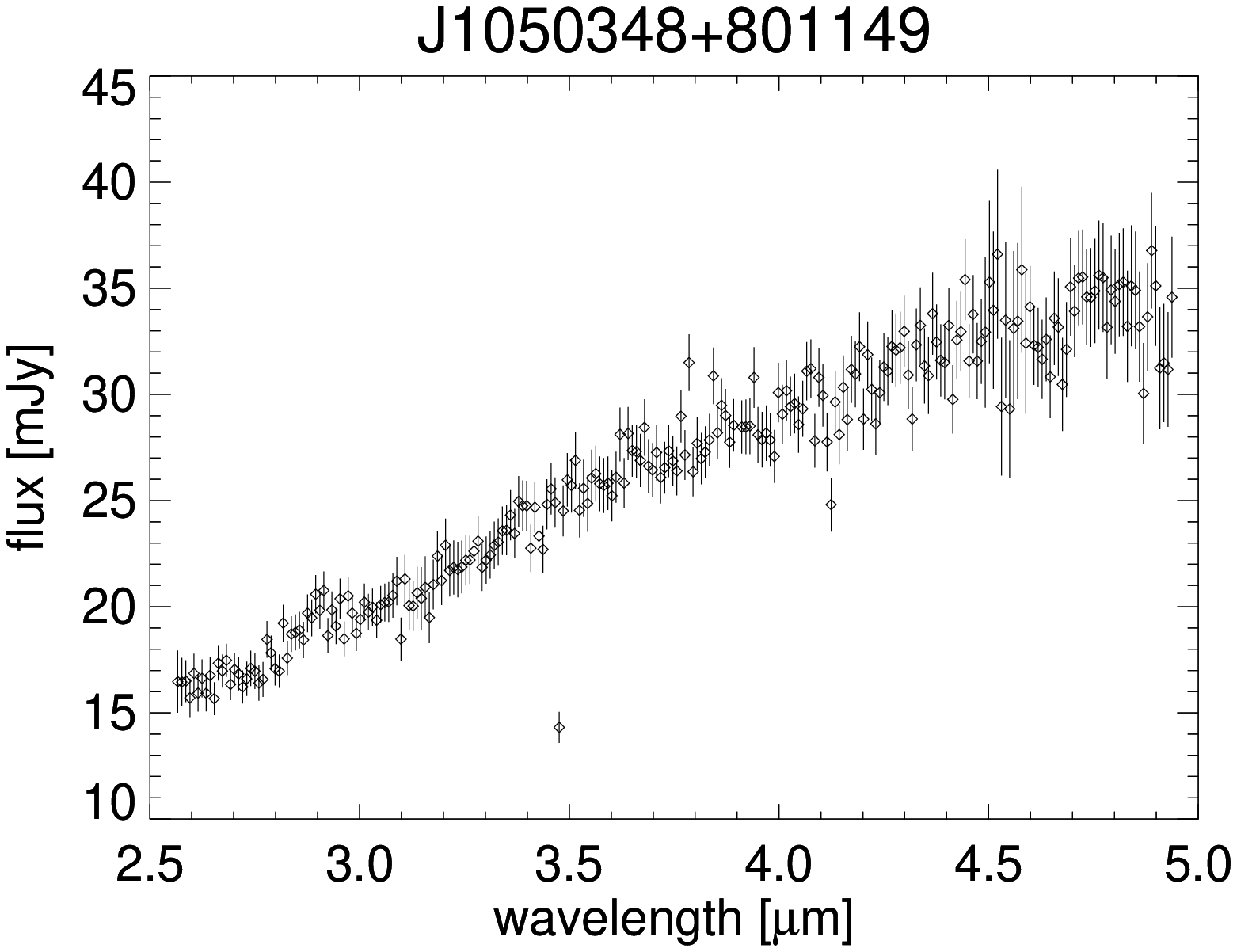}
     \FigureFile(41mm,26mm){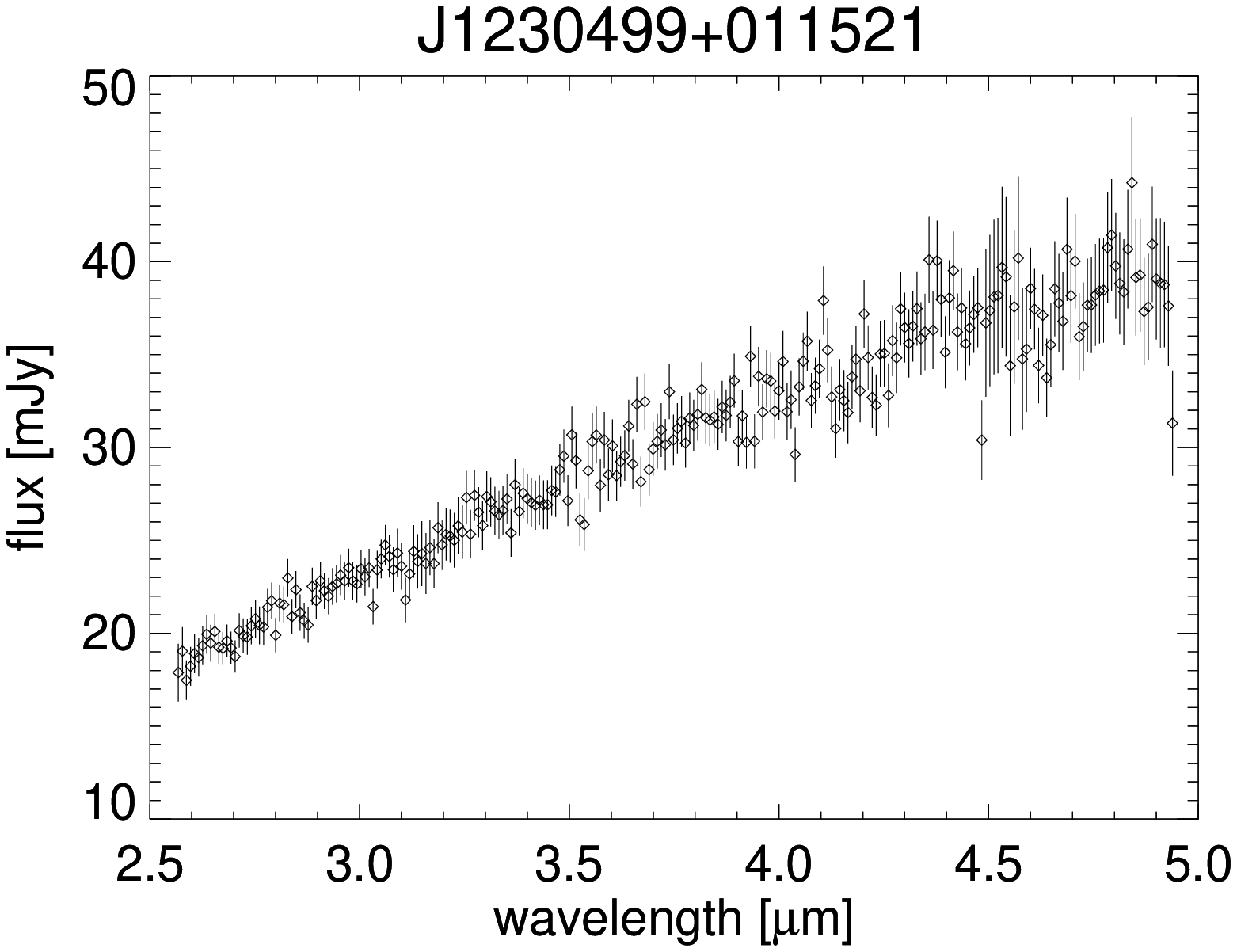}
     \FigureFile(41mm,26mm){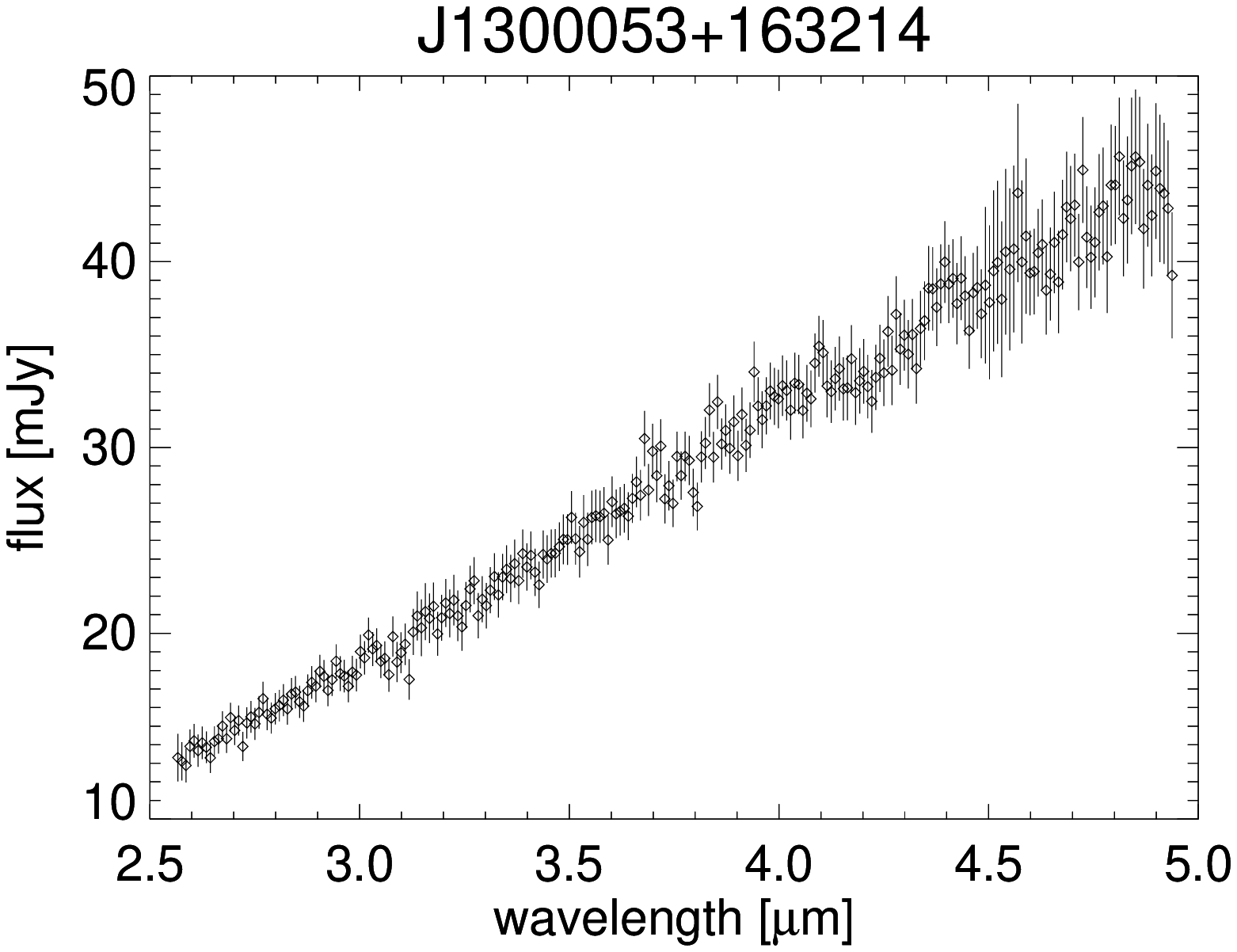}
     \FigureFile(41mm,26mm){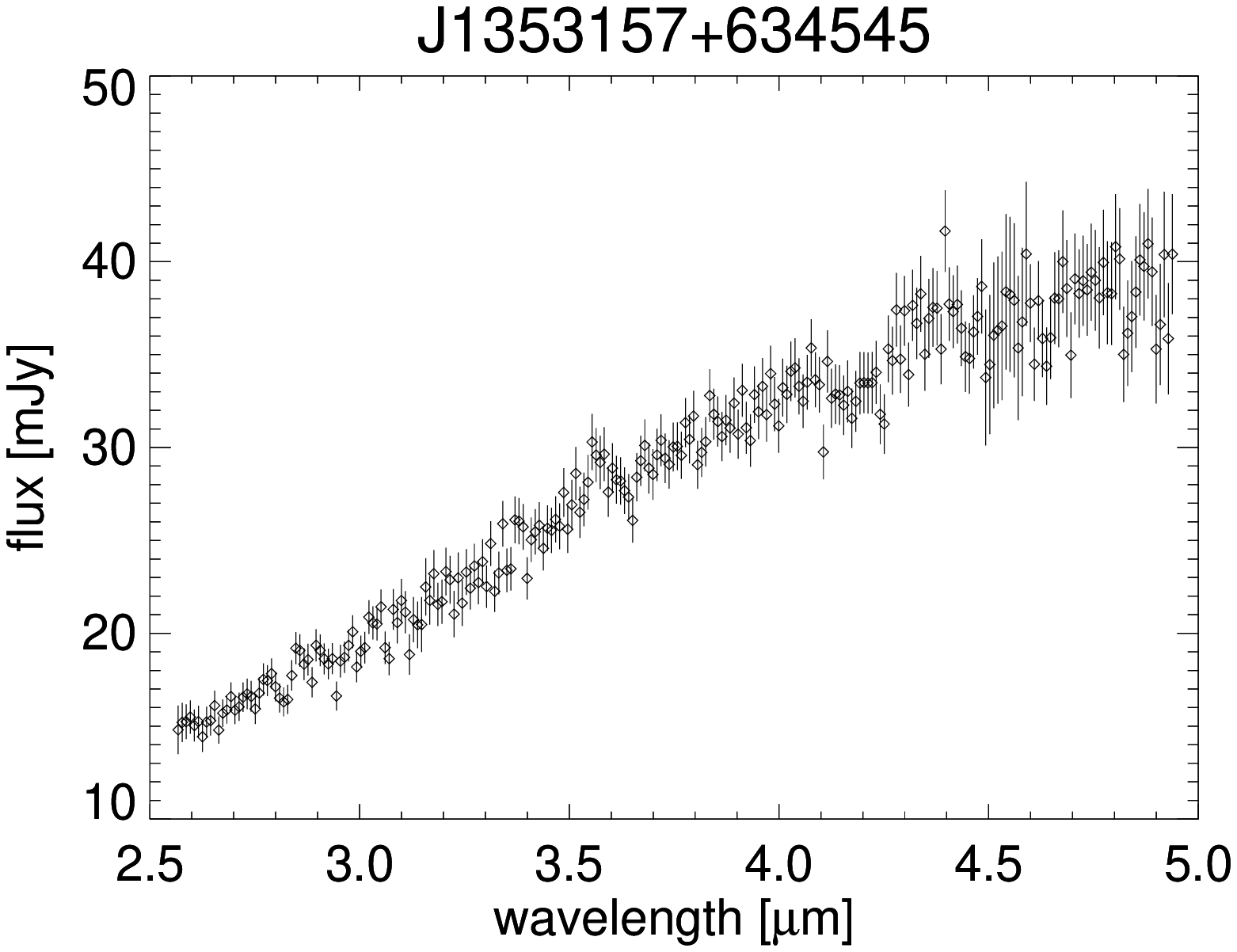}

     \FigureFile(41mm,26mm){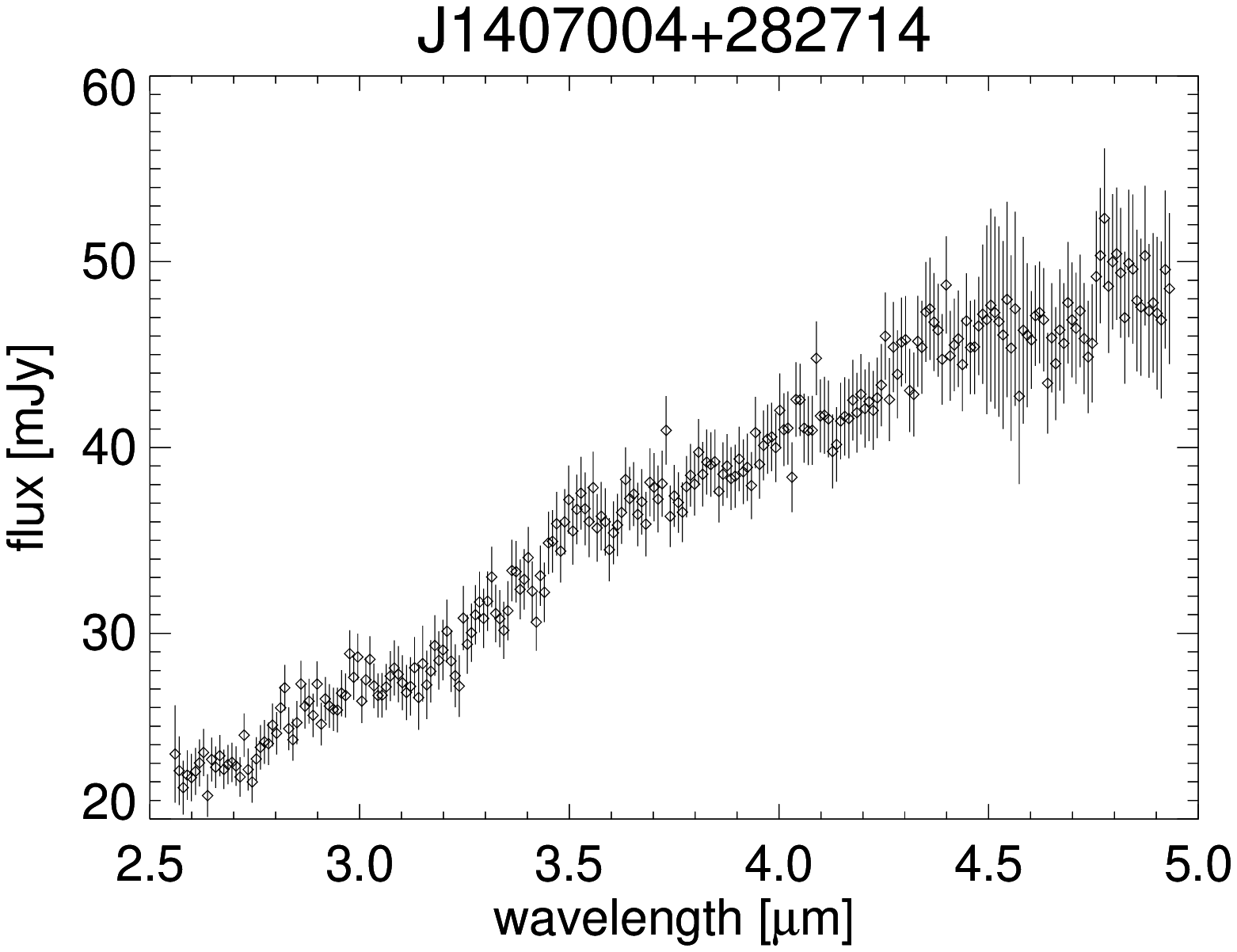}
     \FigureFile(41mm,26mm){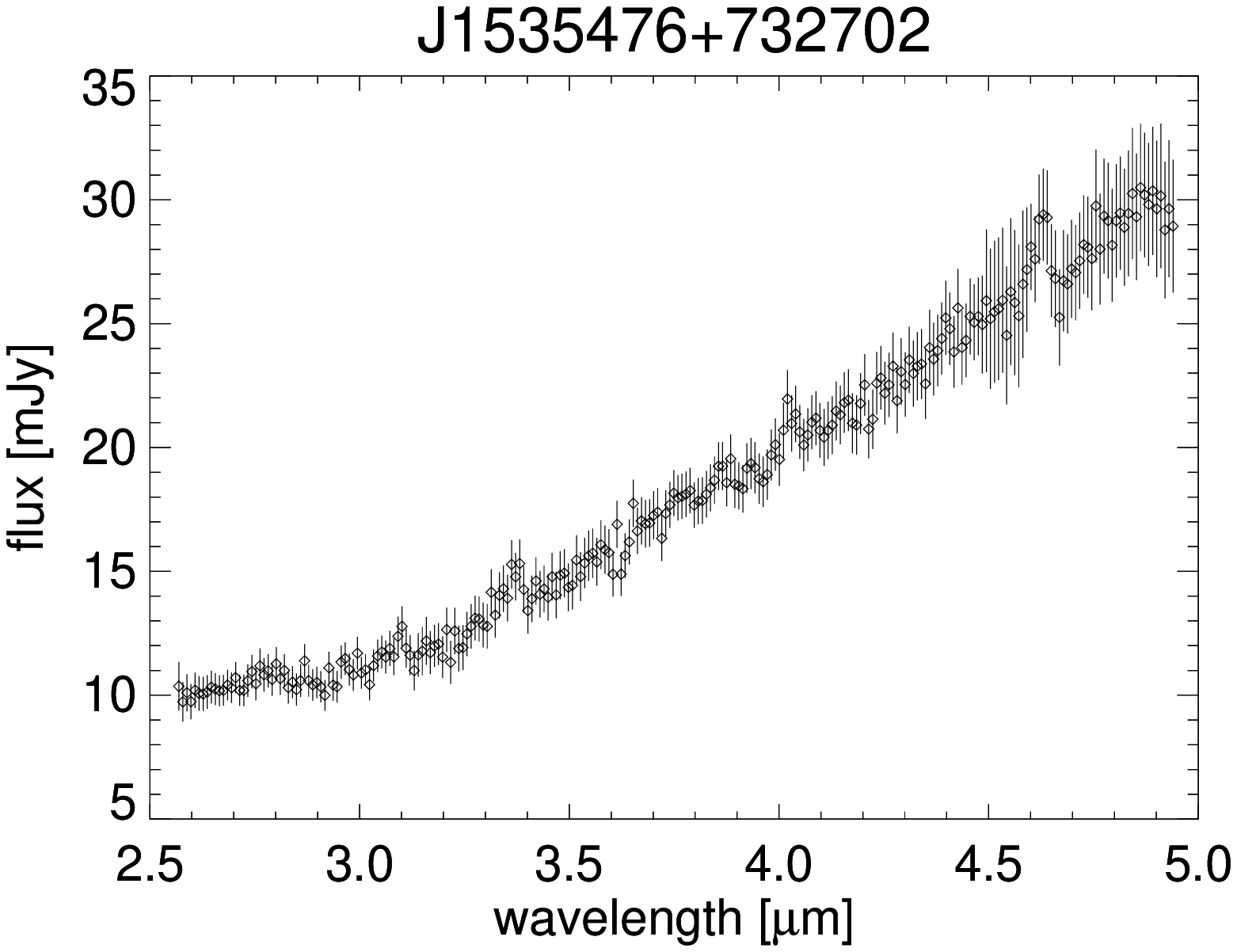} 
     \FigureFile(41mm,26mm){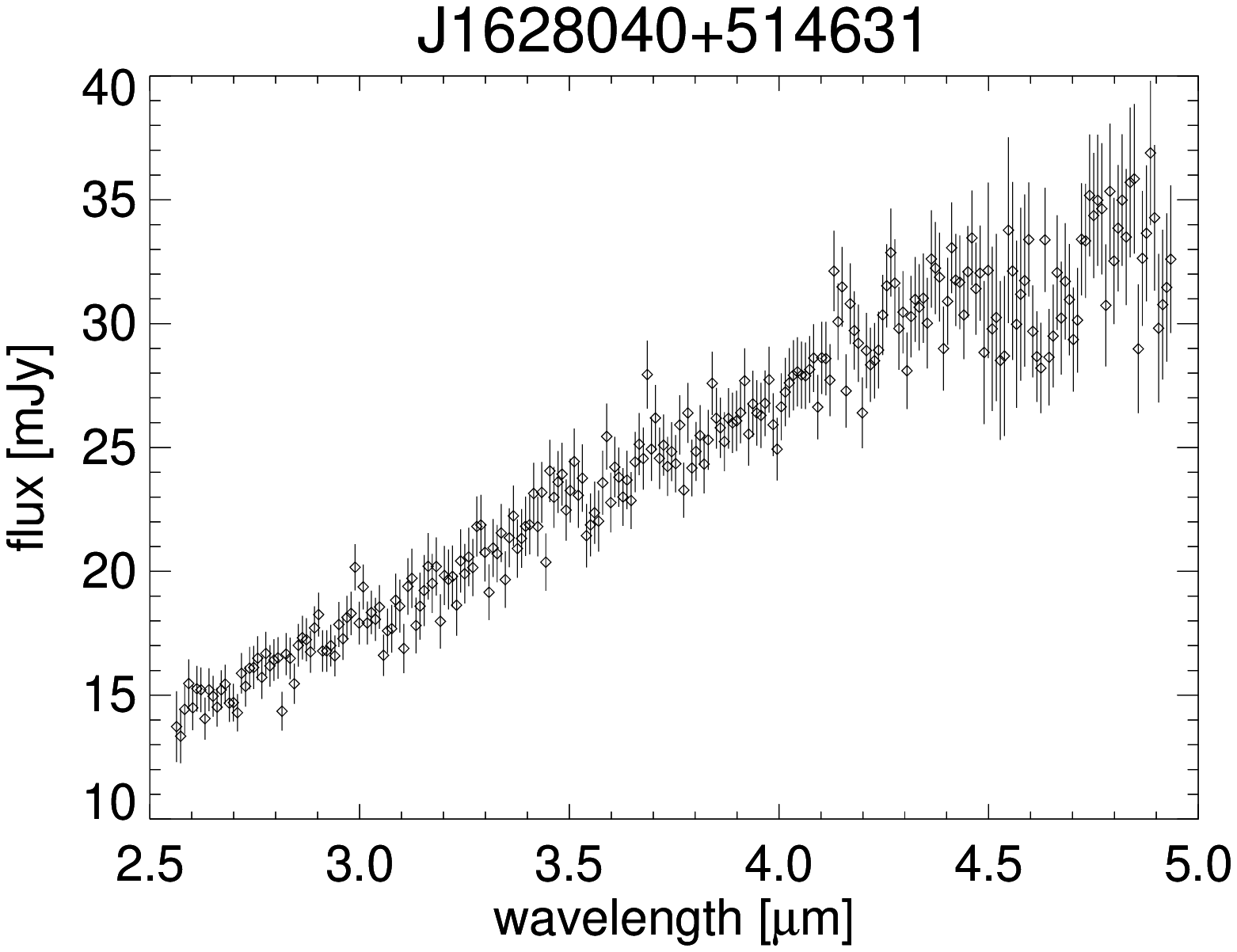}
     \FigureFile(41mm,26mm){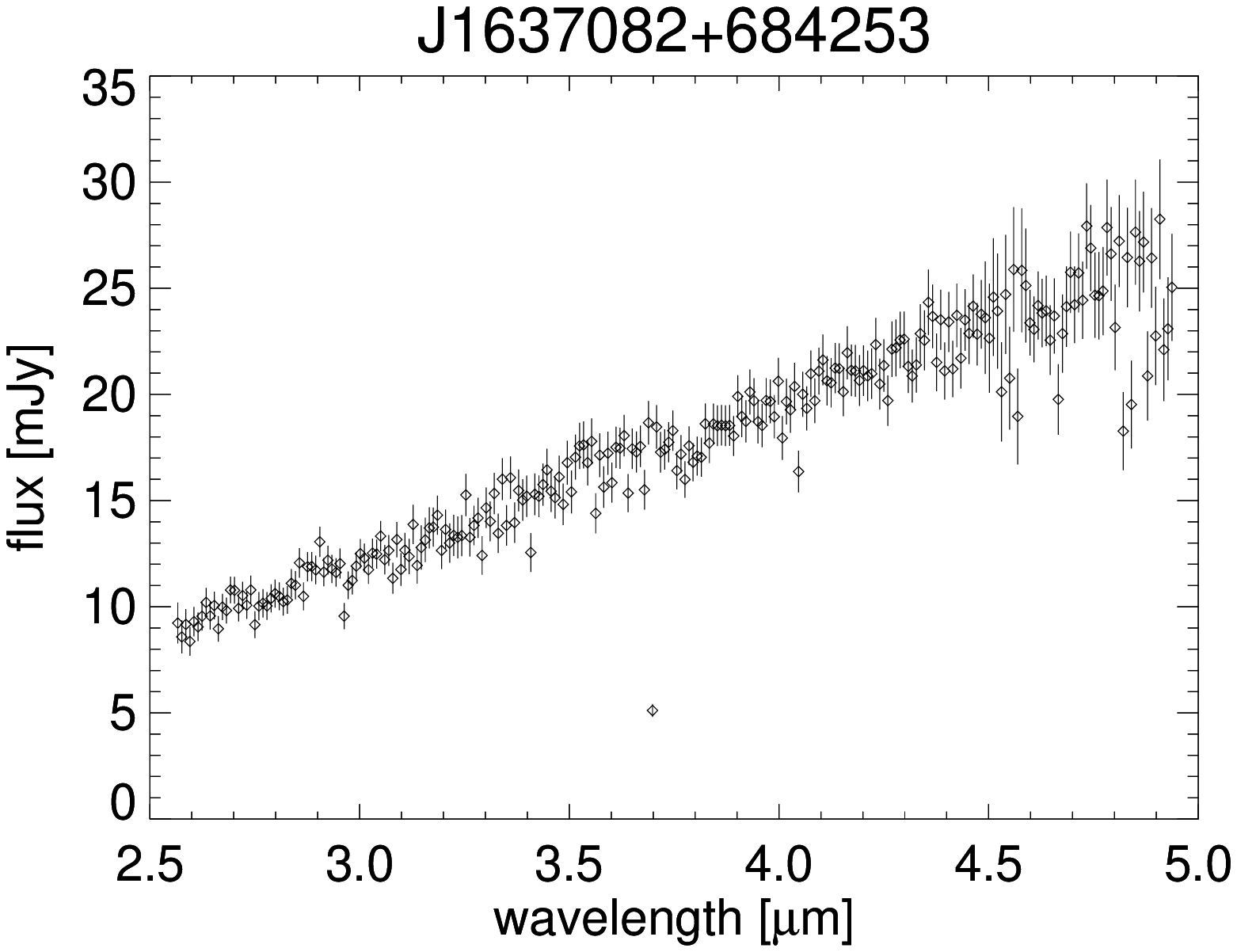}

     \FigureFile(41mm,26mm){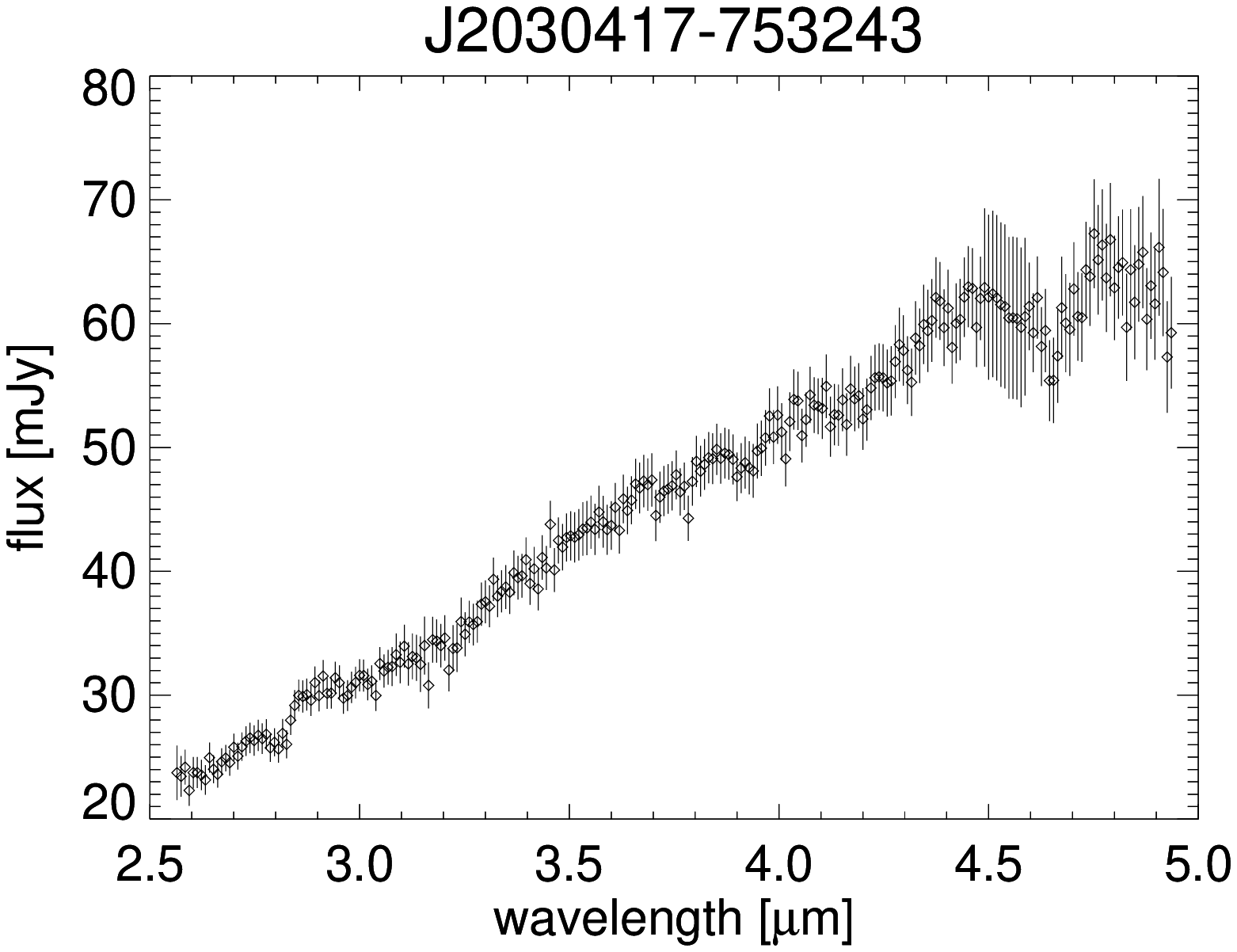}
     \FigureFile(41mm,26mm){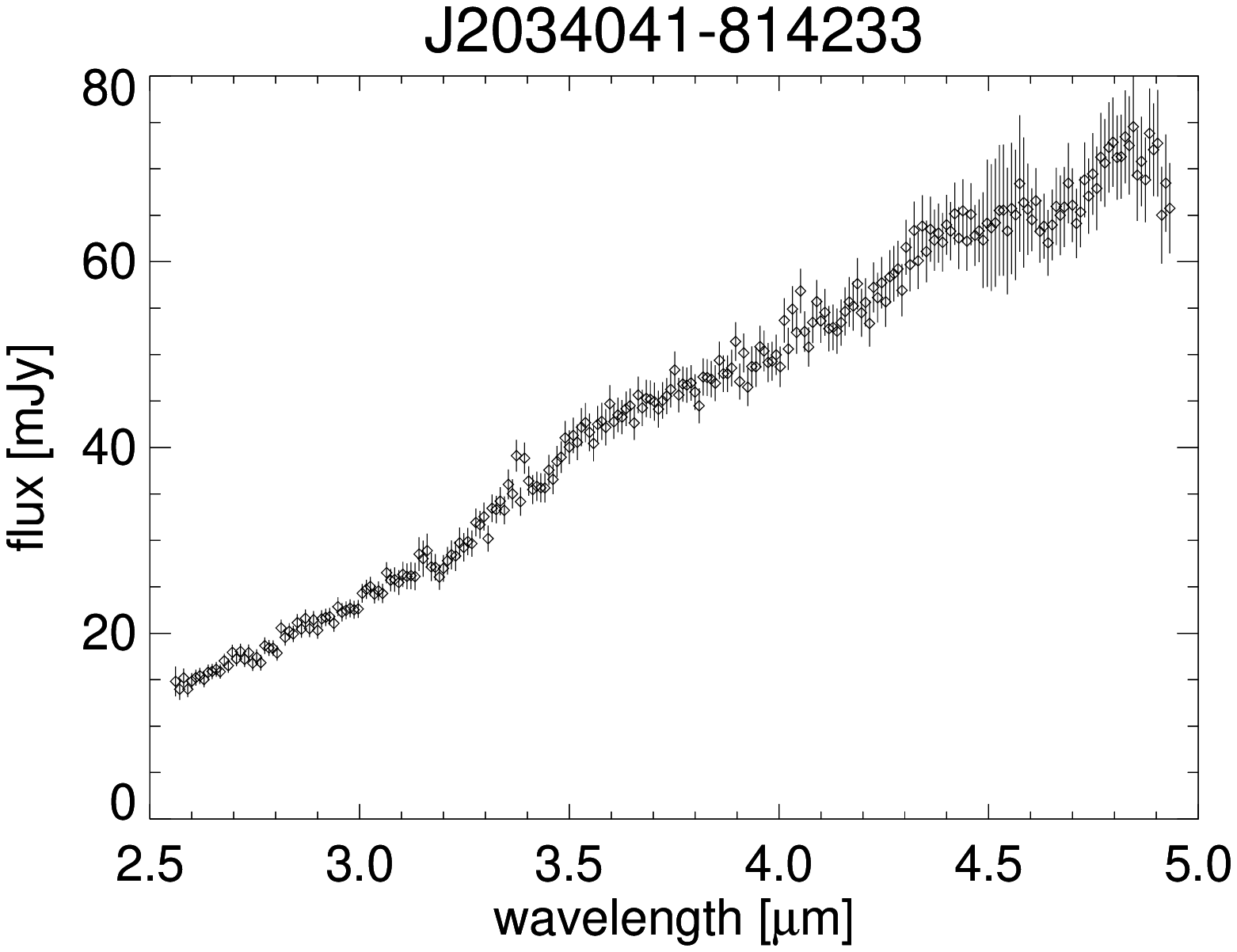}
     \FigureFile(41mm,26mm){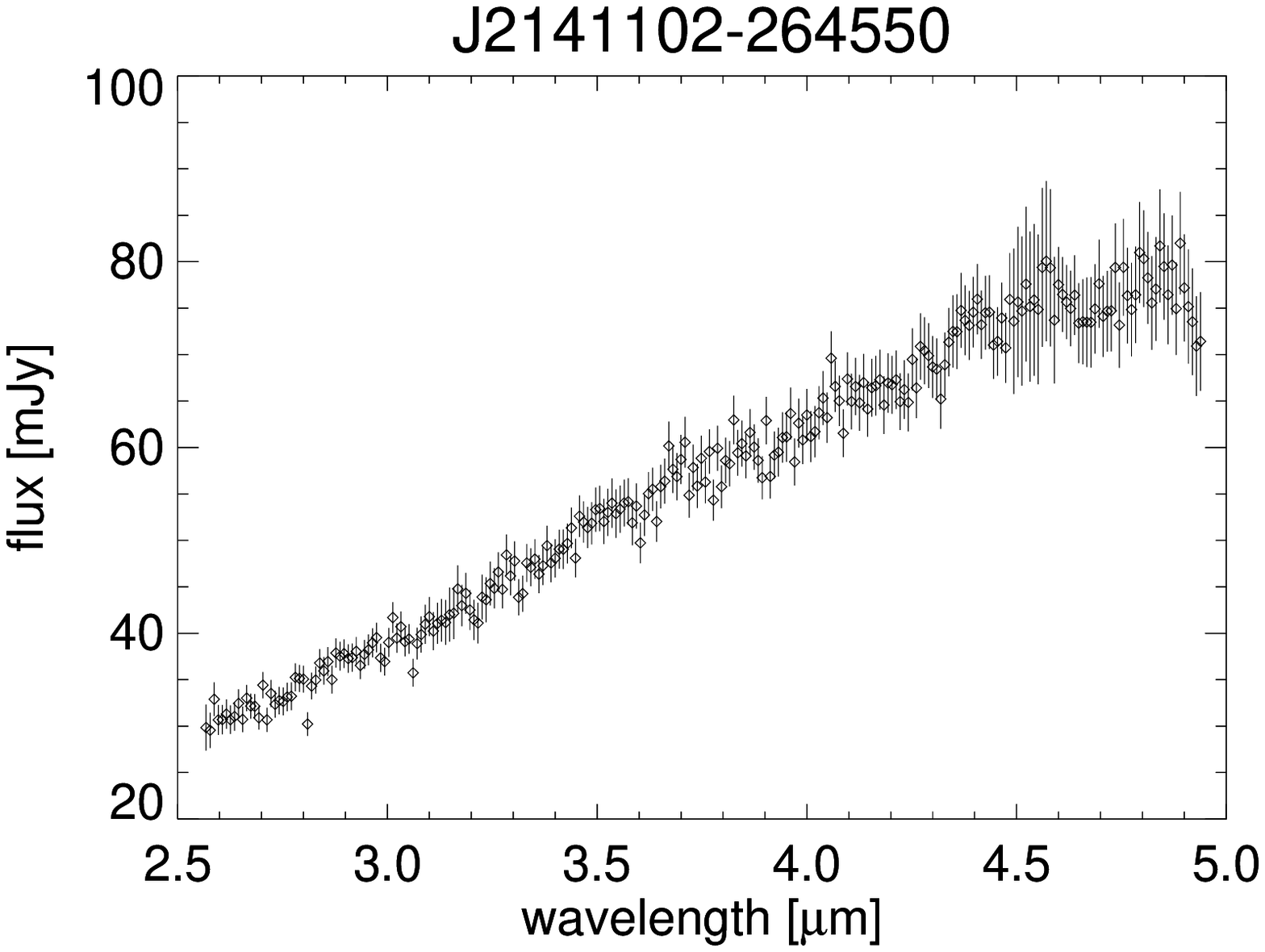}
     \FigureFile(41mm,26mm){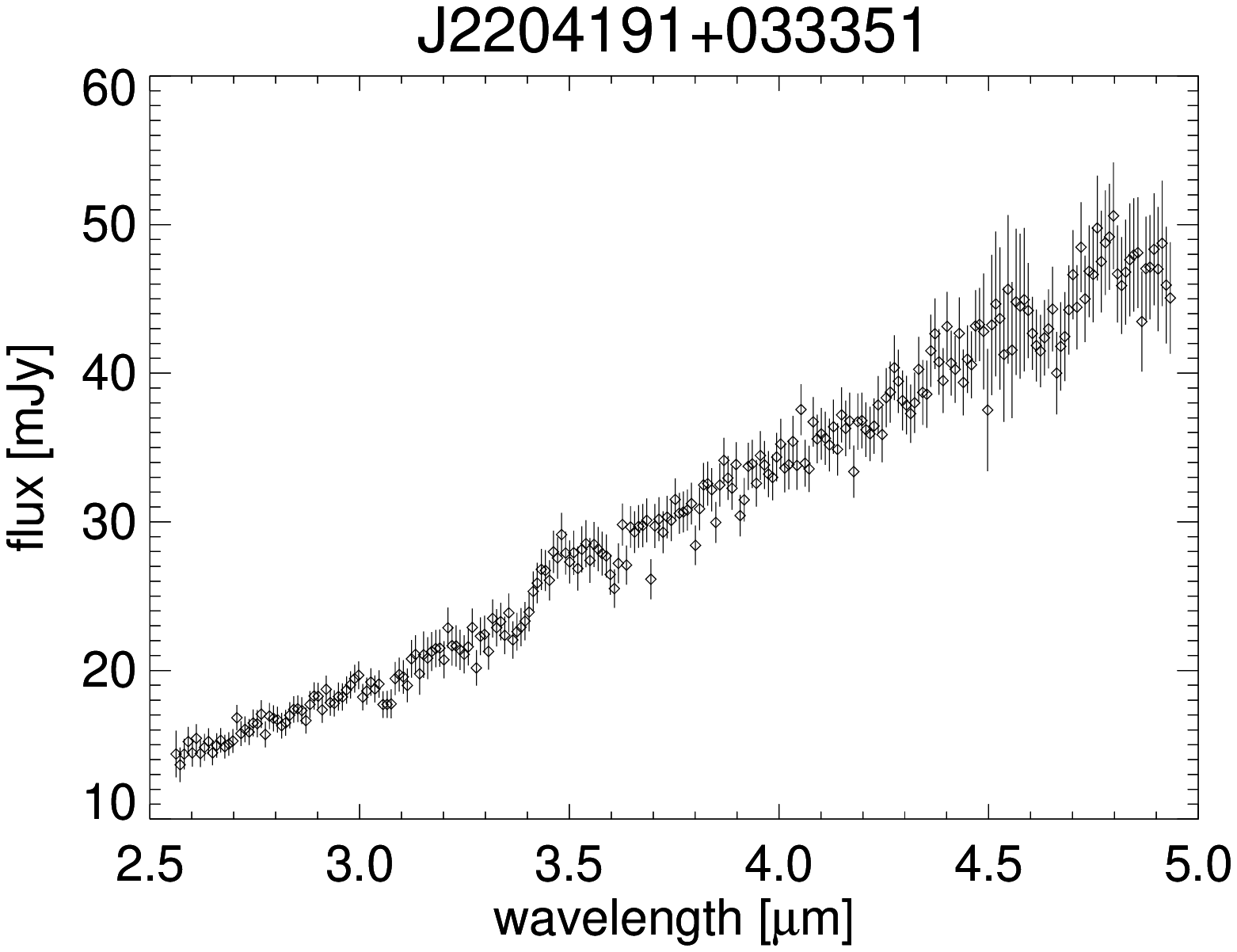}
 \end{center}
\end{figure}
\begin{figure}    
 \begin{center}

     \FigureFile(41mm,26mm){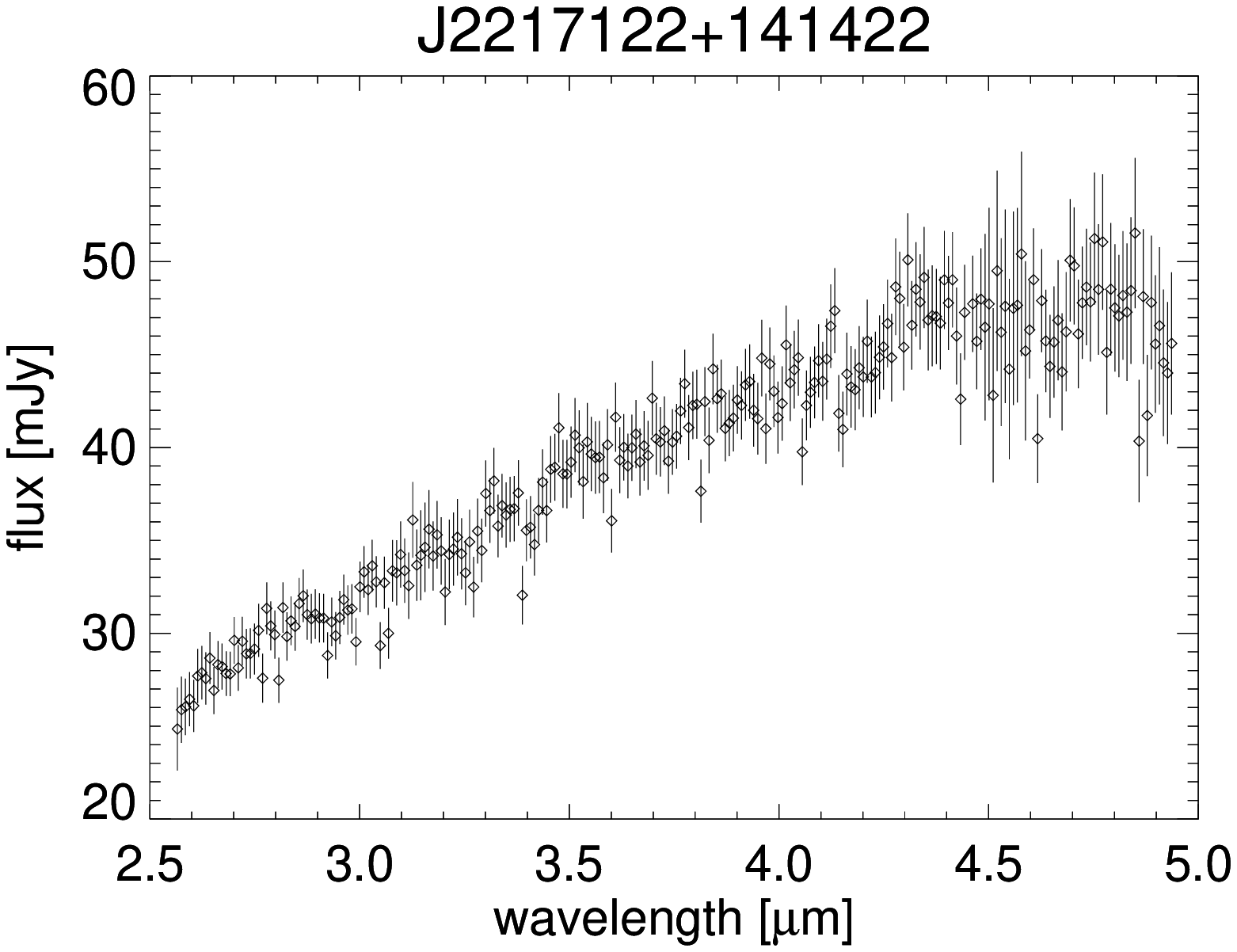}
     \FigureFile(41mm,26mm){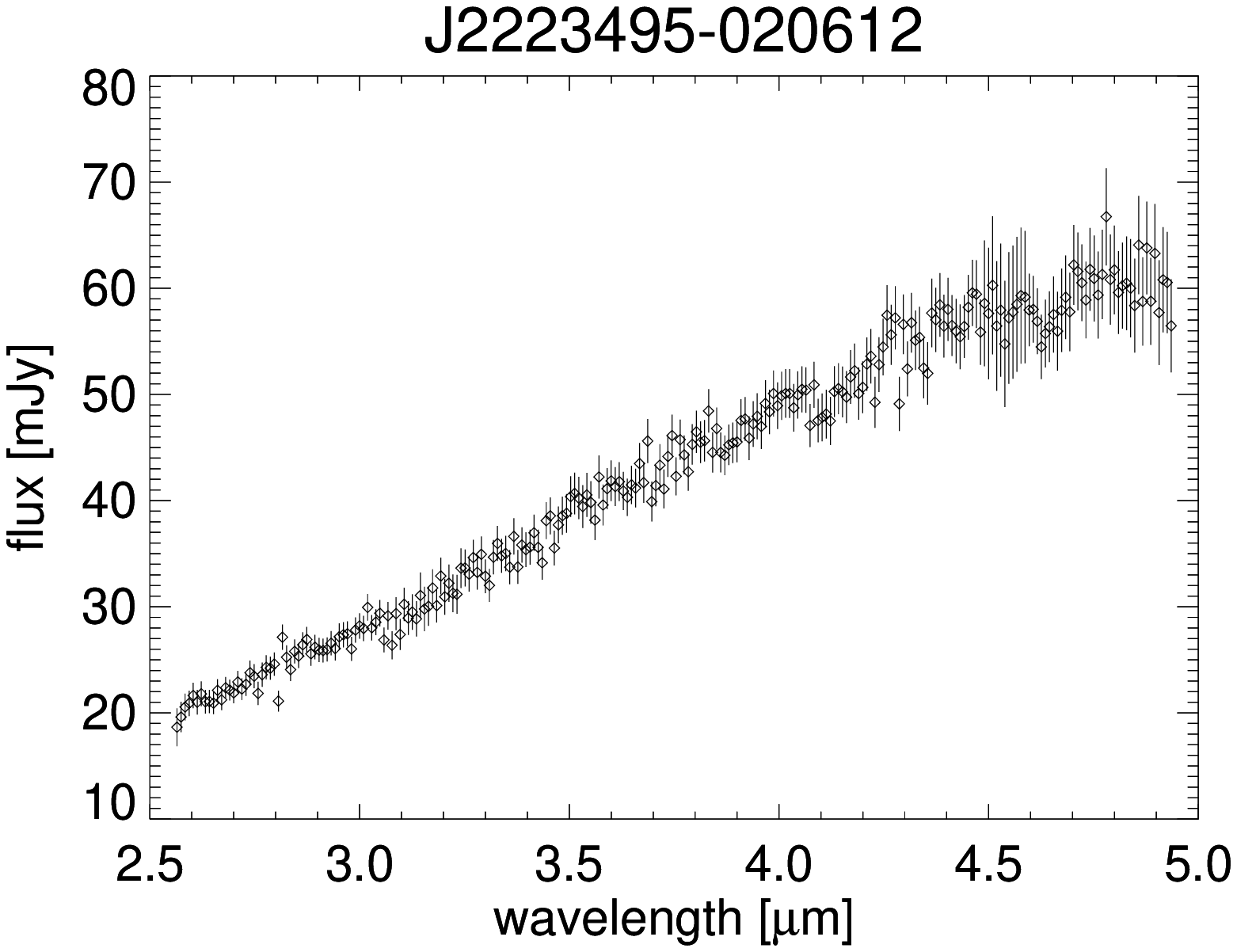}
     \FigureFile(41mm,26mm){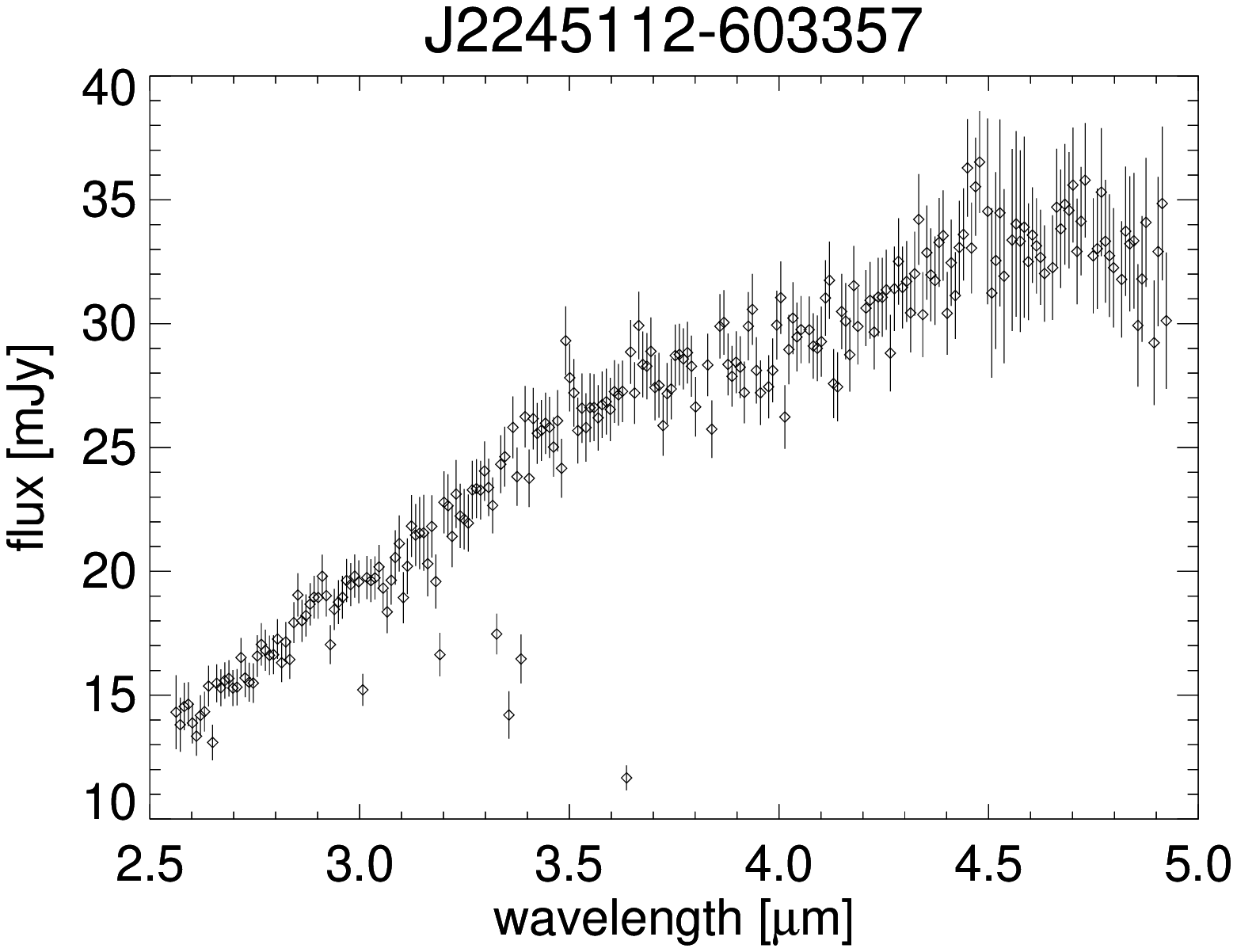}
     \FigureFile(41mm,26mm){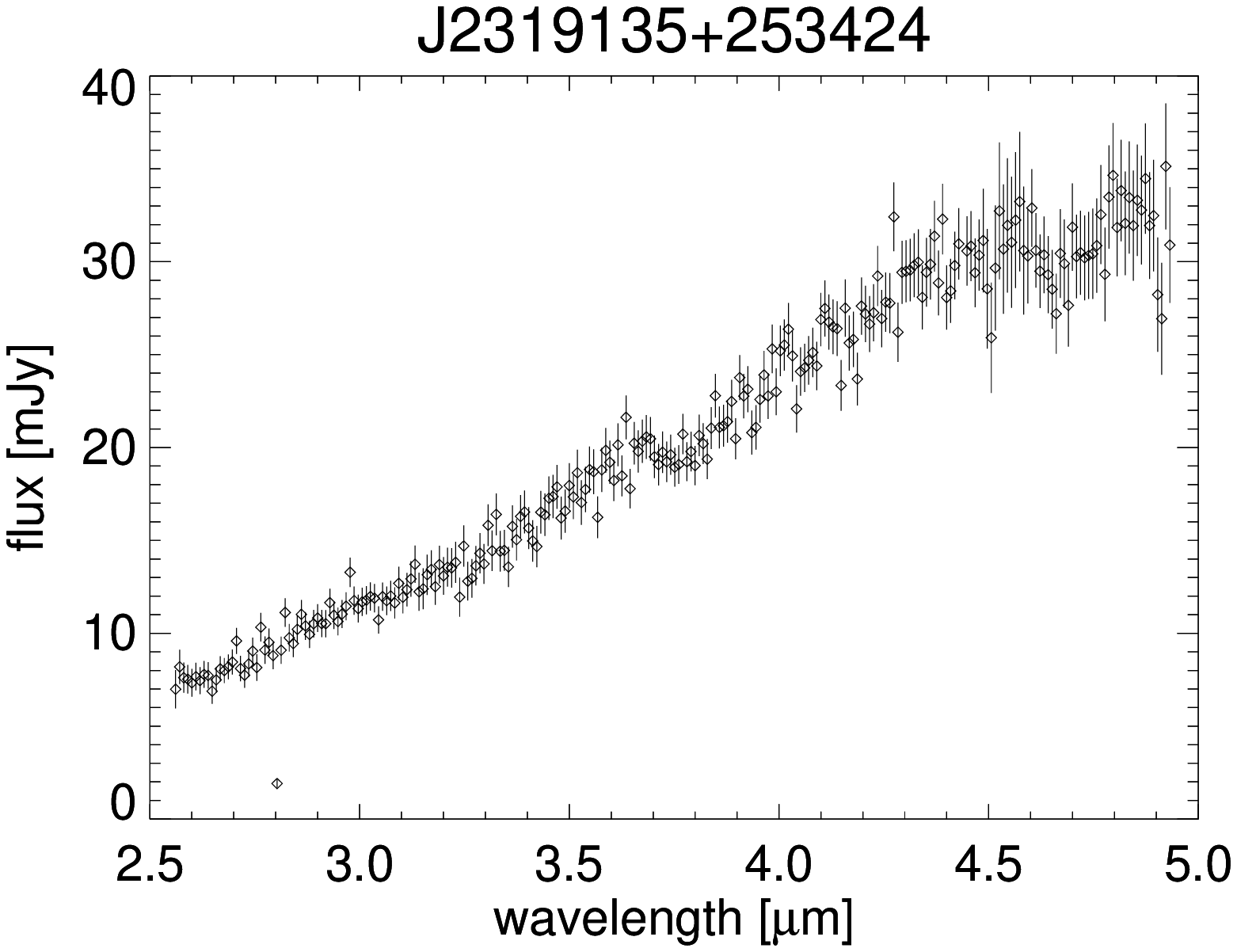}

 \end{center}
   \caption{AKARI IRC 2.5-5 $\mu$m spectra of the galaxies with no significant detection of the PAH 3.3 $\mu$m emission feature.}\label{2} 
\end{figure}

\begin{longtable}{lcccccc}
\caption{Basic properties of the galaxies with a significant detection of the PAH 3.3 $\mu$m emission feature} 
 \hline
object	name\footnotemark[$*$] &	2MASS	coordinates	&	redshift	& 	$F_{\mathrm{9\mu m}}$\footnotemark[$\dagger$]				&  $F_{\mathrm{18\mu m}}$\footnotemark[$\dagger$]				& 	$K_{\mathrm{S}}$\footnotemark[$\ddagger$] & $L_{\mathrm{IR}}$\footnotemark[$\S$]	\\
		&		(J 2000.0) & &	[mJy]				&	[mJy]				&	[mag]	& [$10^{11} \mathrm{L_{\odot}}$]  \\
		\hline
\endhead
\hline
\endfoot
\hline 
\multicolumn{7}{@{}l@{}}{\hbox to 0pt {\parbox{150mm}{\footnotesize 
  \footnotemark[$*$] Object name in the AKARI/IRC all-sky survey point source catalog. 
  \par\noindent
  \footnotemark[$\dagger$] AKARI IRC flux densities at 9 and 18 $\mu$m. 
  \par\noindent
  \footnotemark[$\ddagger$] $K_{\mathrm{S}}$ magnitude taken from the 2MASS catalog. 
  \par\noindent 
  \footnotemark[$\S$] IR (8-1000 $\mu$m) luminosity calculated from the IRAS 12, 25, 60, and 100 $\mu$m bands \citep{s 1996}. The fluxes are from the IRAS Faint Source Catalog \citep{mos 1992}.
  \footnotemark[$\|$] Flux densities not listed in the catalog, but obtained from the original AKARI all-sky image data. 
}}}
\endlastfoot
J	0027496$-$573524	&	00:27:49.76	$-$57:35:25.30	&	0.034	&	74	$\pm$	11\footnotemark[$\|$]		&		225	$\pm$	24	&	12.65	&	...	\\ %
J	0033298$-$564449	&	00:33:29.85	$-$39:22:59.20	&	0.025	&		150	$\pm$	25	&		219	$\pm$	31	&	12.46	&	1.17	\\%
J	0037541$-$062446	&	00:37:54.12	$-$06:24:46.10	&	0.023	&		129	$\pm$	12	&		241	$\pm$	31	&	11.75	&	1.27	\\%
J	0040459$-$791424	&	00:40:46.14	$-$79:14:24.30	&	0.034	&		139	$\pm$	12	&		182	$\pm$	43	&	11.06	&	1.49	\\
J	0055449$-$502749	&	00:55:44.87	$-$50:27:50.70	&	0.016	&		122	$\pm$	21	&		218	$\pm$	19	&	12.20	&	0.429	\\
J	0128477$-$542125	&	01:28:47.69	$-$54:21:25.60	&	0.093	&		75	$\pm$	11	&		225	$\pm$	13	&	11.68	&	9.56	\\
J	0129252$-$523417	&	01:29:25.55	$-$52:34:13.50	&	0.056	&		75	$\pm$	12	&		193	$\pm$	30	&	12.14	&	3.78	\\
J	0137069$-$090857	&	01:37:06.95	$-$09:08:57.50	&	0.073	&	104	$\pm$	14\footnotemark[$\|$]		&		197	$\pm$	57	&	12.36	&	3.98	\\
J	0156538$-$682252	&	01:56:53.88	$-$68:22:52.40	&	0.032	&		179	$\pm$	9	&		176	$\pm$	11	&	11.94	&	1.86	\\
J	0221366$-$643635	&	02:21:36.67	$-$64:36:36.70	&	0.02	&		166	$\pm$	10	&		180	$\pm$	30	&	12.78	&	0.549	\\
J	0225141$-$402526	&	02:25:14.18	$-$40:25:26.70	&	0.021	&		120	$\pm$	6	&		119	$\pm$	1	&	12.41	&	0.861	\\
J	0235220$-$651537	&	02:35:22.19	$-$65:15:37.10	&	0.019	&	$<$	50			&		87	$\pm$	51	&	13.01	&	0.0964	\\
J	0317437$-$572647	&	03:17:43.78	$-$57:26:47.60	&	0.029	&		88	$\pm$	6	&		198	$\pm$	18	&	12.28	&	1.26	\\
J	0324530$-$604418	&	03:24:53.05	$-$60:44:18.70	&	0.018	&		122	$\pm$	6	&		406	$\pm$	13	&	11.77	&	0.342	\\
J	0334478$-$561513	&	03:34:47.80	$-$56:15:12.90	&	0.082	&	104	$\pm$ 15\footnotemark[$\|$]			&		271	$\pm$	15	&	13.40	&	6.33	\\
J	0358184$-$612407	&	03:58:18.44	$-$61:24:07.20	&	0.047	&		94	$\pm$	6	&		299	$\pm$	13	&	12.25	&	2.56	\\
J	0422331$-$244409	&	04:22:33.26	$-$24:44:11.30	&	0.047	&		65	$\pm$	6	&	68	$\pm$	12\footnotemark[$\|$]		&	13.14	&	4.06	\\
J	0451233$-$423403	&	04:51:23.44	$-$42:34:03.90	&	0.059	&	78	$\pm$	11\footnotemark[$\|$]		&		143	$\pm$	20	&	12.74	&	4.38	\\
J	0921062$+$704154	&	09:21:06.36	$+$70:41:53.90	&	0.003	&		163	$\pm$	7	&	159	$\pm$	17\footnotemark[$\|$]		&	11.39	&	0.011	\\
J	1036319$+$022144	&	10:36:31.92	$+$02:21:45.20	&	0.05	&		124	$\pm$	6	&		311	$\pm$	48	&	12.63	&	...	\\
J	1041192$+$574500	&	10:41:19.22	$+$57:44:59.90	&	0.067	&	64	$\pm$	12\footnotemark[$\|$]		&		151	$\pm$	10	&	12.66	&	2.93	\\
J	1050215$+$412750	&	10:50:21.59	$+$41:27:50.60	&	0.024	&		94	$\pm$	6	&		113	$\pm$	6	&	11.67	&	0.751	\\
J	1135490$+$565708	&	11:35:49.07	$+$56:57:08.40	&	0.053	&		114	$\pm$	25	&		343	$\pm$	21	&	11.44	&	2.91	\\
J	1210580$+$635452	&	12:10:58.64	$+$63:54:49.60	&	0.009	&		175	$\pm$	28	&	$<$	90			&	11.76	&	0.134	\\
J	1224360$+$392258	&	12:24:36.20	$+$39:22:59.20	&	0.0035	&		311	$\pm$	39	&		404	$\pm$	32	&	11.37	&	0.0362	\\
J	1227380$+$400938	&	12:27:37.78	$+$40:09:39.20	&	0.037	&	118	$\pm$	15\footnotemark[$\|$]		&		220	$\pm$	6	&	12.26	&	2.59	\\
J	1246568$+$421600	&	12:46:56.78	$+$42:15:59.80	&	0.033	&		161	$\pm$	31	&		219	$\pm$	82	&	12.10	&	1.82	\\
J	1250138$+$073444	&	12:50:13.86	$+$07:34:44.50	&	0.038	&		98	$\pm$	11	&		239	$\pm$	46	&	12.32	&	2.37	\\
J	1255542$+$633641	&	12:55:54.88	$+$02:21:45.20	&	0.009	&		95	$\pm$	17	&		151	$\pm$	36	&	11.64	&	0.124	\\
J	1324352$-$194512	&	13:24:35.28	$-$19:45:11.10	&	0.017	&		164	$\pm$	48	&	151	$\pm$	15\footnotemark[$\|$]		&	11.64	&	0.374	\\
J	1336022$-$171557	&	13:36:01.61	$-$17:15:58.00	&	0.05	&	198	$\pm$15\footnotemark[$\|$]			&		299	$\pm$	51	&	12.37	&	3.93	\\
J	1346494$+$142400	&	13:46:49.45	$+$14:24:01.60	&	0.022	&		116	$\pm$	19	&		429	$\pm$	34	&	11.68	&	0.871	\\
J	1358418$+$350515	&	13:58:41.83	$+$35:05:16.30	&	0.035	&		105	$\pm$	10	&		175	$\pm$	7	&	12.13	&	2.07	\\
J	1427315$+$695610	&	14:27:31.62	$+$69:56:10.60	&	0.031	&		96	$\pm$	32	&	92	$\pm$	9\footnotemark[$\|$]		&	11.81	&	0.991	\\
J	1442347$+$660604	&	14:42:34.91	$+$66:06:04.50	&	0.038	&		84	$\pm$	7	&		344	$\pm$	16	&	13.11	&	1.53	\\
J	1456077$+$833122	&	14:56:10.00	$+$83:31:20.80	&	0.013	&		205	$\pm$	15	&		432	$\pm$	21	&	11.62	&	0.435	\\
J	1458360$+$445300	&	14:58:36.01	$+$44:53:01.00	&	0.036	&		158	$\pm$	15	&		297	$\pm$	103	&	11.97	&	1.46	\\
J	1507080$+$655606	&	15:07:07.83	$+$65:56:05.90	&	0.029	&		96	$\pm$	13	&	117	$\pm$	10\footnotemark[$\|$]		&	12.08	&	0.730	\\
J	1512310$+$655325	&	15:12:31.02	$+$65:53:24.40	&	0.023	&		95	$\pm$	7	&	104	$\pm$	13\footnotemark[$\|$]		&	12.44	&	0.542	\\
J	1525275$+$050031	&	15:25:27.50	$+$05:00:30.00	&	0.036	&		86	$\pm$	11	&		161	$\pm$	31	&	12.66	&	1.55	\\
J	1544218$+$410512	&	15:44:21.51	$+$41:05:11.00	&	0.032	&	142	$\pm$	13\footnotemark[$\|$]		&		125	$\pm$	16	&	13.29	&	1.29	\\
J	1622177$+$502220	&	16:22:17.81	$+$50:22:19.10	&	0.057	&		116	$\pm$	24	&		164	$\pm$	15	&	11.85	&	3.30	\\
J	1642060$+$653507	&	16:42:06.47	$+$65:35:07.80	&	0.025	&		90	$\pm$	4	&		169	$\pm$	45	&	11.87	&	1.08	\\
J	2100169$-$574323	&	21:00:17.05	$-$57:43:22.00	&	0.032	&	72	$\pm$	19\footnotemark[$\|$]		&		177	$\pm$	68	&	12.95	&	1.17	\\
J	2158360$+$120220	&	21:58:35.97	$+$12:02:20.00	&	0.031	&		142	$\pm$	4	&		197	$\pm$	21	&	12.22	&	1.77	\\
J	2218300$+$041628	&	22:18:30.08	$+$04:16:28.60	&	0.035	&		79	$\pm$	12	&	108	$\pm$	12\footnotemark[$\|$]		&	12.58	&	1.31	\\
J	2248041$-$172829	&	22:48:04.33	$-$17:28:30.60	&	0.117	&		72	$\pm$	25	&	134	$\pm$	16\footnotemark[$\|$]		&	13.00	&	11.2	\\
\end{longtable}

\begin{longtable}{lcccccc}
\caption{Basic properties of the galaxies with no significant detection of the PAH 3.3 emission feature} 
 \hline
object	name\footnotemark[$*$] &	2MASS	coordinates	&	redshift	& 	$F_{\mathrm{9\mu m}}$\footnotemark[$\dagger$]				&  $F_{\mathrm{18\mu m}}$\footnotemark[$\dagger$]				& 	$K_{\mathrm{S}}$\footnotemark[$\ddagger$]	& $L_{\mathrm{IR}}$\footnotemark[$\S$] 	\\
		&		(J 2000.0)&			&	[mJy]				&	[mJy]				&	[mag]	&  [$10^{11} \mathrm{L_{\odot}}$]  \\
		\hline
\endhead
\hline
\endfoot
\hline 
\multicolumn{7}{@{}l@{}}{\hbox to 0pt {\parbox{150mm}{\footnotesize 
  Notes. Same as table 1.
}}}
\endlastfoot
J	0018047$-$544909	&	00:18:04.58	$-$54:49:08.10	&	0.069	&	73	$\pm$	1	&	$<$ 90			&	12.3	&	...	\\
J	0021533$-$791007	&	00:21:53.39	$-$79:10:08.00	&	0.073	&	171	$\pm$	6	&	715	$\pm$	15	&	12.5	&	11.0	\\
J	0023139$-$535823	&	00:23:14.01	$-$53:58:23.70	&	0.029	&	97	$\pm$	10	&	278	$\pm$	16	&	12.7	&	...	\\
J	0043041$-$545826	&	00:43:04.11	$-$54:58:25.90	&	0.158	&	96	$\pm$	29	&	345	$\pm$	15	&	12.7	&	...	\\
J	0047001$-$174705	&	00:47:00.11	$-$17:47:02.50	&	0.147	&	119	$\pm$	13\footnotemark[$\|$]	&	120	$\pm$	15	&	12.5	&	...	\\
J	0053562$-$703804	&	00:53:56.23	$-$17:47:02.50	&	0.070	&	200	$\pm$	17	&	595	$\pm$	33	&	11.2	&	5.41	\\
J	0054374$-$083407	&	00:54:37.50	$-$08:34:08.20	&	0.097	&	137	$\pm$	5	&	176	$\pm$	4	&	12.3	&	9.14	\\
J	0842153$+$402532	&	08:42:15.29	$+$40:25:32.80	&	0.055	&	138	$\pm$	10	&	272	$\pm$	15	&	11.8	&	...	\\
J	0109450$-$033233	&	01:09:45.08	$-$03:32:32.60	&	0.054	&	88	$\pm$	12\footnotemark[$\|$]	&	406	$\pm$	31	&	13.2	&	2.99	\\
J	0127539$+$284750	&	01:27:53.94	$+$28:47:51.40	&	0.043	&	105	$\pm$	10	&	182	$\pm$	17	&	14.3	&	...	\\
J	0228152$-$405716	&	02:28:15.25	$-$40:57:14.60	&	0.493	&	63	$\pm$	18	&	135	$\pm$	16	&	11.9	&	...	\\
J	0356199$-$625139	&	03:56:19.97	$-$62:51:39.10	&	0.108	&	111	$\pm$	7	&	278	$\pm$	15	&	12.6	&	11.3	\\
J	0426007$-$571203	&	04:26:00.71	$-$57:12:01.80	&	0.104	&	$<$ 50			&	106	$\pm$	21	&	12.0	&	...	\\
J	0522580$-$362731	&	05:22:57.97	$-$36:27:30.80	&	0.055	&	97.6	$\pm$	0.1	&	216	$\pm$	20	&	11.3	&	...	\\
J	0538503$-$440509	&	05:38:50.36	$-$44:05:09.00	&	0.896	&	154	$\pm$	65	&	356	$\pm$	106	&	11.5	&	1077	\\
J	1040435$+$593409	&	10:40:43.64	$+$59:34:09.30	&	0.148	&	91	$\pm$	8	&	188	$\pm$	58	&	11.8	&	17.9	\\
J	1050348$+$801149	&	10:50:35.71	$+$80:11:50.70	&	0.120	&	64	$\pm$	5	&	120	$\pm$	13\footnotemark[$\|$]	&	11.8	&	...	\\
J	1230499$+$011521	&	12:30:50.04	$+$01:15:22.70	&	0.117	&	66	$\pm$	28	&	160	$\pm$	8	&	11.7	&	...	\\
J	1300053$+$163214	&	13:00:05.31	$+$16:32:14.20	&	0.080	&	88	$\pm$	24	&	175	$\pm$	51	&	11.9	&	...	\\
J	1353157$+$634545	&	13:53:15.83	$+$63:45:45.60	&	0.093	&	108	$\pm$	10	&	307	$\pm$	14	&	11.9	&	7.66	\\
J	1407004$+$282714	&	14:07:00.43	$+$28:27:15.00	&	0.077	&	80	$\pm$	6	&	305	$\pm$	30	&	11.9	&	5.84	\\
J	1535476$+$732702	&	15:35:47.70	$+$73:27:02.50	&	0.025	&	155	$\pm$	18	&	366	$\pm$	19	&	12.1	&	0.68	\\
J	1628040$+$514631	&	16:28:04.05	$+$51:46:31.50	&	0.056	&	67	$\pm$	12	&	214	$\pm$	21	&	11.5	&	...	\\
J	1637082$+$684253	&	16:37:08.03	$+$68:42:53.00	&	0.016	&	73	$\pm$	7\footnotemark[$\|$]	&	103	$\pm$	19	&	12.3	&	...	\\
J	2030417$-$753243	&	20:30:41.71	$-$75:32:43.00	&	0.114	&	145	$\pm$	29	&	340	$\pm$	17	&	11.4	&	...	\\
J	2034041$-$814233	&	20:34:03.90	$-$81:42:33.60	&	0.034	&	228	$\pm$	19	&	440	$\pm$	28	&	12.4	&	1.29	\\
J	2141102$-$264550	&	21:41:10.24	$-$26:45:50.10	&	0.129	&	122	$\pm$	27	&	136	$\pm$	51	&	11.1	&	...	\\
J	2204191$+$033351	&	22:04:19.16	$+$03:33:51.10	&	0.061	&	128	$\pm$	11	&	480	$\pm$	20	&	12.0	&	5.99	\\
J	2217122$+$141422	&	22:17:12.26	$+$14:14:20.90	&	0.066	&	68	$\pm$	14	&	74	$\pm$	12\footnotemark[$\|$]	&	11.3	&	...	\\
J	2223495$-$020612	&	22:23:49.55	$-$02:06:12.90	&	0.056	&	141	$\pm$	13	&	358	$\pm$	6	&	11.0	&	...	\\
J	2245112$-$603357	&	22:45:11.30	$-$60:33:55.80	&	0.113	&	90	$\pm$	14	&	192	$\pm$	27	&	11.9	&	...	\\
J	2319135$+$253424	&	23:19:13.48	$+$25:34:25.60	&	0.113	&	59	$\pm$	4	&	$<$ 90		&	13.3	&	...	\\
\end{longtable}

 \begin{table*}
\scriptsize
\begin{center}
 \caption{Derived properties of the IRGs}
\begin{tabular}{cccccccccc}
\hline
object	& $L_{\mathrm{IR}}$\footnotemark[$*$] & $L_{\mathrm{PAH3.3}}$\footnotemark[$\dagger$] &	$L_{\mathrm{Br\alpha}}$\footnotemark[$\ddagger$] & $\tau_{3.1}$\footnotemark[$\S$] &	$EW_{\mathrm{PAH3.3}}$\footnotemark[$\|$] &  $\it{\Gamma}$\footnotemark[$\#$] & spectral	& $T_{\mathrm{dust}}$\footnotemark[$\dagger\dagger$] & $M_{\mathrm{dust}}$\footnotemark[$\dagger\dagger$] \\
 &	 [$10^{11} \mathrm{L_{\odot}}$] & [$10^{8} \mathrm{L_{\odot}}$] &	[$10^7 \mathrm{L_{\odot}}$]	 & &	[nm]	& & type\footnotemark[$**$] & [K] & [$10^7 \:\mathrm{M_{\odot}}$] \\
\hline
J	0221366$-$643635	&	0.549	&	1.49	$\pm$	0.03	&		0.94	$\pm$		0.10	& 0.12	&	154	&	$-$0.41	&	SF	&	33.9	    &	1.16	\\
J	0235220$-$651537	&	0.0964	&	0.07	$\pm$	0.01	&		0.15	$\pm$		0.04	& 0.26	&	48.6	&	1.11	&	Comp.	&	39.4		&	0.074 \\
J	0055449$-$502749	&	0.429	&	0.61	$\pm$	0.02	&		0.34	$\pm$		0.08	& $<$0.08	&	137	&	$-$0.50	&	SF	&	38.2		&	0.49		\\
J	0225141$-$402526	&	0.861	&	0.81	$\pm$	0.10	&		0.91	$\pm$		0.16	&	0.13	&	117	&	$-$0.19	&	SF	&	36.4		&	1.33	\\
J	0324530$-$604418	&	0.342	&	0.20	$\pm$	0.03	&	$<$			0.52				& 0.077	&	20.3	&	0.95	&	Comp.	&	41.2		&	0.17	\\
J	0921062$+$704154	&	0.011	&0.026	$\pm$	0.002&		0.010	$\pm$	0.003& 0.10	&	158	&	$-$0.30	&	SF	&	33.2		&	0.027 \\
J	1050215$+$412750	&	0.751	&	0.79	$\pm$	0.11	&	$<$			0.59				& 0.14 	&	86.0	&	$-$0.09	&	SF	&	36.4		&	1.13	\\
J	1210580$+$635452	&	0.134	&	0.10	$\pm$	0.02	&		0.11	$\pm$		0.03	& 0.079	&	61.2	&	$-$0.21	&	SF	&	31.9		&	0.44	\\
J	1224360$+$392258	&	0.0362	&	0.06	$\pm$	0.01	&		0.047	$\pm$	0.005& 0.063	&	101	&	$-$0.58	&	SF	&	36.5		&	0.053 \\
J	1255542$+$633641	&	0.124	&	0.12	$\pm$	0.01	&		0.11	$\pm$		0.03	& 0.083	&	62.0	&	$-$0.30	&	SF	&	36.3		&	0.20	\\
J	1324352$-$194512	&	0.374	&	0.47	$\pm$	0.09	&	$<$			0.37				& $<$0.08	&	55.9	&	$-$0.13	&	SF	&	31.5		&	1.2	\\
J	1346494$+$142400	&	0.871	&	0.51	$\pm$	0.10	&		0.84	$\pm$		0.2	& $<$0.08	&	38.2	&	0.91	&	Comp.	&	38.9		&	0.78	\\
J	1427315$+$695610	&	0.991	&	2.2	$\pm$	0.2	&	$<$		 	0.82				& 0.099	&	136	&	$-$0.45	&	SF	&	36.8		&	1.37	\\
J	1456077$+$833122	&	0.435	&	0.52	$\pm$	0.04	&		0.51	$\pm$		0.05	& $<$0.09 &	159  &	$-$0.28	&	SF	&	39.3		&	0.39		\\
J	1507080$+$655606	&	0.730	&	1.11	$\pm$	0.12	&	$<$			0.56				& $<$0.14	&	140	&	$-$0.37	&	SF	&	34.4		&	1.43	\\
J	1512310$+$655325	&	0.542	&	0.81	$\pm$	0.09	&		0.6	$\pm$		0.14	& 0.13 	&	154	&	$-$0.23	&	SF	&	36.5		&	0.77	\\

J	1535476$+$732702	&	0.68		&	$<$	0.068			&	$<$	0.74						& 0.13	&	$<$	5.6	&	1.78	&	AGN	&	39.3	&	0.372 \\ 

NGC	1377                 	&	0.135	&	0.022 $\pm$	0.003 &	$<$	0.033			& $<$0.07	&	32.1      	&	$-$1.25	&	Comp.	&	54.0	&	0.022	\\	\hline

\multicolumn{10}{@{}l@{}}{\hbox to 0pt {\parbox{170mm}{\footnotesize 
\footnotemark[$*$] IR (8-1000 $\mu$m) luminosity calculated from the IRAS 12, 25, 60, and 100 $\mu$m bands \citep{s 1996}. The fluxes are from the IRAS Faint Source Catalog \citep{mos 1992}.
\par\noindent
\footnotemark[$\dagger$] Luminosity of the PAH 3.3 $\mu$m emission. 
\par\noindent
\footnotemark[$\ddagger$] Luminosity of the Br$\alpha$ emission. 
\par\noindent
\footnotemark[$\S$] Optical  depth of the 3.1 $\mu$m $\mathrm{H_2 O}$ absorption feature.
\par\noindent
\footnotemark[$\|$] Rest-frame equivalent width of the PAH 3.3 $\mu$m emission.
\par\noindent 
\footnotemark[$\#$] Power-law index representing the slope of continuum emission.
\par\noindent
\footnotemark[$**$] Near-IR spectral type based on the detection of PAH 3.3 $\mu$m emission feature, $EW_{\mathrm{PAH3.3}}$ and $\it{\Gamma}$. SF and AGN are star-forming galaxy and AGN-dominated galaxy, respectively. Composite is a galaxy which show both star-forming and AGN activity. 
\par\noindent
\footnotemark[$\dagger\dagger$] Dust temperature and mass obtained by the ratio of IRAS 60 to 100 $\mu$m fluxes with an emissivity power-law index $\beta=1$.
}}}
\end{tabular}
\end{center}
\end{table*}

 \begin{table*}
 \scriptsize
 \begin{center}
\caption{Derived properties of the LIRGs}
\begin{tabular}{lccccccccc}
 \hline
object	& $L_{\mathrm{IR}}$\footnotemark[$*$] & $L_{\mathrm{PAH3.3}}$\footnotemark[$\dagger$] &	$L_{\mathrm{Br\alpha}}$\footnotemark[$\ddagger$] & $\tau_{3.1}$\footnotemark[$\S$] &	$EW_{\mathrm{PAH3.3}}$\footnotemark[$\|$] &  $\it{\Gamma}$\footnotemark[$\#$] & spectral	& $T_{\mathrm{dust}}$\footnotemark[$\dagger\dagger$] & $M_{\mathrm{dust}}$\footnotemark[$\dagger\dagger$] \\
 &	 [$10^{11} \mathrm{L_{\odot}}$] & [$10^{8} \mathrm{L_{\odot}}$] &	[$10^7 \mathrm{L_{\odot}}$]	 & &	[nm]	& & type\footnotemark[$**$] & [K] & [$10^7 \:\mathrm{M_{\odot}}$] \\

\hline
J	0033298$-$564449	&	1.17	&	1.3	$\pm$	0.1	&	1.2	$\pm$	0.2	&	0.19	&	171	&	$-$0.1	&	SF	&	37.3	&	1.5	\\
J	0037541$-$062446	&	1.27	&	1.1	$\pm$	0.1	&	0.58	$\pm$	0.17	&	0.17	&	112	&	$-$1.29	&	SF	&	37.7	&	1.6	\\
J	0040459$-$791424	&	1.49	&	2.2	$\pm$	0.2	&	$<$	3.1		&	$<$0.06	&	28.4	&	0.63	&Comp.		&	35.2	&	2.3	\\
J	0128477$-$542125	&	9.56	&	7.4	$\pm$	0.8	&	17	$\pm$	5	&	$<$0.07	&	36.3	&	1.55	&Comp.		&	41.4	&	5.3	\\
J	0129252$-$523417	&	3.73	&	4.1	$\pm$	0.6	&	4.1	$\pm$	1	&	0.17	&	110	&	0.01	&	SF	&	36.2	&	5.6	\\
J	0137069$-$090857	&	3.93	&	5.6	$\pm$	0.5	&	$<$	7.3		&	$<$0.09	&	55.4	&	1.83	&Comp.		&	40.6	&	1.6	\\
J	0156538$-$682252	&	1.86	&	2.6	$\pm$	0.2	&	1.6	$\pm$	0.3	&	0.17	&	156	&	$-$0.11	&	SF	&	34.6	&	3.7	\\
J	0317437$-$572647	&	1.26	&	1.4	$\pm$	0.2	&	1.2	$\pm$	0.3	&	0.19	&	106	&	$-$0.45	&	SF	&	37.5	&	1.6	\\
J	0334478$-$561513	&	6.33	&	4.9	$\pm$	0.3	&	7.2	$\pm$	1.3	&	0.17	&	67.5	&	1.13	&Comp.		&	46.8	&	1.8	\\
J	0358184$-$612407	&	2.56	&	1.4	$\pm$	0.2	&	$<$	2.4		&	0.19	&	41.1	&	0.67	&	SF	&	37.5	&	2.2	\\
J	0422331$-$244409	&	4.06	&	6.0	$\pm$	0.6	&	5.2	$\pm$	1	&	0.14	&	178	&	0.07	&	SF	&	37.2	&	5.1	\\
J	0451233$-$423403	&	4.38	&	3.9	$\pm$	0.4	&	4.5	$\pm$	0.7	&	0.23	&	176	&	$-$0.03	&	SF	&	40.6	&	3.9	\\
J	1041192$+$574500	&	2.93	&	3.2	$\pm$	0.4	&	$<$	5.1		&	$<$0.1	&	44.7	&	1.58	&Comp.		&	42.2	&	1.0	\\
J	1135490$+$565708	&	2.91	&	4.3	$\pm$	0.5	&	$<$	6.2		&	$<$ 0.06	&	34.1	&	1.39	&Comp.		&	42.2	&	1.5	\\
J	1227380$+$400938	&	2.59	&	1.01	$\pm$	0.07	&	1.3	$\pm$	0.2	&	$<$0.1	&	106	&	$-$0.19	&	SF	&	38.1	&	3.0	\\
J	1246568$+$421600	&	1.82	&	2.1	$\pm$	0.2	&	1.7	$\pm$	0.3	&	0.097	&	118	&	0.01	&	SF	&	34.4	&	3.8	\\
J	1250138$+$073444	&	2.37	&	3.4	$\pm$	0.3	&	2.2	$\pm$	0.4	&	0.15	&	191	&	$-$0.05	&	SF	&	39.6	&	2.1	\\
J	1336022$-$171557	&	3.93	&	5.1	$\pm$	0.4	&	3.5	$\pm$	0.7	&	0.21	&	192	&	0.09	&	SF	&	46.3	&	1.6	\\
J	1358418$+$350515	&	2.07	&	2.5	$\pm$	0.2	&	2.4	$\pm$	0.4	&	0.25	&	133	&	0.35	&	SF	&	35.5	&	3.3	\\
J	1442347$+$660604	&	1.53	&	0.78	$\pm$	0.08	&	$<$	1.6		&	0.24	&	40.7	&	2.95	&Comp.		&	53.2	&	0.29	\\
J	1458360$+$445300	&	1.46	&	2.2	$\pm$	0.3	&	$<$	1.2		&	0.18	&	102	&	$-$0.15	&	SF	&	35.4	&	2.3	\\
J	1525275$+$050031	&	1.55	&	1.9	$\pm$	0.2	&	1.5	$\pm$	0.3	&	0.12	&	156	&	0.12	&	SF	&	41.2	&	1.16	\\
J	1544218$+$410512	&	1.29	&	1.8	$\pm$	0.2	&	$<$	0.84		&	$<$0.15	&	198	&	$-$0.03	&	SF	&	32.6	&	3.8	\\
J	1622177$+$502220	&	3.3	&	4.5	$\pm$	0.5	&	4.1	$\pm$	1	&	$<$0.09	&	90.4	&	0.27	&	SF	&	34.3	&	5.9	\\
J	1642060$+$653507	&	1.08	&	0.54	$\pm$	0.04	&	$<$	0.7		&	0.14	&	62.2	&	$-$0.47	&	SF	&	35.4	&	2.04	\\
J	2100169$-$574323	&	1.17	&	1.7	$\pm$	0.1	&	0.15	$\pm$	0.02	&	$<$0.13	&	215	&	0.04	&	SF	&	41.1	&	0.82	\\
J	2158360$+$120220	&	1.77	&	2.0	$\pm$	0.2	&	1.6	$\pm$	0.3	&	$<$0.1	&	145	&	$-$0.18	&	SF	&	39.3	&	1.3	\\
J	2218300$+$041628	&	1.31	&	2.2	$\pm$	0.2	&	0.1	$\pm$	0.03	&	$<$0.13	&	171	&	$-$0.18	&	SF	&	36.3	&	1.9	\\


J	0053562$-$703804	&	5.41	&	$<$	2.3  	&			19  $\pm$ 6	 & 0.07	&	$<$	6.2	&	1.99	&	AGN	&	50.2	&	0.69\\
J	0054374$-$083407	&	9.14	&	$<$	2.2	&	$<$	27	&	$<$ 0.09	&	$<$	9	&	2.46	&	AGN	&	46.1	&	0.18	\\
J	0109450$-$033233	&	2.99	&	$<$	0.4	&	$<$	2.9	&	0.24	&	$<$	14	&	2.54	&	AGN	&	48.6	&	0.45	\\
J	1353157$+$634545	&	7.66	&	$<$	2.3	&	$<$	18	&	$<$ 0.07	&	$<$	7.5	&	1.65	&	AGN	&	45.3	&	0.14	\\
J	1407004$+$282714	&	5.84	&	$<$	2.0	&	$<$	15	&	$<$0.06	&	$<$	7.5	&	1.34	&	AGN	&	41.3	&	1.6	\\
J	2034041$-$814233	&	1.29	&	$<$	0.4	&	$<$	3.6	&	$<$ 0.07	&	$<$	9.1	&	2.61	&	AGN	&	45.5	&	0.28	\\
J	2204191$+$033351	&	5.99	&	$<$	0.97	&	$<$	7.4	&	$<$0.08	&	$<$	10	&	1.95	&	AGN	&	45.4	&	0.89	\\
NGC 34      	&	2.79	&	1.61	$\pm$	0.06	&	0.9	$\pm$	0.2	& 0.25	&	85.2	&	$-$0.07	&	SF	&	47.8	&	1.05		\\
MCG$-$02-01-051	&	2.57	&	2.72	$\pm$	0.03	&	2.3	$\pm$	0.2	& 0.24	&	230	&	0.25	&	SF	&	42.1	&	1.7		\\
NGC 232               	&	2.37	&	1.93	$\pm$	0.04	&	2.1	$\pm$	0.2	& 0.17	&	86.9	&	$-$0.32	&	SF	&	37.7	&	3.2		\\
MCG$-$03-04-014	&	4.29	&	5.62	$\pm$	0.07	&	4.4	$\pm$	0.5	& 0.15	&	165	&	0.03	&	SF	&	40.4	&	3.5		\\
ESO 244-G012 (N)   	&	2.43	&	2.18	$\pm$	0.04	&	2.2	$\pm$	0.2	& 0.17	&	158	&	$-$0.04	&	SF	&	42.4	&	1.4		\\
CGCG 436-030      	&	4.21	&	2.30	$\pm$	0.05	&	3.4	$\pm$	0.4	& 0.36	&	113	&	0.94	&	SF	&	50.2	&	1.2		\\
NGC 695               	&	4.39	&	8.5	$\pm$	0.3	&	2.8	$\pm$	0.9	& $<$0.12	&	212	&	$-$0.21	&	SF	&	37	&	5.4		\\
UGC 2238             	&	1.78	&	3.56	$\pm$	0.03	&	2.2	$\pm$	0.2	& 0.21	&	187	&	0.08	&	SF	&	36.1	&	2.8		\\
IRAS 03359$+$1523	&	2.97	&	1.17	$\pm$	0.06	&	1.6	$\pm$	0.3	& $<$0.16	&	167	&	0.52	&	SF	&	43.2	&	1.8		\\
UGC 2982              	&	1.3	&	1.89	$\pm$	0.02	&	1.0	$\pm$	0.1	&0.14	&	180	&	$-$0.17	&	SF	&	35.5	&	2.2		\\
NGC 1614	&	3.98	&	3.72	$\pm$	0.03	&	4.0	$\pm$	0.3	& 0.16	&	167	&	0.21	&	SF	&	45.9	&	1.45		\\
NGC 2388	&	1.52	&	1.33	$\pm$	0.02	&	0.9	$\pm$	0.1	& 0.17	&	108	&	$-$0.22	&	SF	&	39.9	&	0.84		\\
NGC 2623	&	2.87	&	0.66	$\pm$	0.02	&	1.5	$\pm$	0.1	& 0.50	&	67.9	&	0.15	&	SF	&	45.5	&	1.62		\\
MCG$+$08-18-013	&	1.73	&	1.33	$\pm$	0.04	&	1.4	$\pm$	0.2	& 0.20	&	118	&	$-$0.22	&	SF	&	39.8	&	1.7		\\
NGC 3110	&	1.68	&	2.26	$\pm$	0.03	&	1.2	$\pm$	0.1	& 0.089	&	138	&	$-$0.36	&	SF	&	35.7	&	2.4		\\ \hline
\end{tabular}
\end{center}
\end{table*}

 \begin{table*}
 \scriptsize
 \begin{center}
\begin{tabular}{lccccccccc}
 \hline
object	& $L_{\mathrm{IR}}$\footnotemark[$*$] & $L_{\mathrm{PAH3.3}}$\footnotemark[$\dagger$] &	$L_{\mathrm{Br\alpha}}$\footnotemark[$\ddagger$] & $\tau_{3.1}$\footnotemark[$\S$] &	$EW_{\mathrm{PAH3.3}}$\footnotemark[$\|$] &  $\it{\Gamma}$\footnotemark[$\#$] & spectral	& $T_{\mathrm{dust}}$\footnotemark[$\dagger\dagger$] & $M_{\mathrm{dust}}$\footnotemark[$\dagger\dagger$] \\
 &	 [$10^{11} \mathrm{L_{\odot}}$] & [$10^{8} \mathrm{L_{\odot}}$] &	[$10^7 \mathrm{L_{\odot}}$]	 & &	[nm]	& & type\footnotemark[$**$] & [K] & [$10^7 \:\mathrm{M_{\odot}}$] \\
\hline
ESO 264-G036				&		1.78		&			2.50	$\pm$	0.08		&		0.9	$\pm$	0.3				&	 0.11	 	&			78.7		&		$-$0.57		&		SF		&		34.3			&		3.5		\\
MCG$+$07-23-019		&		3.72		&			2.45	$\pm$	0.06		&		1.2	$\pm$	0.3				&	 0.15	 	&			140		&		$-$0.05		&		SF		&		38.5			&		4.3		\\
CGCG 011-076				&		1.96		&			1.41	$\pm$	0.05		&					$<$	0.82				&	 0.079	 	&			67		&		0.2			&		SF		&		38.9			&		1.9		\\
IC 2810 (NW)					&		3.33		&			0.74	$\pm$	0.07		&					$<$	0.51				&		$<$ 0.16 	&			86		&		$-$0.1		&	  SF		&		37.9			&		4.2		\\
IRAS 12224$-$0624		&		1.55		&				$<$	0.13				&					$<$	0.33			&	 0.20	   	&		$<$	40		&		$-$0.69		&		AGN		&		41.2			&		1.2		\\
ESO 507-G070				&		2.25		&			0.87	$\pm$	0.03		&		0.7	$\pm$	0.1				&	 0.17	 	&			77.6		&		$-$0.32		&		SF		&		43.4			&		1.42		\\
NGC 5010						&		2.17		&			2.01	$\pm$	0.07		&		0.4	$\pm$	0.1				&	 0.15	 	&			64.5		&		$-$0.45		&		SF		&		34.8			&		4.1		\\
IC 860							&		1.12		&			0.11	$\pm$	0.01		&					$<$	0.12				&	 $<$0.04	 	&			26.7		&		$-$0.99		&	Comp.		&		47.4			&		0.48		\\
VV 250 (SE)					&		4.95		&			2.69	$\pm$	0.10		&		3.3	$\pm$	0.3				&	 0.51	 	&			153		&		0.78			&		SF		&		45.5			&		2.1		\\
Arp 193 (UGC 8387)		&		3.76		&			2.24	$\pm$	0.04		&		2.9	$\pm$	0.2				&	 0.44	 	&			157		&		0.6			&		SF		&		40.4			&		3.7		\\
NGC 5104						&		1.29		&			1.14	$\pm$	0.03		&		1.6	$\pm$	0.1				&	 0.12	 	&			65.8		&		$-$0.61		&		SF		&		35.7			&		2.0		\\
MCG$-$03-34-064		&		1.29		&				$<$	0.14				&					$<$	1.1			&	 0.09	 	&		$<$	6.3		&		1.73			&		AGN		&		47.5			&		0.47		\\
NGC 5135						&		1.66		&			0.88	$\pm$	0.05		&		0.7	$\pm$	0.2				&	 0.17	 	&			61.6		&		0.62			&		SF		&		36.6			&		2.2		\\
NGC 5256 (SW)				&		2.75		&			3.22	$\pm$	0.07		&		1.7	$\pm$	0.3				&	 0.21	 	&			100		&		$-$0.26		&		SF		&		40.8			&		2.3		\\
ESO 221-IG010				&		1.00		&			1.45	$\pm$	0.02		&		0.3	$\pm$	0.1				&	 0.054	 	&			124		&		$-$0.38		&		SF		&		37.9			&		0.98		\\
CGCG 247-020				&		1.85		&			1.37	$\pm$	0.04		&		1.3	$\pm$	0.2				&	 0.20	 	&			106		&		$-$0.00		&		SF		&		40.6	 		&		1.54		\\
VV 340							&		3.98		&			5.03	$\pm$	0.10		&		3.7	$\pm$	0.6				&	 0.14	 	&			89.8		&		$-$0.17		&		SF		&		34.4			&		8.9		\\
CGCG 049-057				&		1.4		&			0.21	$\pm$	0.01		&		0.20	$\pm$	0.01				&	 0.11	 	&			70.9		&		$-$0.54		&		SF		&		40.2			&		1.36		\\
VV 705 (S$+$N)				&		6.57		&			4.37	$\pm$	0.09		&		4.9	$\pm$	0.5				&	 0.31	 	&			185		&		0.47			&		SF		&		45.1			&		3.3		\\
IRAS 15250$+$3609		&		9.46		&			1.34	$\pm$	0.19		&		3.1	$\pm$	0.8				&	 0.36	 	&			81.5		&		2.06			&		Comp.		&		52.6			&		2.2		\\
IRAS 15335$-$0513		&		1.65		&			0.83	$\pm$	0.05		&		0.7	$\pm$	0.2				&	 $<$0.05	 	&			55.8		&		$-$0.04		&		SF		&		37.7			&		2.2		\\
NGC 6090						&		2.95		&			3.80	$\pm$	0.05		&		3.8	$\pm$	0.3				&	 0.17	 	&			179		&		$-$0.01		&		SF		&		40.1			&		2.3		\\
CGCG 052-037				&		1.91		&			2.50	$\pm$	0.05		&		1.6	$\pm$	0.2				&	 0.17	 	&			130		&		$-$0.22		&		SF		&		38.6			&		2.3		\\
IRAS 17132$+$5313		&		7.35		&			6.24	$\pm$	0.13		&		5.1	$\pm$	0.6				&	 0.25	 	&			151		&		0.27			&		SF		&		42.0			&		5.2		\\
NGC 6621						&		1.53		&			1.29	$\pm$	0.04		&		1.1	$\pm$	0.2				&	 0.15	 	&			102		&		$-$0.24		&		SF		&		37.2			&		1.95		\\
ESO 339-G011				&		1.15		&			1.00	$\pm$	0.04		&		1.1	$\pm$	0.2				&	 0.14	 	&			88.0		&		0.24			&		SF		&		39.0			&		1.02		\\
NGC 7130						&		2.01		&			1.44	$\pm$	0.04		&		0.9	$\pm$	0.2				&	 0.094	 	&			70.6		&		$-$0.09		&		SF		&		39.1			&		2.0		\\
IC 5179							&		1.24		&			1.14	$\pm$	0.02		&		0.4	$\pm$	0.1				&	 0.21	 	&			174		&		$-$0.35		&		SF		&		36.0			&		1.8		\\
ESO 602-G025				&		1.8		&			3.00	$\pm$	0.06		&		2.0	$\pm$	0.3				&	 0.18	 	&			120		&		0.17			&		SF		&		37.1			&		2.4		\\
UGC 12150					&		1.8		&			1.21	$\pm$	0.04		&		1.2	$\pm$	0.2				&	 0.20	 	&			75.3		&		$-$0.38		&		SF		&		35.9			&		2.7		\\
NGC 7469						&		3.6		&			1.94	$\pm$	0.07		&		2.6	$\pm$	0.6				&	 0.066	 	&			32.0		&		0.55			&	Comp.		&		42.2			&		2.1		\\
CGCG 453-062				&		2.01		&			2.17	$\pm$	0.04		&		1.3	$\pm$	0.2				&	 0.092	 	&			104		&		$-$0.6			&		SF		&		38.3			&		2.2		\\
IC 5298							&		3.3		&			0.32	$\pm$	0.05		&		4.9	$\pm$	0.4				&	 0.46	 	&			10.9		&		0.88			&		Comp.		&		41.7			&		2.0		\\
NGC 7674						&		3.15		&			1.73	$\pm$	0.10		&					$<$	3.2        		&	 0.068	 	&			23.7		&		1.4			&	Comp.		&		39.1			&		2.1		\\
MCG$-$01-60-022		&		1.58		&			2.04	$\pm$	0.05		&		1.5	$\pm$	0.2				&	 0.16	 	&			153		&		$-$0.07		&		SF		&		39.3			&		1.29		\\
IRAS 23436$+$5257		&		3.14		&			1.30	$\pm$	0.10		&		1.2	$\pm$	0.4				&	 0.13	 	&			55.5		&		$-$0.02		&		SF		&		38.7			&		3.0		\\
NGC 7771						&		2.01		&			0.65	$\pm$	0.03		&		0.8	$\pm$	0.1				&	 0.18	 	&			60.0		&		$-$0.48		&		SF		&		35.3			&		3.5		\\
Mrk 331						&		2.58		&			2.19	$\pm$	0.03		&		1.8	$\pm$	0.2				&	 0.18	 	&			122		&		$-$0.24		&		SF		&		42.5			&		1.6		\\
NGC 4418						&		1.00		&		0.020	$\pm$0.001		&					$<$	0.050      		&	 $<$0.05	 	&			11.5		&		0.01			&	Comp.		&		56.3			&		0.16		\\
\hline
\multicolumn{10}{@{}l@{}}{\hbox to 0pt {\parbox{170mm}{\footnotesize 
  Notes. Same as table 3.
}}}
\end{tabular}
\end{center}
\end{table*}

 \begin{table*}
 \scriptsize
\begin{center}
\caption{Derived properties of the ULIRGs}
\begin{tabular}{lccccccccc}
 \hline
object	& $L_{\mathrm{IR}}$\footnotemark[$*$] & $L_{\mathrm{PAH3.3}}$\footnotemark[$\dagger$] &	$L_{\mathrm{Br\alpha}}$\footnotemark[$\ddagger$] & $\tau_{3.1}$\footnotemark[$\S$] &	$EW_{\mathrm{PAH3.3}}$\footnotemark[$\|$] &  $\it{\Gamma}$\footnotemark[$\#$] & spectral	& $T_{\mathrm{dust}}$\footnotemark[$\dagger\dagger$] & $M_{\mathrm{dust}}$\footnotemark[$\dagger\dagger$] \\
 &	 [$10^{11} \mathrm{L_{\odot}}$] & [$10^{8} \mathrm{L_{\odot}}$] &	[$10^7 \mathrm{L_{\odot}}$]	 & &	[nm]	& & type\footnotemark[$**$] & [K] & [$10^7 \:\mathrm{M_{\odot}}$] \\
\hline
	J	2248041$-$172829	&	11.2	&		6.7	$\pm$	0.7	&	$<$		 	2.3		& $<$ 0.11	&		35.8	&	1.24	&	Comp.	&	42.7		&	3.4	\\
   	J	0021533$-$791007	&	11.0	&	$<$	1.3			&		21  $\pm$	5	& 0.10	&	$<$	7.2	&	2.96	&	AGN	&	51.7			&	1.89		\\
	J	0356199$-$625139	&	11.3	&	$<$	3.0			&				$<$	30	& $<$ 0.06	&	$<$	7.3	&	2.29	&	AGN	&	44.3			&	2.78		\\
	J	0538503$-$440509	&	1077 	&			     ...				&				...				&	...	&		...			&	1.52	&	AGN	&	35.2			&	818	\\
	J	1040435$+$593409	&	17.9	&	$<$	6.6			&				...				& $<$ 0.05	&	$<$	5.8	&	1.59	&	AGN	&	42.4			&	5.99		\\ 
	IRAS 00188$-$0856		&		21.8		&		$<$	2.5        				&		2.8	$\pm$	0.4			&	 1.3	 	&		$<$	19		&		2.29  		&		AGN  		&		41.8			&		15.6		\\
	IRAS 04103$-$2838		&		14.7  	&	      5.4	$\pm$	0.4		&					$<$	1.1      		&	 0.27	 	&			63    		&		1.63  		&		Comp.		&		49.1			&		3.8		\\
	IRAS 10378$+$1108		&		18.8		&		$<$	2.1						&					$<$	2.0       		&	 $<$0.7 	 	&		$<$	41   		&		0.84  		&		AGN  		&		54.2			&		4.2		\\
	IRAS 10485$-$1447		&		15.1		&		$<$	1.5      					&					$<$	1.6      		&	 $<$0.7	 	&		$<$	38		&		1.00  		&		AGN   		&		48.5			&		5.0		\\
	IRAS 11095$-$0238		&		15.8		&			2.8	$\pm$	0.3		&					$<$	0.93     		&	 $<$0.6 	 	&			181   		&		2.32  		&		Comp.		&		55.1			&		3.2		\\
	IRAS 11582$+$3020		&		31.7		&			8.5	$\pm$	1.6		&		…             					&	 $<$1.3	 	&			153   		&		0.16  		&		SF    		&		41.7			&		23		\\
	IRAS 12032$+$1707		&		39.8		&		$<$	2.9						&		…              					&	 $<$1.1	 	&		$<$	36  		&		0.57  		&		AGN 		&		44.6			&		18		\\
	IRAS 12112$+$0305		&		18.6		&			3.7	$\pm$	0.2		&					$<$	0.46     		&	 0.47	 	&			128   		&		0.58  		&		SF    		&		43.9	 		&		12		\\
	IRAS 12127$-$1412		&		13.4		&		$<$	5.6        				&					$<$	3.9      		&	 0.39	 	&		$<$	11		&		2.17  		&		AGN   		&		57.3			&		2.1		\\
	IRAS 13335$-$2612		&		11.3		&			5.8	$\pm$	0.6		&					$<$	1.6      		&	 $<$0.8	 	&			 169  		&		0.54  		&		SF    		&		39.6  		&		11		\\
	IRAS 14348$-$1447		&		19.1		&			5.6	$\pm$	0.3		&					$<$	0.48     		&	 0.49	 	&			97   			&		0.59  		&		SF    		&		46.8			&		8.8		\\
	Arp 220                 		&		12.3		&			0.91	$\pm$	0.02		&		0.15	$\pm$	0.02			&	 0.29	 	&			53    		&		$-$0.07		&		SF    		&		45.5			&		6.5		\\
	IRAS 16090$-$0139		&		31.5		&			4.8	$\pm$	0.5		&					$<$	1.8      		&	 0.83	 	&			70  	 		&		1.34  		&		Comp.		&		41.3			&		26		\\
	IRAS 16300$+$1558		&		46.4		&			8.1  	$\pm$	1.5  		&		…              					&	 $<$0.6	 	&			75  			&		0.27  		&		SF    		&		41.4			&		39		\\
	IRAS 00091$-$0738		&		15.6		&			1.7	$\pm$	0.3		&					$<$	0.86      	&	 $<$ 0.8 	&			79  			&		0.48   		&		SF    		&		48.6			&		5.9		\\
	IRAS 00397$-$1312		&		84.7		&		$<$	21      				&		…              						&	 $<$ 0.10 	&		$<$	15		&		4.30    		&		AGN   		&		46.6			&		29.4		\\
	IRAS 01004$-$2237		&		18.0		&			2.6	$\pm$	0.3		&					$<$	1.1      		&	 0.33	 	&			51  			&		2.62  		&		Comp.		&		54.9			&		2.9		\\
	IRAS 08201$+$2801		&		18.0		&			3.0	$\pm$	0.6		&		2.9	$\pm$	0.8			&	 $<$1.3	 	&			87   			&		0.05  		&		SF    		&		43.1			&		11		\\
	IRAS 13509$+$0442		&		16.5		&			4.2	$\pm$	0.4		&					$<$	1.6      		&	 0.36	 	&			64  			&		0.63  		&		SF    		&		38.4			&		18		\\
	IRAS 14060$+$2919		&		11.2		&			7.4	$\pm$	0.4		&		1.3	$\pm$	0.3			&	 $<$0.43	 	&			224   		&		0.59  		&		SF    		&		39.6			&		11		\\
	IRAS 15206$+$3342		&		15.1		&			8.4	$\pm$	0.7		&		2.0	$\pm$	0.6			&	 $<$0.38	 	&			107   		&		1.76  		&		Comp.		&		45.9			&		6.0		\\
	IRAS 15225$+$2350		&		12.7		&		$<$	2.3      					&					$<$	2.2      		&	 0.57	 	&		$<$	35		&		0.68  		&	 AGN  		&		44.5			&		6.5		\\
    IRAS 16474$+$3430(S)	&		13.4		&			8.6 	$\pm$	0.5		&		1.3	$\pm$	0.4			&	 0.41	 	&			118   		&		0.92  		&		SF    		&		42.4			&		9.2		\\
	IRAS 20414$-$1651		&		16.5		&			1.8	$\pm$	0.3		&					$<$	0.66     		&	 0.84	 	&			73   			&		0.66  		&		SF    		&		43.4			&		9.0		\\
	IRAS 22206$-$2715		&		14.5		&			4.9	$\pm$	0.5		&					$<$	1.4      		&	 $<$0.51	 	&			131   		&		$-$0.11		&		SF    		&		41.6			&		12		\\
	IRAS 22491$-$1808		&		12.3		&			2.4	$\pm$	0.2		&					$<$	0.46     		&	 $<$0.43	 	&			117   		&		0.14  		&		SF    		&		53.3			&		3.1		\\
	IRAS 02021$-$2103		&		10.3		&			2.2	$\pm$	0.3		&					$<$	1.1      		&	 $<$0.39	 	&			46    		&		$-$0.46		&		SF    		&		43.7			&		5.4		\\
	IRAS 12018$+$1941		&		27.7		&			7.6	$\pm$	0.8		&		2.6	$\pm$	0.8			&	 $<$0.71	 	&			162   		&		1.79  		&		Comp.		&		47.2			&		10		\\
	IRAS 14197$+$0813		&		11.0		&		$<$	1.4						&					$<$	1.5      		&	 $<$ 0.6	 	&		$<$ 33  		&		$-$0.44		&		AGN    		&		39.5	  		&		10		\\
	IRAS 14485$-$2434		&		11.3		&			5.4	$\pm$	0.8		&		3.7	$\pm$	0.9			&	 0.82	 	&			73    		&		2.11  		&		Comp.		&		46.8	   		&		4.5		\\
	IRAS 08559$+$1053		&		15.2		&			7.6	$\pm$	0.8		&					$<$	2.9       		&	 $<$0.11	 	&			30   			&		1.41  		&		Comp.		&		37.4	  		&		19		\\
	IRAS 11223$-$1244		&		31.8		&		$<$	4.0        				&		…              					&	 0.61	 	&		$<$	40		&		0.32  		&		AGN   		&		39.7			&		32.7		\\
	IRAS 12072$-$0444		&		21.5		&		$<$	3.6						&					$<$	2.1      		&	 0.19	 	&		$<$	29  		&		1.54  		&		AGN  		&		47.4	  		&		7.5		\\
	IRAS 13305$-$1739		&		15.6		&		$<$	2.3	     				&					$<$	2.0      		&	 0.44	 	&		$<$	18		&		1.95  		&		AGN  		&		50.4			&		3.5		\\
	Mrk 273                		&		11.6		&			2.00	$\pm$	0.05		&		0.31	$\pm$	0.03			&	 0.25	 	&			47  			&		0.91  		&		SF    		&		47.9			&		4.6		\\
	IRAS 13443$+$0802		&		13.3		&			5.0	$\pm$	0.5		&					$<$	1.5      		&	 0.22	 	&			52   			&		0.43  		&		SF    		&		41.6			&		10		\\
	PKS 1345$+$12     		&		19.0		&		$<$	3.3  	    				&					$<$	2.1      		&	 $<$ 0.10	 	&		$<$	16		&		2.04  		&		AGN   		&		45.8			&		6.2		\\
	IRAS 15001$+$1433		&		25.0		&			5.4	$\pm$	0.9		&		4.1	$\pm$	1.0			&	 0.24	 	&			32  			&		1.74  		&		Comp.		&		45.4			&		12		\\
	IRAS 15130$-$1958		&		12.3		&		$<$	0.74      				&					$<$	1.7       		&	 $<$0.12	 	&		$<$	5.1		&		1.30  		&		AGN  		&		43.5			&		6.5		\\
	IRAS 16156$+$0146		&		11.0		&			5.0	$\pm$	0.7		&					$<$	2.1       		&	 0.96	 	&			51.3  		&		5.12  		&		Comp.		&		50.8			&		2.6		\\
	IRAS 22541$+$0833		&		17.0		&			2.5	$\pm$	0.5		&					$<$	2.6      		&	 $<$0.57	 	&			45  			&		0.16  		&		SF    		&		42.9			&		11		\\
	IRAS 23060$+$0505		&		28.7		&		$<$	20       				&					$<$	13         		&	 $<$ 0.05 	&		$<$	8.9		&		2.21  		&		AGN  		&		58.0			&		2.7		\\
	IRAS 23233$+$2817		&		10.2		&		$<$	2.4        				&					$<$	1.6       		&	 $<$0.35	 	&		$<$	34		&		1.45  		&		AGN  		&		37.9			&		10.8		\\
	IRAS 23389$+$0300		&		13.1		&			3.7	$\pm$	0.7		&					$<$	1.8      		&	 $<$0.69	 	&			97   			&		0.87  		&		SF    		&		48.8			&		4.0  		\\
	IRAS 00183$-$7111		&		75.7		&		$<$	9.7      					&		…             					&	 0.55	 	&		$<$	9.8		&		3.49  		&		AGN  		&		47.7			&		28.6		\\
	\hline
\end{tabular}
\end{center}
\end{table*}

 \begin{table*}
 \scriptsize
\begin{center}
\begin{tabular}{lccccccccc}
 \hline
object	& $L_{\mathrm{IR}}$\footnotemark[$*$] & $L_{\mathrm{PAH3.3}}$\footnotemark[$\dagger$] &	$L_{\mathrm{Br\alpha}}$\footnotemark[$\ddagger$] & $\tau_{3.1}$\footnotemark[$\S$] &	$EW_{\mathrm{PAH3.3}}$\footnotemark[$\|$] &  $\it{\Gamma}$\footnotemark[$\#$] & spectral	& $T_{\mathrm{dust}}$\footnotemark[$\dagger\dagger$] & $M_{\mathrm{dust}}$\footnotemark[$\dagger\dagger$] \\
 &	 [$10^{11} \mathrm{L_{\odot}}$] & [$10^{8} \mathrm{L_{\odot}}$] &	[$10^7 \mathrm{L_{\odot}}$]	 & &	[nm]	& & type\footnotemark[$**$] & [K] & [$10^7 \:\mathrm{M_{\odot}}$] \\
\hline
	IRAS 06035$-$7102	&	14.3	&		5.49	$\pm$	0.18	&	0.91	$\pm$	0.18		& 0.73	&		52  		&	3.68  	&	Comp.	&	45.2		&	7.0	\\
	IRAS 20100$-$4156	&	37.7	&		7.8	$\pm$	0.8	&				$<$	1.9      	& 1.8	&		121   	&	1.85  	&	Comp.	&	47.8		&	16	\\
	IRAS 20551$-$4250	&	10.0	&		2.36	$\pm$	0.09	&	0.62	$\pm$	0.06		& 0.48	&		98   		&	1.83  	&	Comp.	&	55.1		&	1.9	\\
	IRAS 23128$-$5919	&	10.0	&		3.85	$\pm$	0.08	&	0.58	$\pm$	0.04		& 0.18	&		114   	&	0.62  	&	SF    	&	47.0		&	3.7	\\
	Mrk 231                	&	29.8	&		11	$\pm$	2  	&				$<$	2.5      	& $<$ 0.05	&		12	  	&	1.10  	&	Comp.	&	48.9		&	7.9	\\
\hline
\multicolumn{10}{@{}l@{}}{\hbox to 0pt {\parbox{170mm}{\footnotesize 
  Notes. Same as table 3.
}}}
\end{tabular}
\end{center}
\end{table*}

 \begin{longtable}{ccccccc}
\caption{Derived properties of the galaxies with no IRAS data }
 \hline
object	 & $L_{\mathrm{PAH3.3}}$\footnotemark[$\dagger$] &	$L_{\mathrm{Br\alpha}}$\footnotemark[$\ddagger$] & $\tau_{3.1}$\footnotemark[$\S$] &	$EW_{\mathrm{PAH3.3}}$\footnotemark[$\|$] &  $\it{\Gamma}$\footnotemark[$\#$] & spectral	 \\
 &[$10^{8} \mathrm{L_{\odot}}$] &	[$10^7 \mathrm{L_{\odot}}$]	 & &	[nm]	& & type\footnotemark[$**$] \\
\hline
\endhead
\hline
\endfoot
\hline
\multicolumn{6}{@{}l@{}}{\hbox to 0pt {\parbox{150mm}{\footnotesize 
  Notes. Same as table 3.
}}}
\endlastfoot
J	0027496$-$573524	&	1.47	$\pm$	0.05	&	1.4	$\pm$	0.2	&	0.25	&	139		&	$-$0.32	$\pm$	0.05	&	SF	\\
J	1036319$+$022144	&	3.2	$\pm$	0.1	&	5.7	$\pm$	0.7	&	0.53	&	114		&	1.21	$\pm$	0.06	&	Comp.	\\
J	0018047$-$544909	&		$<$	2.3	&	$<$	6.9		&	$<$0.07	&	$<$	17	&	1.1	$\pm$	0.03	&	AGN	\\
J	0023139$-$535823	&		$<$	0.12	&	$<$	1.3		&	$<$0.07	&	$<$	5.6	&	2.45	$\pm$	0.04	&	AGN	\\
J	0043041$-$545826	&		$<$	12	&	$<$	47		&	$<$ 0.07	&	$<$	17	&	1.87	$\pm$	0.04	&	AGN	\\
J	0047001$-$174705	&		$<$	5.4	&	$<$	63		&	$<$ 0.07	&	$<$	8.3	&	1.84	$\pm$	0.04	&	AGN	\\
J	0842153$+$402532	&		$<$	2  	&	$<$	7.3		&	$<$ 0.07	&	$<$	17	&	2.02	$\pm$	0.02	&	AGN	\\
J	0127539$+$284750	&		$<$	0.73	&	$<$	3.9		&	0.18	&	$<$	21	&	3.91	$\pm$	0.03	&	AGN	\\
J	0228152$-$405716	&		…		&	…			&	…	&	…		&	2.09	$\pm$	0.05	&	AGN	\\
J	0426007$-$571203	&		$<$	4.5	&	17	$\pm$	4	&	$<$ 0.08	&	$<$	19	&	1.31	$\pm$	0.03	&	AGN	\\
J	0522580$-$362731	&		$<$	2.4	&	$<$	6.2		&	$<$ 0.06	&	$<$	18	&	0.78	$\pm$	0.02	&	AGN	\\
J	1050348$+$801149	&		$<$	8.1	&	$<$	26		&	$<$ 0.06	&	$<$	15	&	1.08	$\pm$	0.02	&	AGN	\\
J	1230499$+$011521	&		$<$	9	&	$<$	48		&	$<$ 0.06	&	$<$	16	&	1.19	$\pm$	0.02	&	AGN	\\
J	1300053$+$163214	&		$<$	3.9	&	$<$	14		&	$<$ 0.07	&	$<$	17	&	1.89	$\pm$	0.02	&	AGN	\\
J	1628040$+$514631	&		$<$	1.2	&	$<$	5.2		&	$<$ 0.07	&	$<$	13	&	1.39	$\pm$	0.02	&	AGN	\\
J	1637082$+$684253	&		$<$	0.0069	&	$<$	0.31		&	$<$ 0.08	&	$<$	1.4	&	1.7	$\pm$	0.03	&	AGN	\\
J	2030417$-$753243	&		$<$	12	&	$<$	59		&	$<$ 0.05	&	$<$	14	&	1.68	$\pm$	0.02	&	AGN	\\
J	2141102$-$264550	&		$<$	13	&	$<$	101		&	$<$ 0.05	&	$<$	10	&	1.61	$\pm$	0.02	&	AGN	\\
J	2217122$+$141422	&		$<$	3.4	&	9.5	$\pm$	3	&	$<$ 0.02	&	$<$	15	&	0.68	$\pm$	0.02	&	AGN	\\
J	2223495$-$020612	&		$<$	1.8	&	$<$	8.7		&	$<$ 0.04	&	$<$	11	&	1.85	$\pm$	0.02	&	AGN	\\
J	2245112$-$603357	&		$<$	2.4	&	2.6	$\pm$	0.8	&	$<$ 0.06	&	$<$	5.3	&	0.94	$\pm$	0.02	&	AGN	\\
J	2319135$+$253424	&		$<$	6.4	&	$<$	28		&	$<$ 0.09	&	$<$	20	&	2.73	$\pm$	0.03	&	AGN	\\
\end{longtable}

\begin{figure}
   \begin{center}
      \FigureFile(80mm,50mm){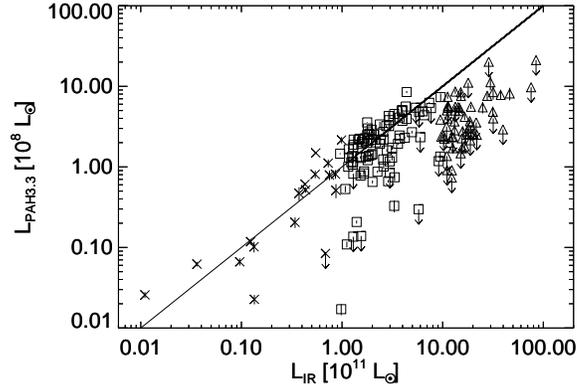}
   \end{center}
   \caption{$L_{\mathrm{PAH3.3}}$ plotted against $L_{\mathrm{IR}}$ for all the sample galaxies i.e. the galaxies with and without AGN signatures. The crosses correspond to the IRGs, while the squares and triangles are the LIRGs and ULIRGs, respectively. The solid line shows $L_{\mathrm{PAH3.3}}/L_{\mathrm{IR}}=10^{-3}$, a ratio typical of starburst galaxies \citep{mou 1990, i 2002}.}\label{3}
\end{figure}

\begin{longtable}{ccccc}
\caption{Classification of the sample galaxies}
\hline 
criteria	&	star-forming	&	star-forming + AGN  & AGN 	& total \\		
\hline
\endhead
\hline
\endfoot
\hline
\multicolumn{5}{@{}l@{}}{\hbox to 0pt {\parbox{140mm}{\footnotesize 
\footnotemark[$*$]  Signatures for AGN activity based on $EW_{\mathrm{PAH3.3}}$ and $\it{\Gamma}$ (see text for details). 
}}}
\endlastfoot
PAH 3.3 $\mu$m emission feature & Yes & Yes & No  & ... \\
AGN signature\footnotemark[$*$] & No & Yes & Yes  & ... \\ \hline
sample & & & &  \\ \hline
IRGs ($L_{\mathrm{IR}}<10^{11}\:\mathrm{L_{\odot}}$)	&	13 &	4  & 1 & 18	\\
LIRGs ($L_{\mathrm{IR}}=10^{11}-10^{12}\:\mathrm{L_{\odot}}$)	&	 67	&  13	& 9 &  89	\\		
ULIRGs ($L_{\mathrm{IR}}>10^{12}\:\mathrm{L_{\odot}}$) &  20 & 15 & 20 & 55 \\
galaxies with no IRAS data & 1 & 1 & 20 & 22 \\
total &  101 &  33 & 50 & 184 \\
\end{longtable}

\begin{figure}
   \begin{center}
      \FigureFile(80mm,50mm){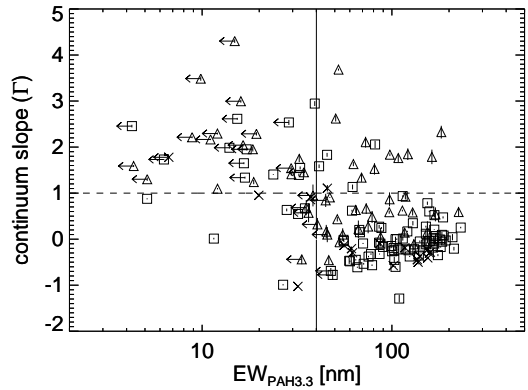}
   \end{center}
   \caption{Continuum slope ($\it{\Gamma}$, a power-law index) vs. the rest-frame equivalent width of the PAH 3.3 $\mu$m emission ($EW_{\mathrm{PAH3.3}}$) for the IRGs (crosses) as well as the LIRGs (squares) and ULIRGs (triangles). The solid line represents $EW_{\mathrm{PAH3.3}}=40$ nm, while the dashed line represents $\it{\Gamma} =$1. The galaxies on the bottom right area possess no AGN signatures, while the galaxies on the other areas show AGN signatures.}\label{4}
\end{figure}

\begin{figure}
   \begin{center}
      \FigureFile(80mm,50mm){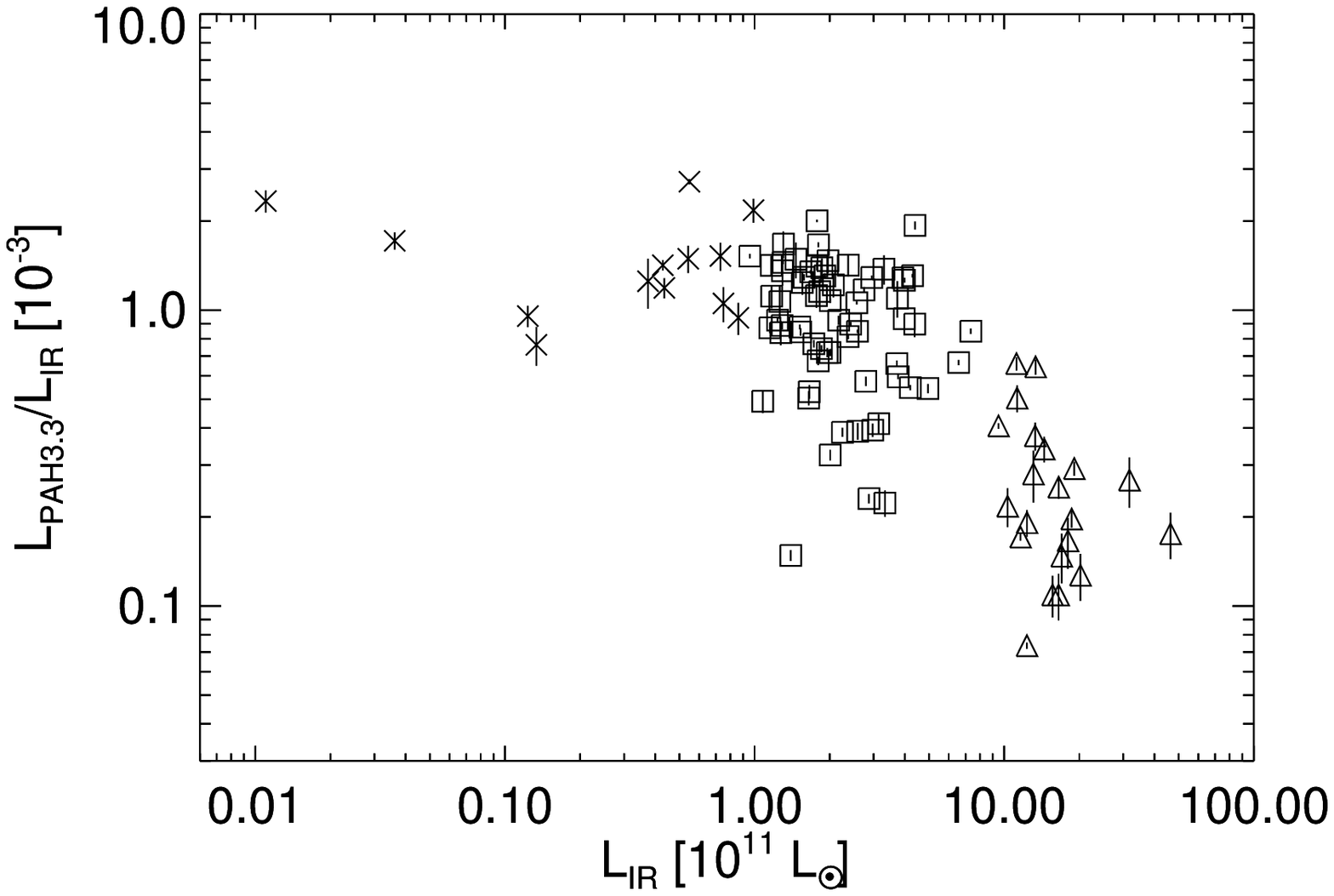}
   \end{center}
   \caption{$L_{\mathrm{PAH3.3}}/L_{\mathrm{IR}}$ plotted against $L_{\mathrm{IR}}$ for the galaxies with no AGN signatures: the IRGs (crosses), the LIRGs (squares), and the ULIRGs (triangles).}\label{5}
\end{figure}

\begin{figure*}
   \begin{center}
      \FigureFile(80mm,50mm){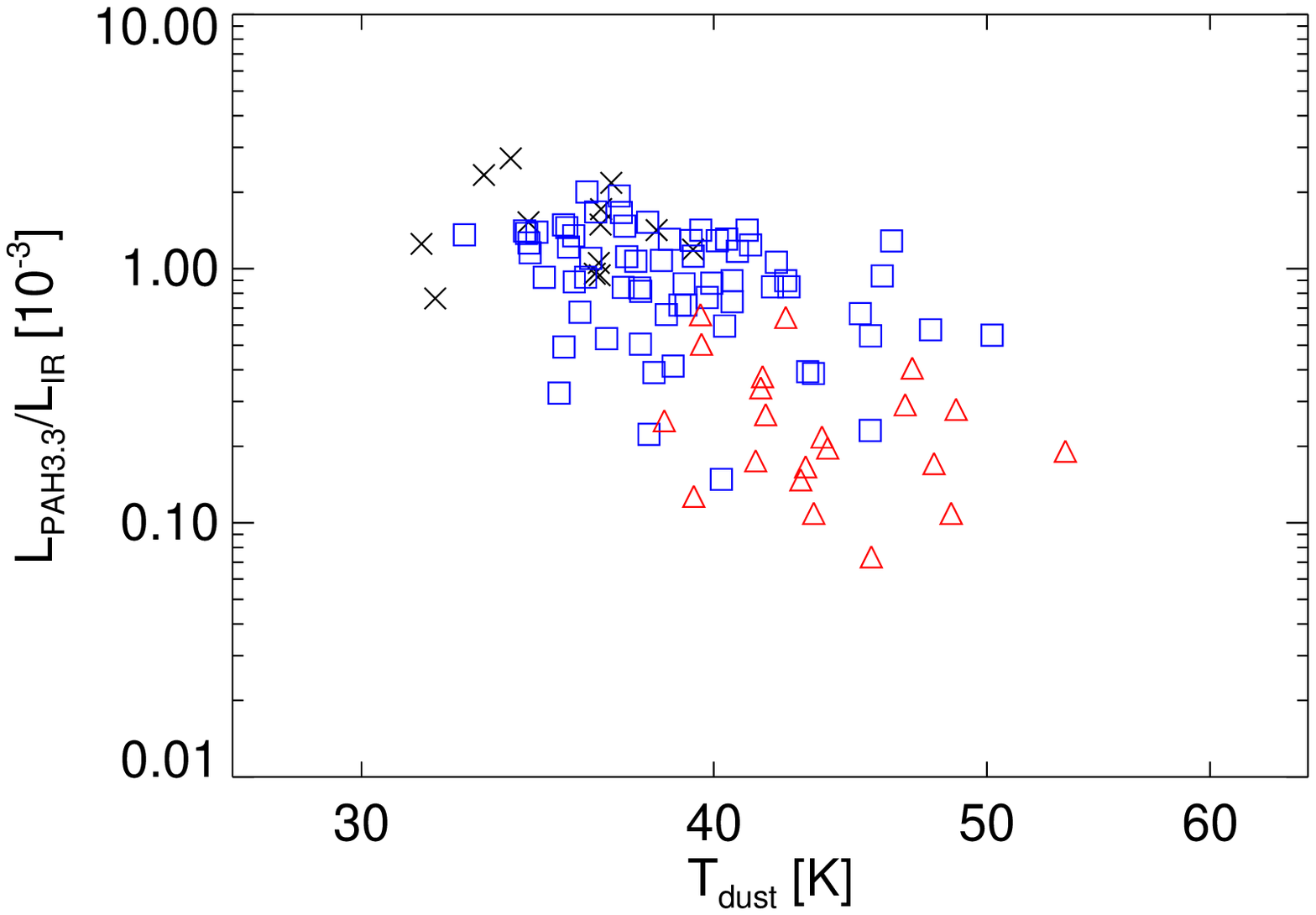}
      \FigureFile(80mm,50mm){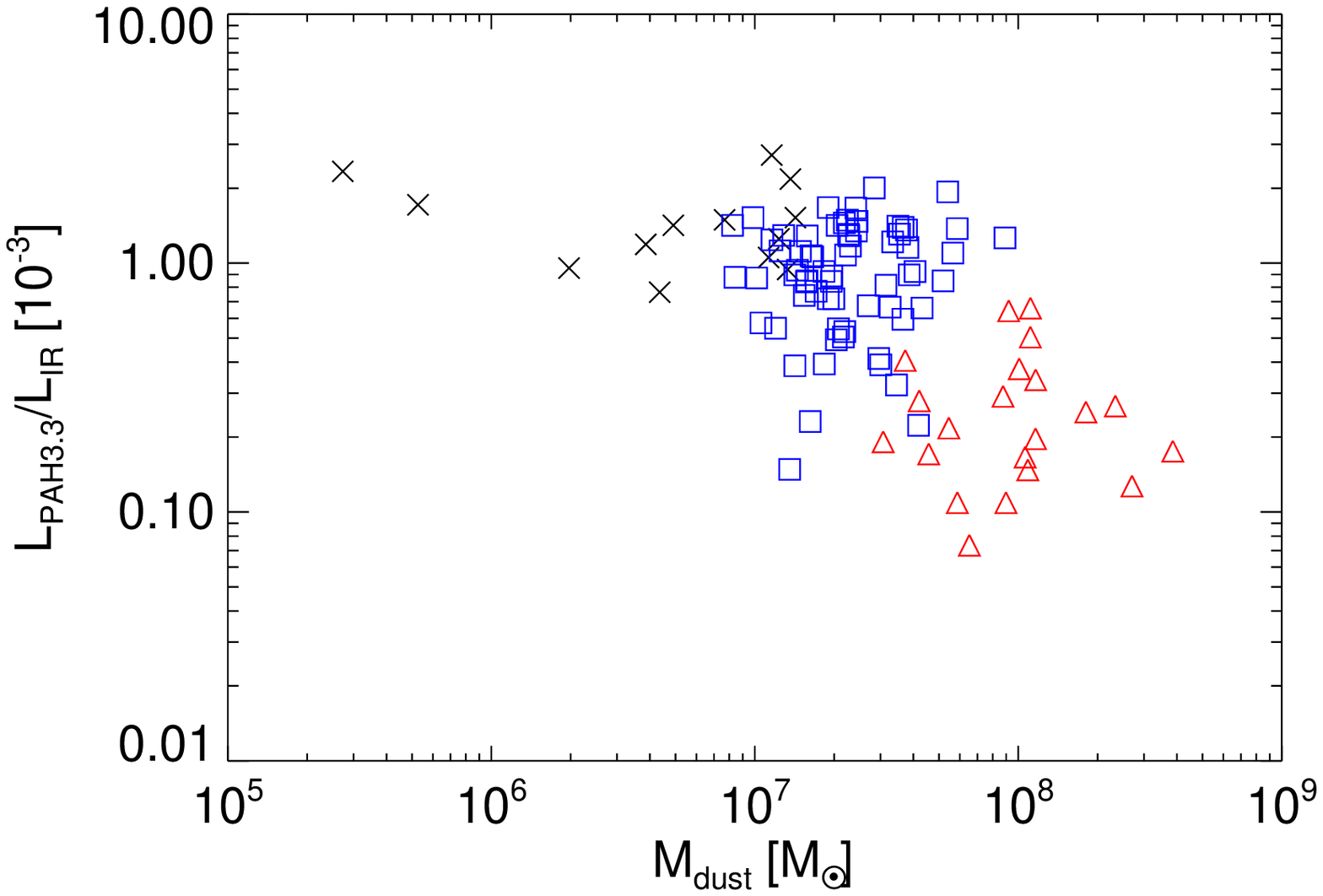}
   \end{center}
   \caption{$L_{\mathrm{PAH3.3}}/L_{\mathrm{IR}}$ plotted against the dust temperature (left) and the dust mass (right) for the galaxies with no AGN signatures: the IRGs (crosses), the LIRGs (blue squares), and the ULIRGs (red triangles).}\label{6}
\end{figure*}

\begin{figure}
   \begin{center}
      \FigureFile(80mm,50mm){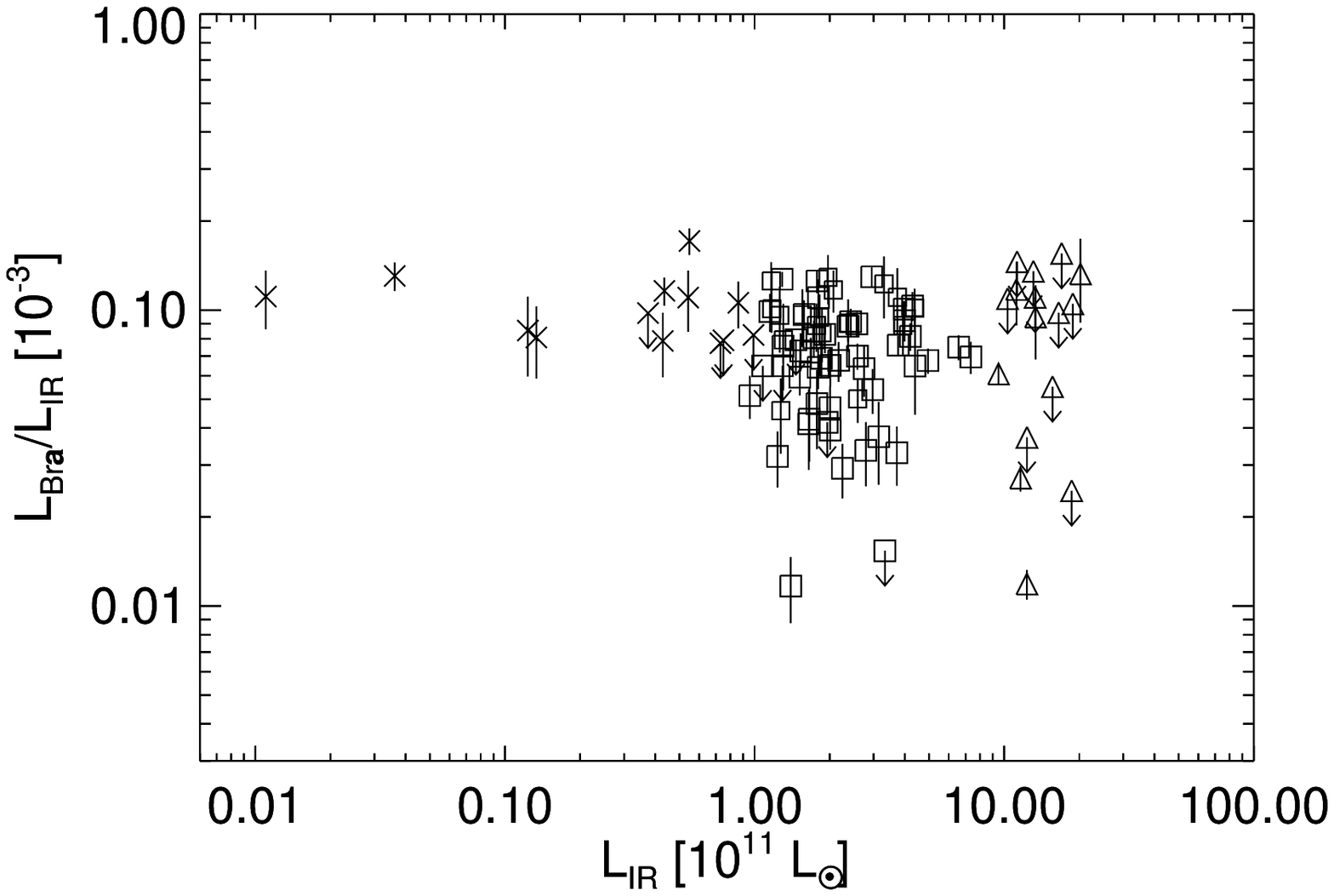}
   \end{center}
   \caption{$L_{\mathrm{Br\alpha}}/L_{\mathrm{IR}}$ plotted against $L_{\mathrm{IR}}$ for the galaxies with no AGN signatures: the IRGs (crosses), the LIRGs (squares), and the ULIRGs (triangles).}\label{7}
\end{figure}

\begin{figure}
   \begin{center}
      \FigureFile(80mm,50mm){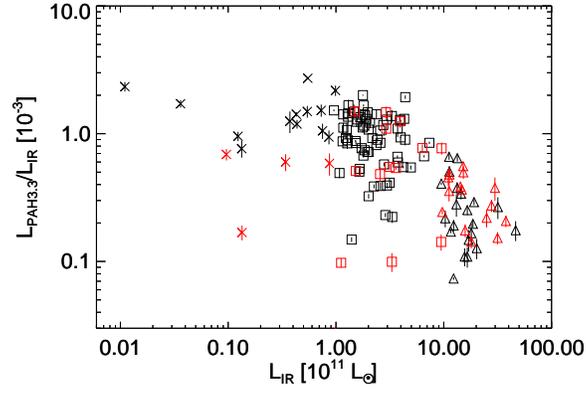}
   \end{center}
   \caption{Same as Fig. \ref{5}, but the data points for the galaxies showing signatures of AGN activity are also plotted in the red symbols. }\label{8}
\end{figure}

\begin{figure}
   \begin{center}
      \FigureFile(80mm,50mm){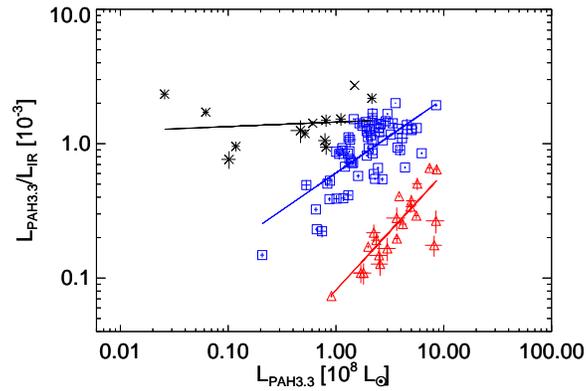}
   \end{center}
   \caption{$L_{\mathrm{PAH3.3}}/L_{\mathrm{IR}}$ plotted against $L_{\mathrm{PAH3.3}}$ for the galaxies with no AGN signatures: the IRGs (crosses), the LIRGs (squares), and the ULIRGs (triangles). The black, blue and red lines show the logarithmic regression lines for the IRGs, LIRGs and ULIRGs, respectively.}\label{9}
\end{figure}


\begin{thebibliography}{32}

\bibitem[Brandl et al.(2006)]{b 2006} Brandl,~B.~R., \etal\ 2006, \apj, 653, 1129 
\bibitem[Casey et al.(2012)]{ca 2012} Casey, C. M., \etal\ 2012, \apj, 761, 140 
\bibitem[Clements et al.(1996)]{c 1996} Clements,~D.~L., Sutherland,~W.~J., McMahon,~R.~G., \& Saunders,~W. 1996, \mnras, 279, 477 
\bibitem[Galliano et al.(2008)]{g 2008} Galliano,~F., Madden,~S.~C., Tielens,~A.~G.~G.~M., Peeters,~E., \& Jones,~A.~P.\ 2008, \apj, 679, 310 
\bibitem[Genzel et al.(1998)]{ge 1998} Genzel, R., \etal\ 1998, \apj, 498, 579 
\bibitem[Goto et al.(2010)]{go 2010} Goto,~T., \etal\ 2010, \aap, 514, A7 
\bibitem[Haas et al.(2003)]{h 2003} Haas,~M., \etal\ 2003, \aap, 402, 87 
\bibitem[Imanishi \& Dudley(2000)]{i 2000} Imanishi,~M., \& Dudley,~C.~C.\ 2000, \apj, 545, 701 
\bibitem[Imanishi(2002)]{i 2002} Imanishi,~M. 2002, \apj, 569, 44 
\bibitem[Imanishi et al.(2008)]{i 2008} Imanishi,~M., Nakagawa,~T., Ohyama,~Y., Shirahata,~M., Wada,~T., Onaka,~T., \& Oi,~N. 2008, \pasj, 60, S489 
\bibitem[Imanishi et al.(2010)]{i 2010} Imanishi,~M., Nakagawa,~T., Shirahata,~M., Ohyama,~Y., \& Onaka,~T. 2010, \apj, 721, 1233 
\bibitem[Ishihara et al.(2010)]{is 2010} Ishihara,~D., \etal\ 2010, \aap, 514, A1 
\bibitem[Kaneda et al.(2005)]{ka 2005} Kaneda,~H., Onaka,~H., \& Sakon,~I. 2005, \apj, 632, L83 
\bibitem[Kaneda et al.(2008)]{ka 2008} Kaneda,~H., Onaka,~T., Sakon,~T., Kitayama,~T., Okada,~Y., \& Suzuki,~T. 2008, \apj, 684, L270 
\bibitem[Kim et al.(2012)]{k 2012}Kim,~J.~H., \etal\ 2012 \apj, 760, 120 
\bibitem[Le Floc'h et al.(2005)]{lf 2005}Le Floc'h E., et al.\ 2005, \apj, 632, 1691 
\bibitem[Lee et al.(2012)]{lee 2012} Lee, J. C., Hwang, H. S., Lee, M. G., Kim, M., Lee, J. H. 2012, \apj, 756, 95 
\bibitem[Li \& Draine(2001)]{l 2001} Li,~A., \& Draine,~B.~T. 2001, \apj, 554, 778 
\bibitem[Moorwood(1986)]{moo 1986} Moorwood,~A.~F.~M.\ 1986, \aap, 166, 4 
\bibitem[Moshir et al.(1992)]{mos 1992}Moshir, M., Kopman, G., \& Conrow, T. A. O. 1992, in IRAS Faint Source Survey, Explanatory supplement version 2, ed. M. Moshir, G. Kopman, T. A. O. Conrow (Pasadena: Infrared Processing and Analysis Center, California Institute of Technology) 
\bibitem[Mouri et al.(1990)]{mou 1990} Mouri,~H., Kawara,~K., Takeushi,~Y., \& Nishida,~M. 1990, \apj, 356, L39 
\bibitem[Murakami et al.(2007)]{mu 2007} Murakami, H., et al. 2007, \pasj, 59, 369 
\bibitem[Ohyama et al.(2007)]{oh 2007} Ohyama, Y., et al. 2007, \pasj, 59, 411 
\bibitem[Onaka et al.(2007)]{on 2007}Onaka, T., et al. 2007, \pasj, 59, 401 
\bibitem[Oyabu et al.(2011)]{o 2011}Oyabu, S., et al. 2011, \aap, 529, 122 
\bibitem[Peeters et al.(2004)]{p 2004} Peeters,~E., Spoon,~H.~W.~W., \& Tielens, A.~G.~G.~M.\ 2004, \apj, 613, 986 
\bibitem[Risaliti et al.(2006)]{r 2006}Risaliti, G., et al. 2006 \mnras, 365, 303 
\bibitem[Rodighiero et al.(2010)]{ro 2010}Rodighiero, G., et al.\ 2010, \aap, 515, A8 
\bibitem[Rodr{\'{\i}}guez-Ardila \& Viegas(2003)]{rv 2003}Rodr{\'{\i}}guez-Ardila, A., \& Viegas, S.~M.\ 2003, \mnras, 340, L33 
\bibitem[Sajina et al.(2009)]{sa 2009} Sajina, A., Spoon, H., Yan, L., Imanishi, M., Fadda, D., \& Elitzur, M. 2009, \apj, 703, 270 
\bibitem[Sanders \& Mirabel(1996)]{s 1996}Sanders, D. B., \& Mirabel, I. F. 1996, ARA\&A, 34, 749 
\bibitem[Schweitzer et al.(2006)]{sc 2006} Schweitzer, M., \etal\ 2006, \apj, 649, 79 
\bibitem[Siana et al.(2009)]{si 2009} Siana, B., \etal\ 2009, \apj, 698, 1273 
\bibitem[Smith et al.(2007)]{sm 2007}Smith, J. D. T., et al.\ 2007, \apj, 656, 770 
\bibitem[Voit(1992)]{v 1992}Voit, G. M., 1992, \mnras, 258, 841 
\bibitem[Woo et al.(2012)]{w 2012}Woo, J. H., Kim, J. H., Imanishi, M., \& Park, D.\ 2012, \aj, 143, 49 
\bibitem[Yamagishi et al.(2011)]{y 2011}Yamagishi, M., Kaneda,~H., Ishihara,~D., Oyabu,~S., Onaka,~T., Simonishi,~T., \& Suzuki,~T. 2011, \apj, 731, L20 

\end{thebibliography}
\end{document}